\documentclass{aa}  

\usepackage{graphicx}
\usepackage{txfonts}
\usepackage[breaklinks=True, colorlinks, citecolor=blue]{hyperref}
\usepackage{bbold}

\newcommand{\dd}{\mathrm{d}}
\newcommand{\pof}[1]{p\left(#1\right)}
\newcommand{\pofc}[2]{p\left(#1\mathrel{}\middle|\mathrel{}#2\right)}
\newcommand{\tmodel}{\vec\Theta_{\rm model}}
\newcommand{\tsel}{\vec\Theta_{\rm sel}}
\newcommand{\tbin}{\vec\Theta_{\rm bin}}
\newcommand{\tmes}{\vec\Theta_{\rm meas}}
\newcommand{\tselr}{\vec\Theta_{{\rm sel}, R}}
\newcommand{\tselt}{\vec\Theta_{{\rm sel}, T}}
\newcommand{\ftroiscents}{\texttt{F\_300kpc}}
\newcommand{\fluxerassco}{\texttt{F500\_0520}}

\newcommand{\zetasptpol}{\zeta_{\rm SPTpol}}
\newcommand{\zetasptecs}{\zeta_{\rm ECS}}
\newcommand{\zspt}{z_{\rm SPT}}
\newcommand{\zefeds}{z_{\rm eFEDS}}
\newcommand{\zerass}{z_{\lambda}}
\newcommand{\isdet}[1]{I_{\rm {#1}}}
\newcommand{\sbright}{\Sigma}
\newcommand{\sbrightmod}{\eta}
\newcommand{\normaldist}[2]{\mathcal{N}\left(#1\mathrel{};\mathrel{}#2\right)}
\newcommand{\fluxunit}{erg\,s$^{-1}$\,cm$^{-2}$}
\newcommand{\luminbahar}{\texttt{Lx500}}
\newcommand{\extlike}{\mathcal{L}_{\rm ext}}
\newcommand{\detlike}{\mathcal{L}_{\rm det}}

\usepackage{xcolor}

\begin{document}

   \title{The SRG/eROSITA All-Sky Survey}

   \subtitle{X-ray selection function models for the eRASS1 galaxy cluster cosmology}

   \author{N.~Clerc\inst{1}\fnmsep\thanks{email: \url{nicolas.clerc@irap.omp.eu}},
          J.~Comparat\inst{2},
          R.~Seppi\inst{2},
          E.~Artis\inst{2},
          Y.~E.~Bahar\inst{2},
          F.~Balzer\inst{2},
          E.~Bulbul\inst{2},
          T.~Dauser\inst{3},
          C.~Garrel\inst{2},
          V.~Ghirardini\inst{2},
          S.~Grandis\inst{4},
          C.~Kirsch\inst{3},
          M.~Kluge\inst{2},
          A.~Liu\inst{2},
          F.~Pacaud\inst{5},
          M.~E.~Ramos-Ceja\inst{2},
          T.~H.~Reiprich\inst{5},
          J.~Sanders\inst{2},
          J.~Wilms\inst{3},
          X.~Zhang\inst{2}
          }

   \institute{IRAP, CNRS, UPS, CNES, 14 avenue Edouard Belin, 31400 Toulouse, France\\
              \email{nicolas.clerc@irap.omp.eu}
            \and
            Max Planck Institute for Extraterrestrial Physics, Giessenbachstrasse 1, 85748 Garching, Germany
            \and
            Dr.\ Karl-Remeis-Sternwarte and ECAP, Friedrich-Alexander-Universit\"at Erlangen-N\"urnberg, Sternwartstr. 7, 96049 Bamberg, Germany 
             \and
            Universit\"at Innsbruck,  Institut f\"ur Astro- und Teilchenphysik, Technikerstr. 25/8, 6020 Innsbruck, Austria
            \and
            Argelander-Institut f\"ur Astronomie (AIfA), Universit\"at Bonn, Auf dem H\"ugel 71, 53121 Bonn, Germany
}

   \date{Received -; accepted -}

% \abstract{}{}{}{}{} 
% 5 {} token are mandatory
 
  \abstract
   {}
   {Characterising galaxy cluster populations from catalog of sources selected in astronomical surveys requires knowledge of sample incompleteness, known as selection function. The first All-Sky Survey (eRASS1) by eROSITA onboard Spectrum Roentgen Gamma (SRG) has enabled the collection of large samples of galaxy clusters detected in the soft X-ray band over the Western Galactic hemisphere. The driving goal consists in constraining cosmological parameters, which puts stringent requirements on accuracy, flexibility and explainability of the selection function models.}
   {We use a large set of mock observations of the eRASS1 survey and we process simulated data identically to the real eRASS1 events. We match detected sources to simulated clusters and we associate detections to intrinsic cluster properties. We train a series of models to build selection functions depending only on observable surface brightness data. We develop a second series of models relying on global cluster characteristics such as X-ray luminosity, flux, and expected instrumental count-rate as well as on morphological properties. We validate our models using our simulations and we rank them according to selected performance metrics. We validate the models with datasets of clusters detected in X-rays and via the Sunyaev-Zeldovich effect. We present the complete Bayesian population modelling framework developed for this purpose.}
   {Our results reveal the surface brightness characteristics most relevant to cluster selection in the eRASS1 sample, in particular the ambiguous role of central surface brightness at the scale of the instrument resolution. We have produced a series of user-friendly selection function models and demonstrated their validity and their limitations. Our selection function for bright sources reproduces well the catalog matches with external datasets. We discuss potential inconsistencies in the selection models at low signal-to-noise revealed by comparison with a deep X-ray sample acquired by eROSITA during its performance verification phase.}
   {Detailed modelling of the eRASS1 galaxy cluster selection function is made possible by reformulating selection into a classification problem. Our models are used in the first eRASS1 cosmological analysis and in sample studies of eRASS1 cluster and groups. These models are crucial for science with eROSITA cluster samples and our new methods pave the way for further investigation of faint cluster selection effects.}


\titlerunning{eRASS1 galaxy cluster selection function}
\authorrunning{N.~Clerc et al.}

   \maketitle

%
%-------------------------------------------------------------------

\section{Introduction}

Galaxy clusters sit at the most massive nodes of the cosmic web. They form last in the cosmic evolution by accreting groups and smaller structures. Their distribution is sensitive to the underlying cosmological model and for this reason they are recognised as a key cosmological probe \citep{2023ClercFinoguenov}. Being rare objects, their census requires surveys spanning a large fraction of the sky with sensitive instrumentation. The diffuse X-ray emitting gas, filling the Megaparsec-wide intracluster medium (ICM), signposts virialized halos and enables their discovery up to large distances.
In the context of X-ray cluster studies, accurate knowledge of a selection function has been required for cosmological abundance studies \citep[e.g.][]{2015BoehringerChon, 2015Mantz, 2020Finoguenov, 2022Garrel}, cluster clustering studies \citep[e.g.][]{2018Marulli, 2021Lindholm}, scaling relation studies \citep[e.g.][]{2006Pacaud, 2019Mantz, 2022Bahar}, investigation of extreme objects \citep[e.g.][]{2012Hoyle}.

The eROSITA instrument onboard the Spectrum Roentgen Gamma Mission \citep[SRG/eROSITA,][]{2021Predehl} has surveyed the entire sky during its first six months of operations \citep{2024Merloni}, collecting enough photons to discover several thousands of galaxy clusters in the Western Galactic hemisphere \citep{2024Bulbul}. Those extended sources are identified in multi-band optical surveys and their redshift is measured, hence providing a distance to us observers. The first catalogs of clusters discovered in the first eROSITA All-Sky Survey (eRASS1) Western Galactic hemisphere \citep{2024Bulbul, 2024Kluge} are the support of detailed individual cluster studies \citep[e.g.][]{2023Veronica, 2023Liu} and population studies revealing for the first time the large-scale evolution of clusters and groups up to $z \simeq 1$ and beyond. Among them the cosmological analyses particularly stand out, being a driver of the eROSITA mission design and of the construction of cluster catalogs. The first cosmological results based on cluster number counts are presented in \citet{2024Ghirardini, 2024Artis}. In this article, we explain how the selection function model supporting these results is constructed. In fact, virtually any cluster population study based on the published eRASS1 catalogs needs to account for incompleteness at some stage. It is indeed inefficient, if not ill-defined, to require complete samples for population analyses \citep{2021Rix}; modelling incompleteness is in general more fruitful and in fact, rather inevitable. The current blooming of large astronomical surveys fosters development of accurate selection models with powerful statistical methods. The recent Gaia survey is an illustrative instance where selection functions for the parent catalog \citep{2020BoubertEverall} and for its subsets \citep{2021Rix, 2021BoubertEverall} are modelled using detailed accounting for the magnitudes, position and parallaxes (and uncertainties) of the stars.

This paper is organised as follows. We first present detailed motivation for this work in Sect.~\ref{sect:motivation}, where we also introduce our notations.
We present the twin eRASS1 simulations in Sect.~\ref{sect:datasets}, in particular we discuss our procedure matching detections and simulated clusters. We describe a first class of selection function models relying uniquely on X-ray counts profiles in Sect.~\ref{sect:logistic_profile}, highlighting some salient features of cluster selection in the observable space. In Sect.~\ref{sect:gp_models} we introduce a second class of models relying on intermediate variables.
We show the outcome of a validation of our intermediate models using external catalogs (X-ray and millimeter-band detections) in Sect.~\ref{sect:external_validation}. We discuss our results in Sect.~\ref{sect:discussion} and summarize our findings in Sect.~\ref{sect:conclusion}.

Unless stated differently, we assume a flat $\Lambda$CDM cosmological model with parameters from \citet{planckparam}. We use $\ln$ for natural logarithm, base 10 logarithm writes $\log_{10}$. The notation $\normaldist{\mu}{\sigma}$ indicates a normal distribution centred on $\mu$ with standard deviation $\sigma$. We use symbol $\arcsec$ for arcsecond units.

%--------------------------------------------------------------------

\section{Motivation}\label{sect:motivation}
Let us follow an illustrative example based on cluster count cosmology studies, although this demonstration may be generalized to other kinds of population analyses.
Most cosmology analyses involve a numerical likelihood:
\begin{equation}\nonumber
L({\rm model}) \propto \pofc{\rm data}{\rm model}
\end{equation}
This quantity represents the probability that a set of observed data derives from a certain model. One assumes the model to be dependent on a set of parameters $\tmodel$ fully describing it (see Table~\ref{table:summary_notations}):
\begin{equation}\label{eq:likelihood}
    L(\tmodel) \propto \pofc{\rm data}{\tmodel}
\end{equation}

Posterior distributions $\pofc{\tmodel}{\rm data}$ are final products of cosmology analyses, obtained by application of the Bayes theorem involving the likelihood (Eq.~\ref{eq:likelihood}). Most often the required integrals are computed by repeated samplings of the parameter space, and by formulating as many queries to the likelihood function value $L$. A practical consequence is that likelihood evaluation must be computationally efficient.

In experiments involving counts of galaxy clusters -- which include cluster abundance and cluster clustering studies -- objects are grouped in bins of one or several measured quantities. The bins are drawn in a predefined parameter space $\tbin$: measured mass, redshift, flux, pairwise distances, etc. The likelihood functions combine observed numbers $\hat{N}_j$ and model-predicted numbers $n_j$ by means of a statistics ($j$ is an index for the bins):
\begin{equation}\nonumber
L \equiv L \left( \left\{\hat{N}_j\right\} , \left\{n_j\right\}, \dotsc \right)
\end{equation}
A famous example is the Poisson log-likelihood \citep[see also][]{1979Cash}, adequate for shot-noise dominated bins:
\begin{equation}\nonumber
\ln L = \sum_j  \hat{N}_j \ln(n_j) - \sum_j n_j  \ \ (+ {\rm constant})
\end{equation}
The size of the bins is part of the likelihood design and it is usual practice to consider infinitesimally small bins containing $\hat{N}_j = 0$ or $1$ object. In such a case, $n_j$ transforms into $\dd n/ \dd \tbin$ and the sum signs into continuous integrals in the expression above.
In presence of additional variance terms, this likelihood must be modified. For instance, \citet{2022Garrel} present a likelihood accounting for sample variance \citep[e.g.][]{2003HuKravtsov}. This component is an important source of variance when dealing with large numbers of objects (relatively to Poisson noise that affects small samples).

A model should then predict the value of $n_j$.
In the ideal case of unrestricted computational power, excellent fidelity may be achieved by means of end-to-end models of the eRASS1 sky, and by retrieving the number $n_j$ from the simulated catalogs.
However it is prohibitive to run one new end-to-end simulation each time a value is queried by the likelihood function (Eq.~\ref{eq:likelihood}). The cosmological analysis process necessarily resorts to approximations, interpolations and emulators. For instance, a parameterized halo mass function (HMF) model would replace series of full N-body simulations; intra-cluster medium (ICM) scaling relations replace detailed hydrodynamical simulations; while a selection function model replaces mock image generation and processing.

We write $\pofc{I}{\tsel}$ the probability of selecting an object in the sample given the value of parameters $\tsel$. These values must be a prediction of the model. The expression for the modelled counts $n_j$ may write formally:
\begin{multline}\label{eq:number_counts_equation}
    n_j = \int \dd \tbin
    \dd \tsel
    \dd \tmodel
    \mathbb{1}(\tbin \in j) \\
    \times 
    \pofc{I}{\tsel}
    \pofc{\tbin, \tsel}{\tmodel}
    \frac{\dd n}{\dd \tmodel}
\end{multline}
In the latter expression we have introduced $\dd n/\dd {\tmodel}$, the predicted number distribution of galaxy clusters under a certain model assumption. The indicator function $\mathbb{1}$ ensures only those objects with $\tbin$ taking values in bin indexed by $j$ are counted. The three sets of parameters may share part or all of their members.

An important preparatory task in the cosmological analysis consists in carefully selecting $\tbin$ and $\tsel$ so the model is accurate and the modelling effort is well balanced between $\pofc{I}{\tsel}$ and $\pofc{\tbin, \tsel}{\tmodel}$. Depending on how close $\tsel$ is from purely observable quantities, selection function models involve a varying amount of astrophysical modelling. It is also important to recognize that some aspects of selection may result from human intervention and thus they are difficult to model with analytic formulae. A certain degree of empiricism may then be introduced in building selection models. We will discuss several options in this paper and compare their advantages and shortcomings.

\begin{table*}
\caption{\label{table:summary_notations}Glossary of symbols and conventions used throughout this paper.}
\centering
\begin{tabular}{p{0.07\linewidth}p{0.83\linewidth}}
\hline\hline
Symbol	&	Signification\\
\hline
$\tmodel$	&	Variables governing the population of clusters, e.g.~cosmological parameters, or $(M_{500}, z)$	\\
$\tsel$	&	Variables describing the selection of clusters, e.g.~count-rate	\\
$\tmes$	&	Variables describing measurement associated to clusters, e.g.~measured flux	\\
$\isdet{catalog}$	&	Boolean\tablefootmark{a} random variable indicating the presence of a cluster in a catalog, e.g.~$I=1$	\\
\hline
$\isdet{eRASS1}$	&	A source appears in the X-ray source detection list, it is detected    		\\
$\isdet{main}$	&	A source is detected and classified as extended in the eRASS1 primary cluster catalog ($\extlike > 3$)  	\\
$\isdet{cosmo}$	&	A source is detected and classified as extended in the eRASS1 cosmology catalog ($\extlike > 6$) 	\\
\hline
$L_X$	&	True cluster rest-frame 0.5-2\,keV luminosity integrated in a cylinder of radius $R_{500}$     \\
$z$	&	True cosmological redshift of a cluster (no peculiar motion)  \\
$f_X$	&	True, noiseless absorbed 0.5-2\,keV flux of a cluster within an aperture of radius $R_{500}$	 \\
$CR$	&	True, noiseless unabsorbed 0.2-2.3\,keV count-rate of a cluster within $R_{500}$ aperture		\\
$T_{\rm exp}$	&	Local exposure time in the eRASS1 survey	 \\
$N_H$	&	Hydrogen column density along the line-of-sight direction   \\
bkg	&	Local background surface brightness in the eRASS1 survey in the 0.3-2.3\,keV band  \\
$\mathcal{H}$	&	Abridged notation for $(N_H, T_{\rm exp}, {\rm bkg})$		\\
\hline
\end{tabular}
\tablefoot{\tablefoottext{a}{Boolean random variables values may be mapped to integers 0 and 1, following usual practice.
%An overline indicates negation, e.g.~the notation $\overline{I}$ is equivalent to $I=0$.
}\\}
\end{table*}

%--------------------------------------------------------------------
\section{The eROSITA galaxy cluster survey and its simulated twin\label{sect:datasets}}

    \subsection{The eRASS1 galaxy cluster samples}
   
The primary eRASS1 cluster catalog in the Western Galactic hemisphere (hereafter eRASS1-primary or eRASS1-main) is described in detail in \citet{2024Bulbul} and \citet{2024Kluge}. This catalog builds upon the eRASS1 X-ray source detection list in the soft band ($0.2-2.3$\,keV) presented in \citet{2024Merloni}. The source detection is achieved with the eROSITA Science Analysis Software (eSASS in version 010), and it incorporates two essential steps in its final stages \citep[see][for further information]{2022Brunner}. First, sources are detected by searching counts significantly above the background level; those sources are individually characterized using a point-spread function (PSF) fitting algorithm. The detection likelihood (\texttt{DET\_LIKE\_0} in the parent catalog, hereafter shortened to $\detlike$) and the extent likelihood (\texttt{EXT\_LIKE}, hereafter $\extlike$) are among the most relevant parameters for selecting extended sources. In particular, the eRASS1-primary catalog is compiled based on a low threshold of $\extlike>3$ to maximize completeness. A rigorous cleaning process is applied to the X-ray source catalog to ensure clean and uncontaminated images are suitable for extended source identification. A series of additional cleaning steps remove spurious split detections; we will assume their impact on completeness is null, as suggested by the number of secondary matches found in the digital twin matching the fraction of split sources in real data \citep[][and next section]{2022Seppi}.

The description of the eRASS1 cosmology sample (hereafter eRASS1-cosmo) is detailed in \citet{2024Bulbul}. It is compiled with a more conservative cut on the measured parameter $\extlike > 6$ compared to the eRASS1-primary sample to maximize the purity. Exclusively, galaxy clusters with measured photometric redshifts between 0.1 and 0.8 are selected, keeping only areas of the sky where the photometry can be uniformly applied in redshift measurements in the Legacy DR10-south area \citep{2024Kluge}. The cosmology sample comprises 5\,259 identified clusters with low contamination levels below 5\%.

This paper focuses only on X-ray selection function models. However, both the eRASS1-primary and eRASS1-cosmo catalogs are derived from extensive optical follow-up designed for identification and redshift measurement. This identification may act as an additional selection filter impacting completeness and purity estimates and should be accounted for in science analyses with these catalogs \citep[see][]{2024Ghirardini}. Most of the discussion in the paper neglects this effect, and we refer to \citet{2024Kluge} for a detailed description of the optical selection function.
    
    \subsection{The eRASS1 digital twin}

Understanding selection effects requires mock observations reproducing as many of the characteristics of the actual data as possible. 
We briefly summarize here the main features of the eRASS1 digital twin depicted in \citet{2022Seppi}. The parent halo catalog originates from UNIT1i dark-matter simulations \citep{ChuangYepesKitaura_2019MNRAS.487...48C}. A full sky light cone is created by replicating shells of the individual snapshots of the simulation \citep{2020Comparat}. The large, albeit finite, box size prevents simulating the very nearby, massive halos that constitute a tiny fraction of the sources in eRASS1 and require a separate treatment.
X-ray sources are associated to the light cone using models for the emission of AGN \citep{ComparatMerloniSalvato_2019MNRAS.487.2005C} and galaxy clusters and groups \citep{2020Comparat}. The cluster model is generated from a set of real clusters by accounting for the covariances between the surface brightness profile, halo mass, temperature, and redshift. The baryon profiles are painted onto halos based on observations; therefore, the predicted emissivity profiles are representative of observations present in the literature. Low-mass groups $M_{500c} < 5 \times 10^{13} M_{\odot}$ require a flux rescaling correction \citep[see appendix of][]{2022Seppi}. 
It replaces simulated objects along the observed stellar mass--X-ray luminosity relation of groups and clusters \citep[][Zhang Y. et al. in preparation]{AndersonGaspariWhite_2015MNRAS.449.3806A, ComparatTruongMerloni_2022A&A...666A.156C}. 
A simulated half-sky twin contains more than $10^6$ simulated halos up to $z \sim 1.5$. Besides AGN and clusters, foreground X-ray emitting stars and galactic foregrounds also have their share of simulated photons.

The SIXTE software \citep{2019Dauser} generates mock event lists and images of these photons, as would be seen by eROSITA in the eRASS1 survey. 
In particular, instrumental characteristics and scanning laws enable reproducing the response of the telescopes to the X-ray simulated sky. In order to examine variability due to stochastic Poisson noise, several tens of simulations are produced (using the same X-ray parent population).
Processing of event files in tiles of size $3.6 \times 3.6 \deg^2$ took place in a similar way as for real data. In particular the eSASS (version eSASSusers\_201009) source finding algorithm runs over each tile and delivers a list of detections with associated measurements $\detlike$ and $\extlike$.

All clusters simulated in a twin mock are associated to a set of properties, such as mass $M_{500}$, true redshift $z$, X-ray luminosity $L_X$, flux $f_X$ or count-rate $CR$ (all within $R_{500}$, see Table~\ref{table:summary_notations}). These `true' properties will serve in establishing selection models in Sect.~\ref{sect:gp_models}. However, we consider important at this stage to recall that such properties are `labels' associated to simulated sources. The link between those labels and the mock events depends obviously on the models imprinted in the twin simulation.

    \subsection{Matching input and detected sources\label{sect:matching_catalogs}}

The next step consists in attributing a binary flag to each simulated halo stating whether it is selected in the cluster sample. This operation relies on matching the input catalog to entries in the detection list. We explore and compare two different procedures. The first method, hereafter `photon-based', takes advantage of the SIXTE simulator as it individually tracks simulated CCD events back to the source that emitted the photon.
The second method, hereafter `position-based', uses positions and sizes of sources on sky, aided with prior knowledge of the flux distribution of selected sources.

		\subsubsection{Photon-based matching}

The photon-based matching is in fact our baseline method. We refer the reader to \citet{2022Seppi} for a comprehensive explanation of the procedure.
The input object contributing most photons to the counts making a detection is identified as the best matching counterpart. If a detection is split into several sources by the detection software, again preference is given to the source comprising most photons originating from the identified counterparts. This enables assigning a unique match (ID\_Uniq) to a given detection and to a given simulated object. This is the definition we take for matches in the following. Sources with a unique match are flagged as selected ($\isdet{\rm main}=1$). We note that this procedure involves sources of all types: simulated AGN, clusters, stars; and detected point-like and extended sources.

		\subsubsection{Positional matching}
		
The positional matching technique is a two-way match between the input and detection catalogs. It was employed in the context of establishing a selection function for the eROSITA Final Equatorial-Depth Survey \citep[eFEDS, see][]{2022LiuA}. This technique does not assume a fixed cross-matching radius; instead it takes into account the sizes of sources. Each simulated halo is listed with a central coordinate and an angular extent (we take 10\% of the virial radius). Each extended detection is reported with a coordinate, a 1-$\sigma$ error circle, and an angular extent (we take the core-radius of the best-fit $\beta$-model). We combine source extents with the source positional uncertainty; effectively it is similar to spreading location of the source over a small region. We use the NWAY algorithm \citep{2018Salvato}, since it is well-suited to cross-matching catalogs in presence of positional uncertainties.
We first take the input catalog as reference and look for matches in the detection catalog. For each reference source, the algorithm returns a list of all matches located within a large buffer region of radius 3 arcmin. Each of the matches is assigned a probability $p_i$ that it is indeed a valid association, based on its proximity and positional uncertainties. A second value $p_{\rm any}$ is computed, representing the probability for the input source to have at least one counterpart among the detections. The probabilities $p_i$ and $p_{\rm any}$ account for chance associations, through the source density estimated over a degree-scale region. Sorting matches by the value of $p_i$ provides a ranking of most likely counterparts. More often than not our catalogs enclose complex configurations with multiple input sources projected along neighboring line-of-sights, and detections split in multiple sources. To deal with these effect we update the probabilities with a prior distribution of the flux of selected sources, as enabled by the NWAY formalism. The exact shape of this prior is unimportant in this context, and we simply take it from the simpler pre-launch selection functions derived in \citet{2018Clerc}. As a consequence, NWAY will preferentially up-weight the probability of brighter sources.
Converting probabilities into binary flags requires setting thresholds on $p_i$ and $p_{\rm any}$. Our experiments showed that the result is rather insensitive to their exact values. For this first pass (input catalog as reference) we kept only pairs with $p_{\rm any}>0.9$ and $p_i>0.1$. Pairs with the highest $p_i$ among all associations are called primary matches.
We reiterate the above procedure after exchanging the role of the input and detection catalogs. For this second pass (detection catalog as reference) we do not input any prior and we chose $p_{\rm any} > 0.5$ and $p_i>0.1$ to select valid pairs and primary matches. Finally, matches that are primary in both runs were selected as solid matches. For each input source, it is flagged as a selected cluster ($\isdet{\rm main}=1$) if it belongs to a solid match. Such a double matching procedure is more conservative than a single-pass procedure; as a primary matched source will not be available for a second pair.

		\subsubsection{Comparison of matching techniques}

\begin{figure}
   \centering
   \includegraphics[width=\linewidth]{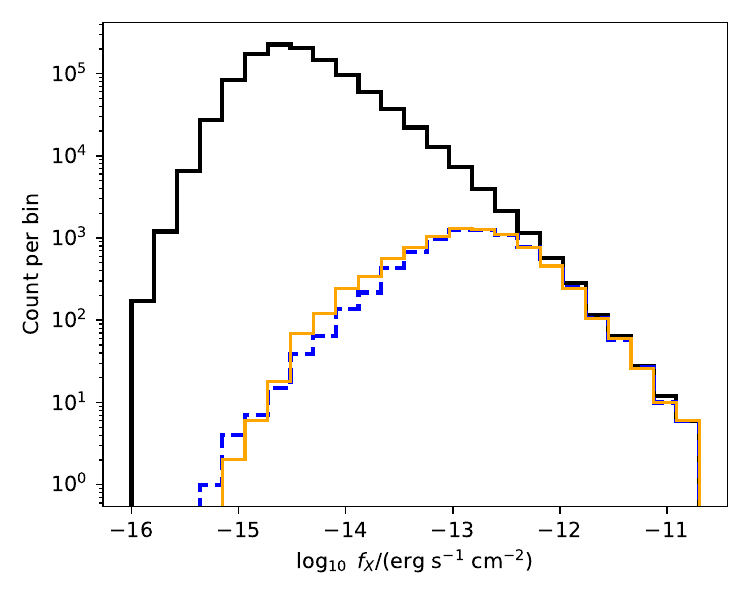}
   \caption{Comparing the outcome of two matching procedures relating simulated halos (black plain histogram) to detected sources (photon-based matching as a blue dashed line, position-based matching as a thin orange line). A simulated halo is flagged as selected ($\isdet{\rm main}=1$) whenever it is (solidly) matched to an extended source in the detection catalog. The details of the matching algorithms impact the flux distribution of detected sources, especially at flux below $\sim 10^{-14}$\,\fluxunit{}, with deviations up to a factor 2 in certain bins.}
    \label{fig:compa_phbased_nway}
\end{figure}

\begin{figure}
   \centering
   \includegraphics[width=\linewidth]{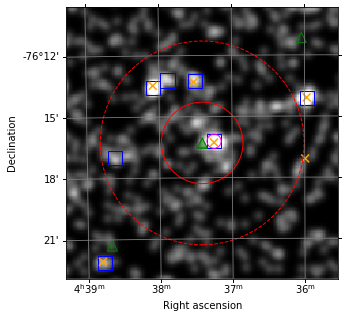}
   \caption{Example case of blending, where a bright AGN drives the detection and the presence of a faint extended source favors its classification as extended. This figure is a cut-out of a simulated count image, smoothed with a Gaussian. The red 2- and 5-arcmin circles are centred on a faint simulated cluster (filled triangle symbol) with flux $\sim 5\times10^{-14}$\,\fluxunit{}. Detected sources appear with blue squares. A detection close to the cluster centre is classified as extended (magenta circle). It also coincides with a bright simulated AGN (orange `x'). The position-based matching algorithm does consider the simulated cluster as selected ($\isdet{\rm main}=1$), contrary to the photon-based matching procedure. Two other faint simulated clusters are shown with open triangles, they have no impact on the detection.}
    \label{fig:compa_matching_example}
\end{figure}

Figure~\ref{fig:compa_phbased_nway} illustrates the comparison between both matching procedures. They agree well on a wide range of fluxes. Deviations are visible at low flux, i.e. below a few $10^{-14}$\,\fluxunit{}. One possible cause (not unique) stems from the presence of a bright AGN near a relatively dim cluster. An example is shown in Fig.~\ref{fig:compa_matching_example}. In photon-based matching, even if there is an extended detection nearby, it is matched to the input simulated AGN due to the large amount of photons (and events) it deposits in the vicinity of the extended detection. In the position-based procedure, the extended source is matched to the input cluster, since it does not include input AGN sources in the analysis. The problem of characterising a source as detected in such case is ill-posed.
Another subtle difference in the treatment of the catalogs plays a role, due to the definition of an extended detection. The position-based matching takes as input the list of sources with extent likelihood greater than a predefined threshold (6 for the cosmology sample). The photon-based matching takes as input the entire list of sources, associates a detection in the list and only then flags as selected those halos with a detection having extent likelihood above threshold. Reversing the order of operations has an impact in crowded regions with multiple split sources, or multiple halos along neighouring line-of-sights.

The comparison exercise described in this section serves in bracketing the uncertainty associated to the matching procedure. In practice, model uncertainties in the low count regime are expected to overcome such variations. The photon-based matching is taken as a baseline for the rest of the paper.

    \subsection{Training and test samples}

Each realization of the twin eRASS1 sky comprises about $10^6$ halos, less than 4\% being detected ($\isdet{\rm eRASS1}=1$) and less than 1\% being selected ($\isdet{\rm main}=1$) according to the matching algorithm. The sample is split into two parts: two thirds are saved for training a model, the other third is left untouched and will serve to test the validity of the model. Splitting is performed after shuffling the list of halos at random.

We also create supersets of the simulated sky by concatenating training sets of eleven realizations, and randomly shuffling their content. Although the realizations are not independent stricto sensu (they share the exact same population of objects), this procedure helps in reducing the impact of photon noise. Each super-training set is $11 \times 2/3 \simeq 7$ times larger than the actual eRASS1 sky.

%--------------------------------------------------------------------
\section{Selection models with surface brightness profiles\label{sect:logistic_profile}}

Equation~\ref{eq:number_counts_equation} involves three important factors that require modelling. The task of modelling the selection function consists in building $\pofc{I}{\tsel}$, this is the main interest of this paper.
The choice of $\tsel$ usually results from a compromise between precision of the selection function and complexity of the model $\pofc{\tbin, \tsel}{\tmodel}$.
        
A natural choice for the selection parameters $\tsel$ is the collection of pixel values that form the images of clusters. Then $\pofc{I}{\rm image}$ is obtained by feeding eSASS with an image and applying thresholds on the values ($\extlike$, {\tt EXT}) returned by the algorithm; because eSASS is deterministic, $\pofc{I}{\rm image}$  takes either value 0 or 1. The selection function model is thus very precise -- it is actually the most precise model since it exactly reproduces the actual data processing. However, analysing a typical $10\arcmin \times 10\arcmin$ image with eSASS takes about three seconds with standard CPUs. Computing a likelihood (Eq.~\ref{eq:likelihood}) for about 12\,000 clusters would then amount to about three CPU$\cdot$hours, which is unrealistically long for a cosmological analysis. To this cost must be added the time to compute a model cluster image given $\tmodel$.

Realizing that galaxy cluster images are almost circularly symmetric, we may accept to lose precision and to reduce the complexity of modelling a cluster.
Let us consider the number of photon hits (counts) deposited by sources onto the detectors, split in several sky annular apertures around the centre of a putative dark matter halo.
We emphasize that the number of counts should be a prediction of the model $\pofc{\rm counts}{\tmodel}$. It is not the outcome of a measurement process -- that is available for detected sources only. It is not the scope of this paper to discuss how such a generative model is constructed, we simply assume it exists.
For constructing our selection model we rely on the twin simulations, that have kept a record of the origin of each count deposited in the image.

    \subsection{Illustration with a single feature: the $90\arcsec$ cluster counts}

\begin{figure}
   \centering
   \includegraphics[width=\linewidth]{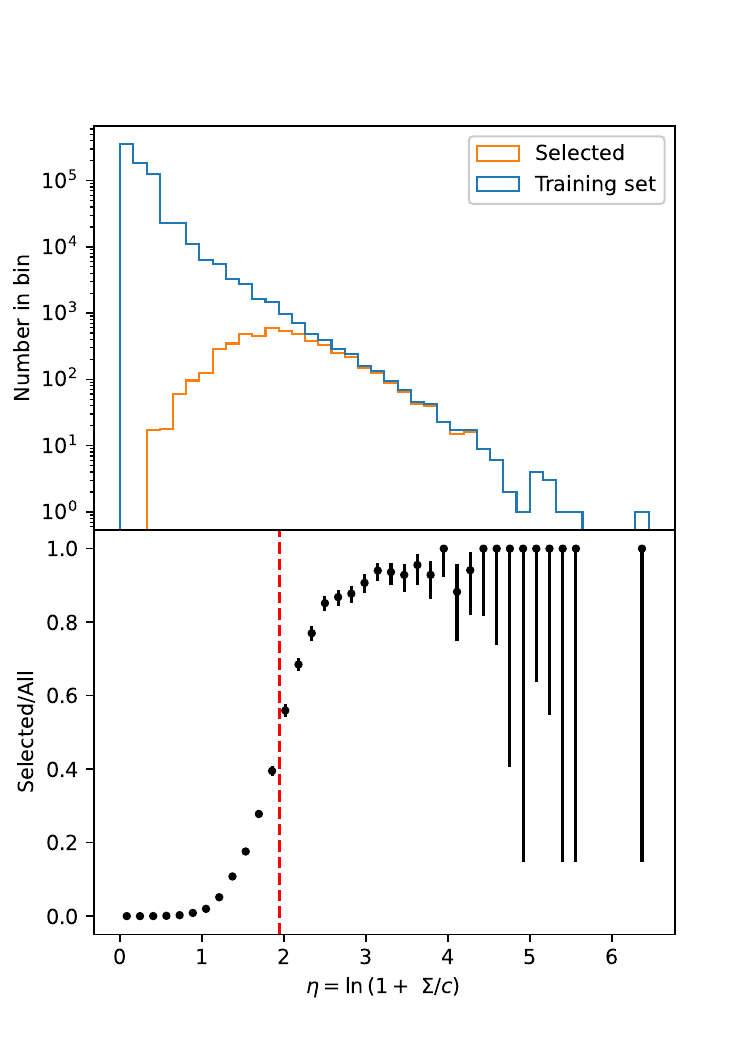}
   \caption{Distribution of clusters in the training set as a function of $\sbright$ (units counts arcsecond$^{-2}$), the average surface brightness in the $90\arcsec$ radius around their centre. The x-axis is rescaled with $c=2\times 10^{-4}$\,arcsecond$^{-2}$. \emph{Top panel:} histogram of all simulated clusters (blue) and histogram of the subset of those found as extended by eSASS (orange). \emph{Bottom panel:} dots indicate the ratio of the histograms (empirical selection rate). Error bars are the 68\% confidence range estimated according to App.~\ref{app:binomial_unc}.
   The vertical dashed line indicates the transition $\hat\sbrightmod=1.94$ in the logistic model described by Eqs.~\ref{eq:N_to_zeta} and~\ref{eq:expit_formula} -- corresponding to $N \simeq 30$ counts.
   }
    \label{fig:empirical_sf_aperture_counts}%
\end{figure}

Let us initiate our procedure by involving only one feature, namely the number of cluster counts received in a circular aperture around the (RA, Dec) centre of the simulated halo. The aperture radius is set to $R=90\arcsec$. 
We transform the counts (integer values $N$) with:
\begin{equation}
\label{eq:N_to_zeta}
\sbrightmod = \ln\left(1+ \frac{\sbright}{c}\right)
\end{equation}
We set $\sbright=N/(\pi R^2)$ and $c=0.0002$\,counts\,arcsecond$^{-2}$. In our training set, $N$ ranges from 0 to 3194 counts within $90\arcsec$ aperture.
Fig.~\ref{fig:empirical_sf_aperture_counts} displays the histogram of $\sbrightmod$ associated to all halos in the training set, and for those flagged as selected ($\isdet{\rm main}=1$). The ratio of the two histograms provides an empirical estimate of the probability of detecting a halo given the number of counts deposited on the detectors. The empirical probability follows a characteristic "S-shaped" curve.
We perform logistic regression to fit a model to the probability of detecting a cluster as extended ($\extlike > 3$) given the value $\sbrightmod$. We use the {\tt scikit-learn} implementation \citep{scikitlearn} with the Broyden-Fletcher-Goldfarb-Shanno optimizer and fit two coefficients, namely the intercept $w_0$ and the slope $w_{\sbrightmod}$ of the linear model $f(\sbrightmod) = w_0 + w_{\sbrightmod} \sbrightmod$. The model probabilities are such that
\begin{equation}\label{eq:expit_formula}
\pofc{\isdet{main}}{\sbrightmod} = \left(1+e^{-f(\sbrightmod)}\right)^{-1}.
\end{equation}
The cost function to minimize during the fitting procedure is the log-loss $\mathcal{C}(w_0, w_{\sbrightmod})$, its expression is shown in Eq.~\ref{eq:log_loss}.

Thanks to the one-to-one relation between $\sbrightmod$ and $N$ (Eq.~\ref{eq:N_to_zeta}), we thus obtain a model $\pofc{\isdet{main}}{N}$. We can interpret the values of the best-fit coefficients as follows: $w_{\sbrightmod}$ governs the sharpness of the transition in the S-shaped curve, $\hat \sbrightmod=(-w_0/w_{\sbrightmod})$ is the value at which this transition occurs. In our experiment, we find $w_{\sbrightmod}=4.59 \pm 0.04$ and $w_0=-8.92 \pm 0.06$, leading to $\hat\sbrightmod = 1.94$. This corresponds to a transition of the "S-shaped" curve taking place at $N \simeq 30$ cluster counts in the $90\arcsec$ circular region. Positiveness of $w_{\sbrightmod}$ reflects the fact that detection probability increases with number of counts $N$. The computation of coefficient uncertainties is detailed in App.~\ref{app:uncertainties_logistic}.

	\subsection{A model using light profiles from all sources}

The model presented so far involves only one feature, it is too simplistic to explain subtle variations in the probability of selecting a given cluster. The model improves by involving more of the quantities available from the twin simulations, namely the number of detector hits deposited by each of the five sky source components: the cluster of interest, Active Galactic Nuclei (AGN), stars, back- and foreground and other clusters. Each of these contribute to the count image and deposit photons into 7 annular regions around the central coordinate of the cluster. Radial boundaries (units arcseconds) are $0-20$, $20-40$, $40-60$, $60-90$, $90-120$, $120-150$ and $150-180$. For each simulated cluster, we thus construct a $5\times7=35$-element feature vector $\vec{\sbrightmod}=\{\sbrightmod_{j}\}_{j=1..35}$ and perform logistic regression, involving a total of 36 coefficients $\{w_j\}_{j=0..35}$ (one for each feature, plus the intercept). Introducing $\sbright_j$ the average surface brightness in an annulus, we write:
\begin{equation}\label{eq:f_zeta}
f(\sbrightmod) = w_0 + \sum_{j=1}^{35} w_j \sbrightmod_j {\ \rm, with\ \ } \sbrightmod_j = \ln\left( 1 + \frac{\sbright_j}{c}\right)
\end{equation}

One may interpret the value of the coefficients $w_j$ as the sensitivity of the detection rate $p \equiv \pofc{\isdet{main}}{\vec{\sbrightmod}}$ to a small variation of the surface brightness of one component in one annulus. Indeed we have:
\[
\left.\frac{\partial p}{\partial \sbrightmod_j}\right|_{\sbrightmod_k}  = p(1-p) w_j
\]
where the derivative is performed at constant $\sbrightmod_k$ ($k \neq j$). It yields:
\begin{equation}
    \label{eq:marginal_probability_logistic}
    \left.\frac{\partial p}{\partial \sbright_j}\right|_{\sbright_k} = \frac{p(1-p)}{\sbright_j+c} w_j
\end{equation}

\begin{figure}
   \centering
   \includegraphics[width=\linewidth]{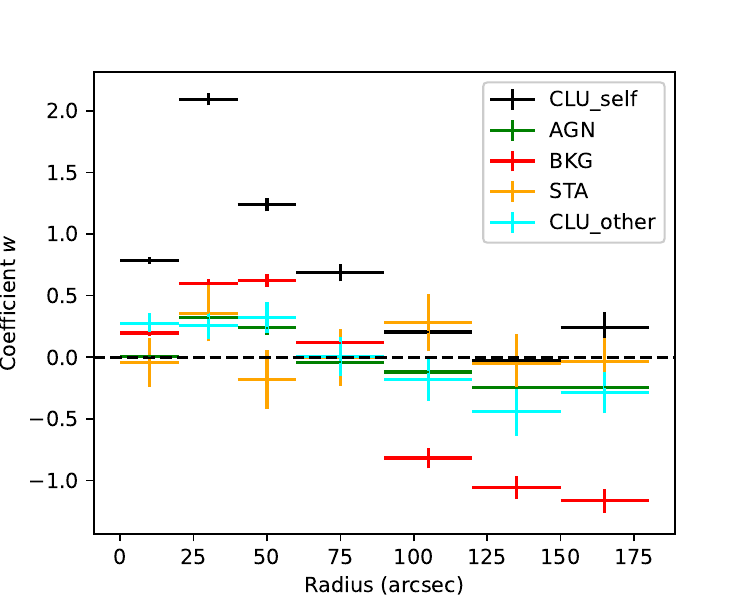}
   \caption{Representation of the 35 coefficients $w_j$ of a logistic regression model $\pofc{\isdet{main}}{\rm counts}$, trained to predict whether a cluster is selected in the primary cluster sample. The 35 features are surface brightness in 7 radial annuli, associated to the 5 components indicated in legend (counts from the galaxy cluster of interest, from neighbouring AGN, fore- and background counts, counts from foreground stars, counts from other neighbouring galaxy clusters.) High absolute value of a coefficient indicates high importance of the associated feature (Eq.~\ref{eq:marginal_probability_logistic}).}
    \label{fig:coeff_profile}%
\end{figure}

In other terms, everything else maintained fixed, a small relative variation $\epsilon = \Delta\sbright_j/\sbright_j$ of the surface brightness of one component in one given annulus leads to a variation of the detection probability proportional to $w_j \epsilon$. In particular, positive coefficients relate to those features that marginally contribute to selecting a cluster, and conversely for negative coefficients. Fig.~\ref{fig:coeff_profile} conveniently represents the value of the 35 coefficients $w_j$ $(j\neq 0)$ split into each component and annulus; the last coefficient not shown on this figure is $w_0= -8.82 \pm 0.08$.

Keeping in mind that $w_j$ are proportional to the marginal increase of the detection probability, comparing together values of coefficients provides an approximate way of assessing the `importance' of a feature in promoting detectability of a cluster. For instance, it appears that a small increase of surface brightness in the $20-40\arcsec$ annulus from a given cluster has the strongest impact on the detectability of this cluster. The fact that it is twice as important as the $0-20\arcsec$ surface brightness is not surprising. A detected source ($\isdet{\rm eRASS1} = 1$) must be classified as extended in order to be selected ($\isdet{\rm main} = 1$); from this perspective, it is much more profitable to increase the counts beyond the eRASS point-spread function radius ($\sim 30\arcsec$).
This argument may appear more clearly from Fig.~\ref{fig:coeff_profile_onlydet}, which shows the coefficients $w_j$ of a model trained to predict the detection (not the selection) of a cluster. Clearly in this case the central $20\arcsec$ surface brightness is the most relevant feature contributing to detectability of a cluster.

The effect of star-emitted photons is barely constrained, mostly due to their paucity in the simulation. AGN photons within one arcmin of the centre tend to marginally increase the detection rate, while those located beyond 1 arcmin tend to decrease the detection rate. This is because the source detection algorithm would put a mask and remove photons useful for cluster detection.
An excess of photons from instrumental and astrophysical fore- and background within $1.5\arcmin$ of a cluster tend to increase its probability of being selected; while at larger distances they decrease its probability by a large factor. The impact of neighbouring clusters is similar, although less pronounced.

	\subsection{Internal validation of the selection models}

\begin{figure}
   \centering
   \includegraphics[width=0.9\linewidth]{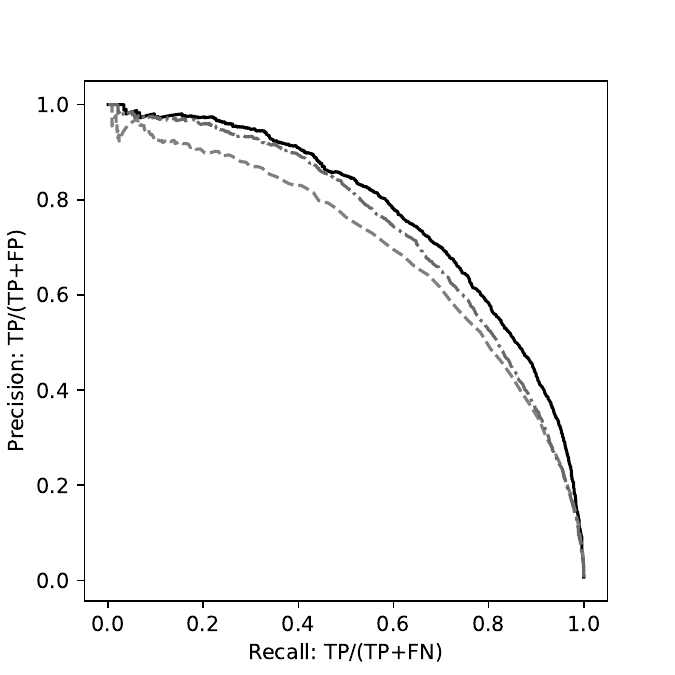}
   \caption{Performance of three different logistic models $\pofc{\isdet{main}}{\rm counts}$ to predict a cluster as selected given the values of surface brightness in radial bins. The plain black line is obtained with a model using surface brightness values in 7 annuli for each of the 5 sky components. The dot-dashed line is for a model taking as input both cluster and background counts within $90 \arcsec$. The dashed grey line stands for a model using as input only the $90\arcsec$ cluster counts. Each precision-recall curve results from model evaluations on the test sample. It is obtained by varying the threshold over which a cluster should be considered as selected and counting the number of true positives (TP), false negatives (FN) and false positives (FP).}
    \label{fig:logistic_precision_recall}%
\end{figure}

The previous interpretation of the $w_j$ coefficients in terms of marginal probabilities should not hide it is only relevant to a specific model, here a logistic model, that is not a perfect classifier. We assess performance of this model on the left-apart test sample. Fig.~\ref{fig:logistic_precision_recall} represents the so-called `precision-recall' curve. This curve is obtained by scanning values of a threshold set to convert probabilities $\pofc{\isdet{main}}{\rm counts}$ into a binary classification $\isdet{main}$. Once this threshold is set, one retrieves the number of true positives (TP), false positives (FP) and false negatives (FN) by comparison with the actual detection flag in the test sample. The higher the recall\footnote{Recall (also known as sensitivity) and precision are respectively equivalent to completeness and purity, defined with respect to the true classification in the test sample. However we prefer using the former terminology in order to avoid possible confusion with the usual measures of completeness and purity of a certain sample, defined with respect to the entire population of clusters.} $TP/(TP+FN)$ and the precision $TP/(TP+FP)$, the better performance of a model.
For any value of the threshold, the model using 35 surface brightness indicators performs significantly better than the model using only the 1.5-arcmin cluster counts as input. We have also displayed the curve obtained from a logistic model using two features, namely the 1.5-arcmin cluster and background counts. Its performance lies between the simple and the sophisticated ($7 \times 5$) models. This agrees with common sense, in that adding more detailed information provides more precise models.

\begin{figure}
   \centering
   \includegraphics[width=0.9\linewidth]{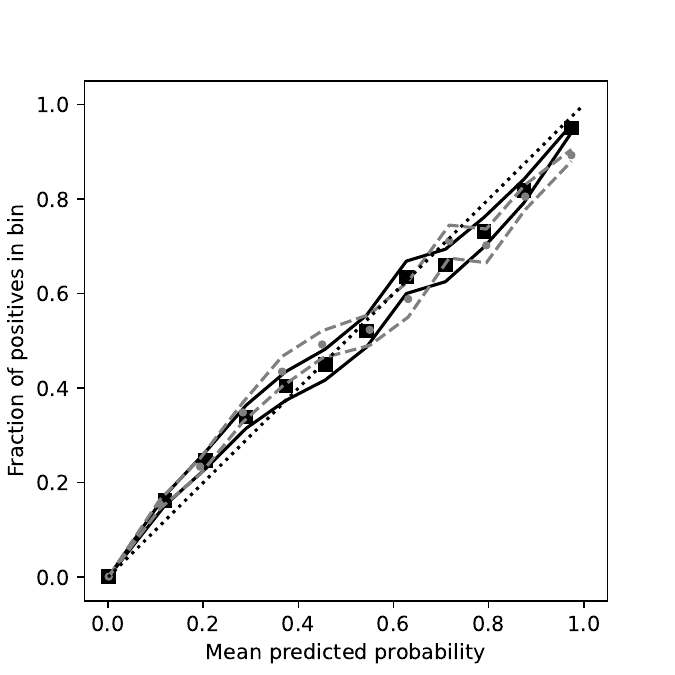}
   \caption{Reliablity of two different logistic models $\pofc{\isdet{main}}{\rm counts}$ to predict the probability of a cluster being selected given the values of surface brightness in radial bins. Each series of points is obtained with the test sample, by comparing probabilities predicted by the model (horizontal axis) with the actual fraction of selected objects (vertical axis) in each bin of probability. The envelopes materialise the 68\% uncertainties (App.~\ref{app:binomial_unc}). An ideal model would align along the 1:1 curve (thin dotted line). Black squares and plain lines correspond to the model taking $7 \times 5$ parameters as input, grey dots and dashed lines for the model using only cluster counts within $90\arcsec$ to make its prediction.}
    \label{fig:logistic_calibration}%
\end{figure}

A second test is shown on Fig.~\ref{fig:logistic_calibration}. This evaluation is performed by binning test clusters by their value of $\pofc{\isdet{main}}{\rm counts}$ returned by the model. The actual fraction of objects selected in the sample is reported on the vertical axis. The 35-feature model produces the plain envelope, in satisfactory agreement with the one-to-one line. This result indicates reliable probabilities delivered by the model in a statistical sense. The probabilities from the model using only one cluster count indicator (round dots and dashed envelope) are less reliable. This is reflected in Fig.~\ref{fig:empirical_sf_aperture_counts} where the "S"-shaped curve does not reach unity around values of $\sbrightmod \simeq 2.5-4$ (corresponding to $N \simeq 60-270$ counts in the $90\arcsec$ aperture).

Despite its apparent simplicity, a linear logistic model using circularly averaged surface brightness as training features does output probabilities that are representative of the actual selection probabilities. It is therefore well-suited to selection function problems involving populations of clusters. However, an average precision of $77\%$ (Fig.~\ref{fig:logistic_calibration}) indicates that it is not well-suited to the classification on an individual basis. This floor performance is due to using circularly averaged profiles, hence neglecting two-dimensional effects (e.g.~blending, masking, gradients) in the selection process. Furthermore, logistic regression is a simple linear model. An interesting development would involve more sophisticated models such as neural networks (e.g.~building upon the multi-layer perceptron). Their use may be able to add the right level of non-linear feedback and complexity to accurately reproduce the selection process, at the expense of explainability and simplicity.

A selection function model based on surface brightness as above is attractive due to its use of observable features. In the context of cosmological studies (as idealized in Eq.~\ref{eq:number_counts_equation}), it is necessary to construct a second, independent model able to predict the radial surface brightness profile of five components: clusters, AGN, stars, background and neighbouring clusters. Such a complex model $\pofc{\rm counts}{\tmodel}$ represents a computational bottleneck.

%--------------------------------------------------------------------
\section{Selection models with intermediate variables\label{sect:gp_models}}

We may instead choose $\tsel$ such as to reduce this complexity, at the cost of a less precise selection model. We consider variables (i.e.~selection model features) such as cluster mass, redshift, luminosity, size, central emissivity, etc. We also consider exposure time, background level, galactic hydrogen column density. For some of these variables (e.g. flux, luminosity) it is more convenient to manipulate their logarithmic values.

Having fixed the set of selection variables $\tsel$, we build a model assuming that the detection probability varies slowly over the range of interest of these (rescaled) features. However, we do not make prior assumption on the exact shape, nor on monotonicity of the function. A kernel-based model is well-suited to describe this problem, as it reflects the fact that similarly-looking clusters should have similar detection probabilities. We use Gaussian Process classification (GPC) to build our model, using the SVGP implementation in the \texttt{GPy} package \citep{gpy2014}. Details on the procedure and algorithm are discussed in App.~\ref{app:gaussian_process_classifier}. The very nature of GPC allows to issue not only an estimate of $\pofc{I}{\tsel}$, but also a range of statistical uncertainty. Our implementation does not provide extrapolation properties and a model returns a default value ($p=0.5$) when $\tsel$ is outside the domain where the training set lives. For the purpose of cosmological analysis it is not deemed an issue, as long as care is taken in handling the integration bounds in Eq.~\ref{eq:number_counts_equation}.

In the following we highlight three models relevant to this paper. In particular we demonstrate the interest of a model based on count-rate $CR$, which was selected for the eRASS1 cosmological analysis \citep{2024Ghirardini}. In App.~\ref{app:more_models} we describe several other models, notably those including cluster morphology in their input parameters.

		\subsection{Mass and redshift}

\begin{figure}
   \centering
   \includegraphics[width=\linewidth]{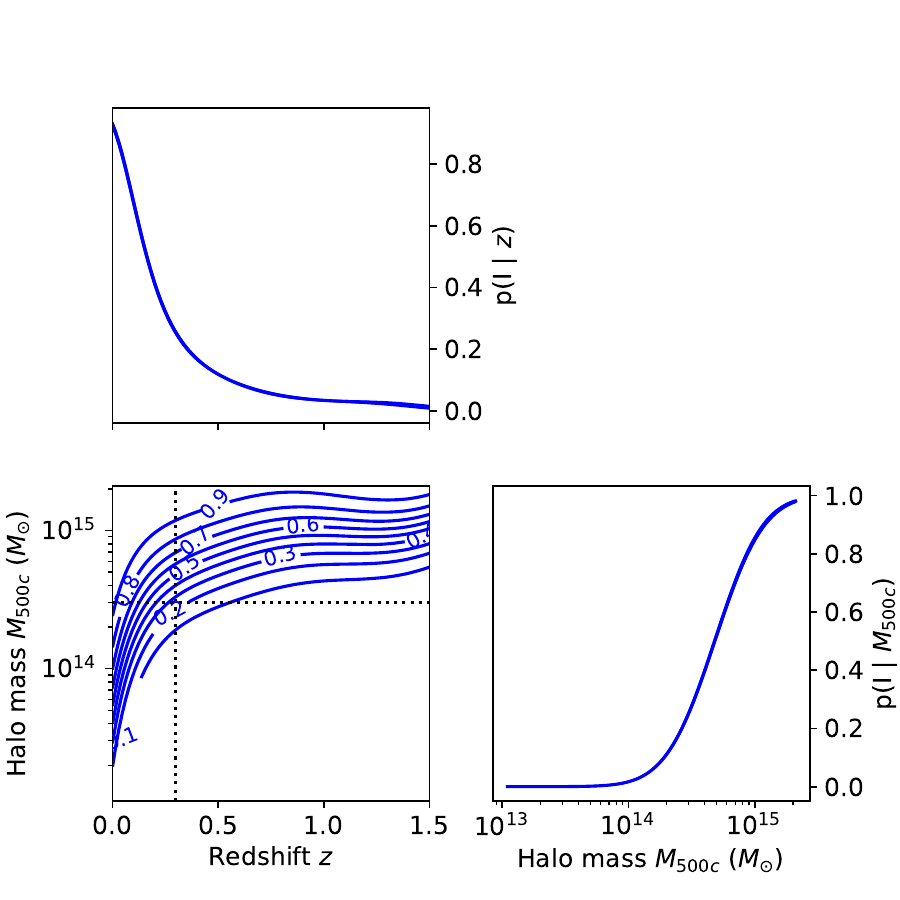}
   \caption{Representation of the model $\pofc{\isdet{main}}{M_{500c}, z}$ predicting a cluster to be detected and selected with an extent likelihood above 3. The explanatory variables (features) are labels attached to simulated clusters in the twin simulation, standing for galaxy cluster $M_{500c}$ mass and cosmological redshift. The bottom-left panel represents contours of equal detection probability, labelled in steps of 0.1. Both one-dimensional curves (top-left and bottom-right panels) are slices through the function displayed in the bottom-left corner, at fixed $z=0.3$ and $M_{500c}=3\times 10^{14}$\,$M_{\odot}$ (indicated with dotted lines).}
    \label{fig:GP_M500_z_AllXGOOD_cut_0-0_SEED0189addM}
\end{figure}

Setting $\tsel = \{M_{500c}, z\}$ allows to gain insights on the selection process in the fundamental parameter space of cluster cosmology. Fig.~\ref{fig:GP_M500_z_AllXGOOD_cut_0-0_SEED0189addM} is a visualization of a model adjusted to predict the selection of a cluster with extent likelihood above 3. The  probability of selecting a cluster located at $z=0.3$ monotonically increases with mass, with a transition taking place at $M_{500c} \simeq 5\times 10^{14}$\,$M_{\odot}$. 

Despite its formal interest, this model is not of practical use in context of cosmology studies. The two features entering this model (mass and redshift) are mere labels attached to clusters in the twin simulation. The model is therefore valid only under assumption of physical models implemented in the simulation. In particular, it is only valid for the set of scaling relations and cosmological parameters \citep{planckparam} imprinted in the simulation.

		\subsection{Luminosity and redshift}

\begin{figure}
   \centering
   \includegraphics[width=\linewidth]{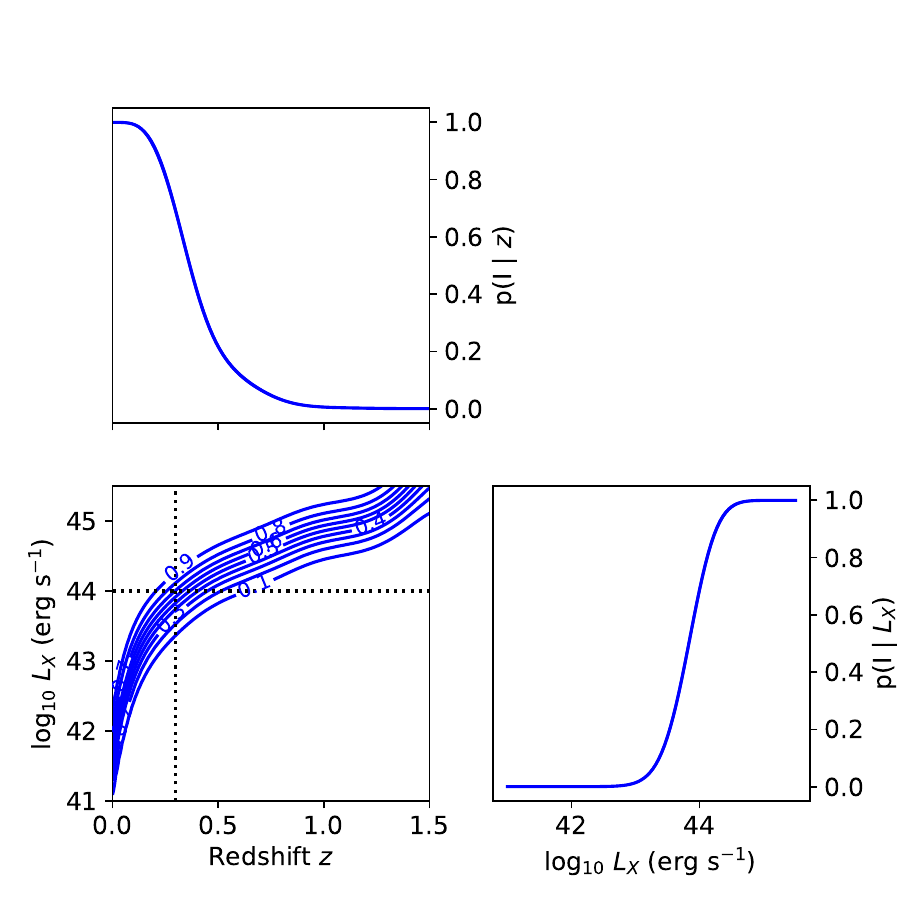}
   \caption{Representation of the model $\pofc{\isdet{main}}{L_{X}, z}$ predicting a cluster to be detected and selected with an extent likelihood above 3. The explanatory variables (features) are labels attached to simulated clusters in the twin simulation, standing for galaxy cluster luminosity $L_{X}$ measured in in the 0.5--2\,keV energy band at the cluster rest-frame (units erg\,s$^{-1}$, rescaled in $\log_{10}$) and for cosmological redshift $z$. Other details are similar as in Fig.~\ref{fig:GP_M500_z_AllXGOOD_cut_0-0_SEED0189addM}.}
    \label{fig:GP_lx_z_AllXGOOD_cut_0-0_SEED0189}
\end{figure}

Setting $\tsel= \{L_{X}, z\}$ advantageously removes the strong dependence of the model upon the $M_{500} \rightarrow L_{X}$ relation that is imprinted in the twin simulation. It is therefore up to the cosmologist to model the relation $\pofc{L_{X}}{\tmodel}$ with (most likely) parameterized scaling laws. A visual representation of the model is shown in Fig.~\ref{fig:GP_lx_z_AllXGOOD_cut_0-0_SEED0189}. As expected, clusters brighter and closer to us are more likely to be selected.

\begin{figure}
   \centering
   \includegraphics[width=\linewidth]{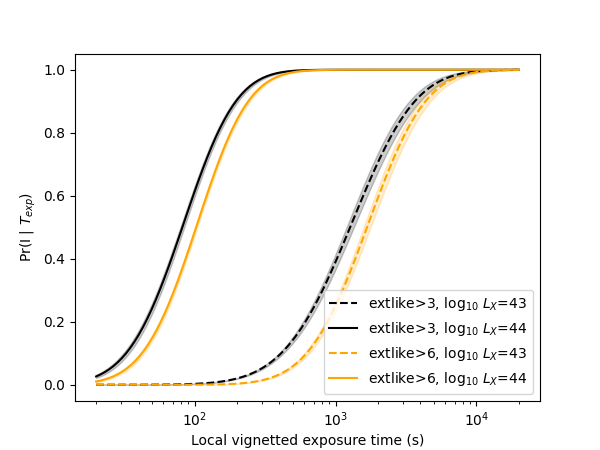}
   \caption{Representation of the models $\pofc{\isdet{main}}{L_{X}, z, N_H, T_{\rm exp}, {\rm bkg}}$ and $\pofc{\isdet{cosmo}}{L_{X}, z, N_H, T_{\rm exp}, {\rm bkg}}$ for fixed values of cluster redshift $z$, two different values for the luminosity $L_X$, and local galactic absorption $N_H$, a nominal background level and a range of exposure times $T_{\rm exp}$ (along the x-axis). The shaded regions indicate the approximate 68\% confidence range output of the model.}
    \label{fig:GP_nH_Texp_simbkg_lx_z_AllXGOOD_cut_0-0_SEED0189}
\end{figure}

Although useful, this model cannot account for variations of the detection probability as a function of sky position: galactic absorption, background and exposure time values all have an impact on cluster detection. Offering these degrees of freedom in the selection function allows one to detach from the assumptions imprinted in the twin simulation. We have trained a second model with those three features, using the GPC formalism. Fig.~\ref{fig:GP_nH_Texp_simbkg_lx_z_AllXGOOD_cut_0-0_SEED0189} is a partial visualisation of the model output, all following quantities but exposure time $T_{\rm exp}$ being fixed at $z=0.3$, $N_H=3 \times 10^{20}$\,cm$^{-2}$, background brightness $5.2 \times 10^{-15}$ erg\,s$^{-1}$\,cm\,$^{-2}$\,arcmin$^{-2}$ in the 0.3--2.3\,keV energy band, $L_{X} = 10^{43}$ or $10^{44}$\,erg\,s$^{-1}$. The dependence of the model output on the exposure time appears clearly on this figure and thus the model is more precise in delivering a selection probability for a given cluster.

		\subsection{Unabsorbed count rate and redshift}

\begin{figure}
   \centering
   \includegraphics[width=\linewidth]{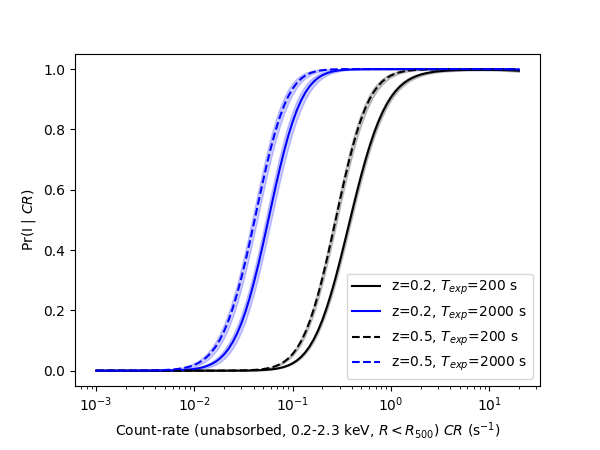}
   \caption{Representation of a model $\pofc{\isdet{cosmo}}{CR, z, N_H, T_{\rm exp}, {\rm bkg}}$ for two values of cluster redshift $z=0.2$ and $z=0.5$, fixed local galactic absorption $N_H$, a nominal background level, two values of exposure times $T_{\rm exp}$ (units s) and a range of count-rate values (x-axis). The shaded regions indicate the approximate 68\% confidence range output of the model.}
    \label{fig:GP_RateFLwoNH0223_nH_simbkg_Texp_z_Cosmo_cut_6-0_SEED000}
\end{figure}

In the course of the development of the eROSITA cosmology pipeline, trade-off discussions led to setting $\tsel$ as a five-component feature, namely: the cluster redshift $z$, sky position-dependent quantities $\mathcal{H} \equiv$ ($N_H$, $T_{\rm exp}$ and background) and the cluster $CR$. Here $CR$ represents the unabsorbed, 0.2--2.3\,keV survey-average count-rate collected within a $R_{500}$ aperture and unaffected by shot noise. It derives from the formula:
\begin{equation}
CR = \frac{L_{X}}{4 \pi d_L(z)^2 ECF(z, k_B T, N_H=0)} \ \ \ {\rm (units\ s}^{-1} {\rm )}
\end{equation}
In this expression, $ECF$ is the Energy Conversion Factor from the $0.5-2$\,keV band in the cluster rest-frame to the $0.2-2.3$\,keV band in the observer frame. It is computed for a APEC emission spectrum at temperature $k_B T$, redshift $z$, unabsorbed (hence $N_H=0$) and folded through the survey-averaged instrumental response of the 7 telescopes. This quantity is readily available for each simulated cluster in the twin simulation, since we know its redshift and temperature. The luminosity-distance $d_L$ is computed using the same cosmology as for the twin simulation.
Fig.~\ref{fig:GP_RateFLwoNH0223_nH_simbkg_Texp_z_Cosmo_cut_6-0_SEED000} represents a few slices through the multi-dimensional model $\pofc{\isdet{cosmo}}{CR, z, N_H, T_{\rm exp}, {\rm bkg}}$, fixing the value of $N_H$ and background level as previously, for two different redshift values and a range of count-rate ($CR$) values. The dependence on count-rate shows the usual "S"-shaped curve and larger exposure times imply higher sensitivities. The slight dependence on redshift is less intuitive: all other quantities being fixed, a source at higher redshift is more compact, thus its surface brightness is concentrated over a smaller area, making its detection comparatively easier.

The selection model obtained this way is quasi-independent of the cosmology imprinted in the twin simulation. It is also capable of reflecting the variations of selection depth over the sky. The cluster model is relatively simple (one cluster is represented by one count-rate and one redshift), which makes the computation of $\pofc{CR, z}{\tmodel}$ fast and easy in likelihood (Eq.~\ref{eq:number_counts_equation}). However this simplicity comes at a cost, since all variations of the selection due to e.g., cluster morphology, are marginalized over, relying on the distribution of morphologies in our twin simulation.

	\subsection{Internal validation of the selection models\label{sect:gp_internal_validation}}

We now quantify the absolute performance of models relying on intermediate variables $\tsel$. By doing so we will also assess the relative performance between models involving various parameters as input. We use our test sample to make such tests and compare the predicted model outcomes to the actual detection flags in the sample. Similarly as in Sect.~\ref{sect:logistic_profile} we will highlight two performance tests: the precision-recall curve and the calibration curve. 

\begin{figure}
   \centering
   \includegraphics[width=\linewidth]{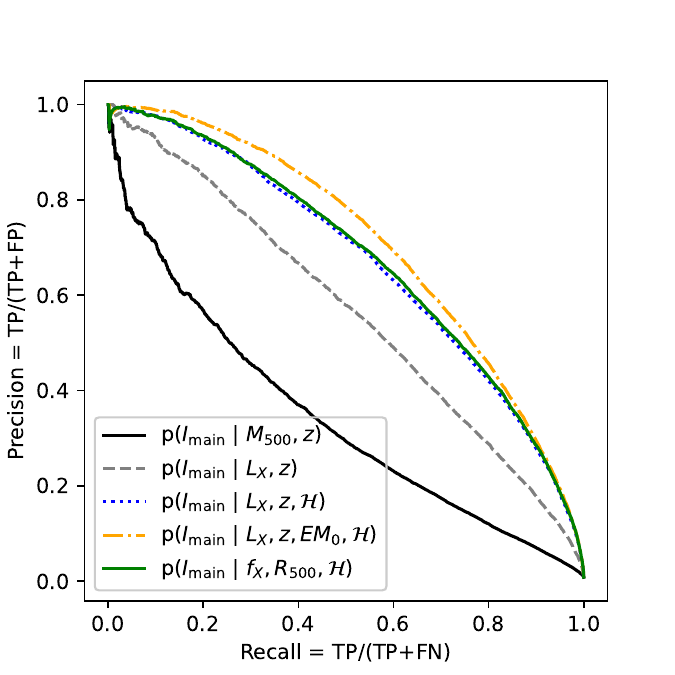}
   \caption{Precision-recall curves obtained with selection models trained to predict the selection of a cluster in the primary sample. The input parameters vary from one model to the other, as indicated in legend. A model involving luminosity, redshift, morphological features and local sky information performs best (yellow dot-dashed curve).}
    \label{fig:mainmetric_recprec}
\end{figure}

Fig.~\ref{fig:mainmetric_recprec} compares five of our models predicting the presence of a cluster in the eRASS1-main sample. They differ from each other by their input parameters (features $\tsel$). The model involving only mass and redshift has quite a low precision and recall (average precision of $34\%$). Predicting the detectability of a cluster requires indeed more detailed characterisation. Changing to a model taking luminosity and redshift notably improves the overall performance, since luminosity is more tightly linked to photon counts than mass. However, it still has an average precision of $56\%$, unsatisfactory enough for detailed analyses. Including sky position-dependent parameters $\mathcal{H} = (N_H, T_{\rm exp}, {\rm bkg})$ dramatically boosts the classification performance (average precision of $67\%$). As expected, adding information on the local background, exposure time and foreground absorption increases the capability of the model. An additional performance gain is obtained by adding a morphological parameter, here the $EM_{0}$ parameter (as presented in App.~\ref{app:more_models}). The average precision increases to $71\%$, closer to that obtained with the $7 \times 5$ surface brightness model depicted in Fig.~\ref{fig:logistic_precision_recall}. Finally, a model involving (absorbed) flux, apparent $R_{500}$ radius and $\mathcal{H}$ shows good performance with an average precision reaching $68\%$.

\begin{figure}
   \centering
   \includegraphics[width=\linewidth]{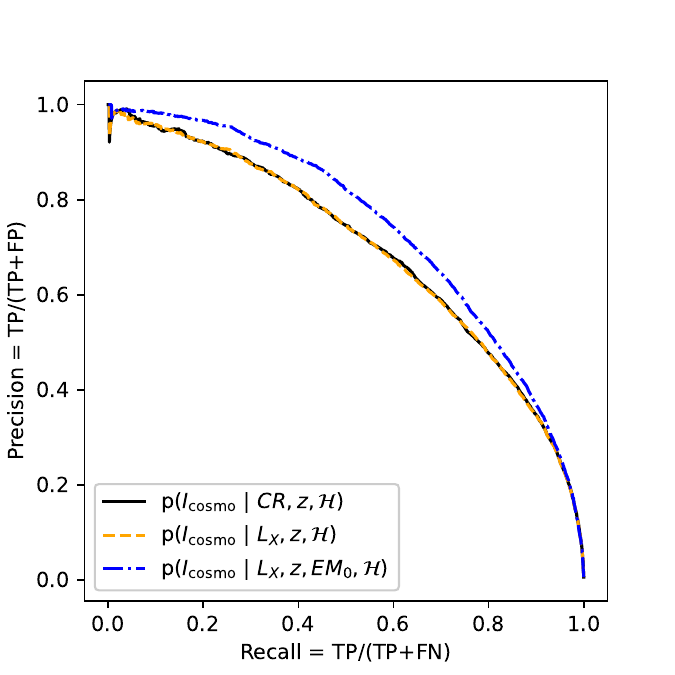}
   \caption{Similar as Fig.~\ref{fig:mainmetric_recprec}, comparing three models predicting the presence of a cluster in the cosmology. The model represented with the black curve corresponds to the baseline selection model used in the analysis of \citet{2024Ghirardini}.}
    \label{fig:cosmometric_recprec}
\end{figure}

A similar assessment is provided in Fig.~\ref{fig:cosmometric_recprec}, now for models predicting the presence of a cluster in the eRASS1 cosmology sample. The baseline model used for the eRASS1 cosmological analysis relies on $CR$ (count-rate) and redshift, as well as on sky local information. Its good performance (average precision 70\%) is comparable to a model involving luminosity and redshift. Again, adding morphological information (here via $EM_0$, see App.~\ref{app:more_models}) provides a noticeable enhancement of the model precision (average precision $74\%$).

\begin{figure}
   \centering
   \includegraphics[width=\linewidth]{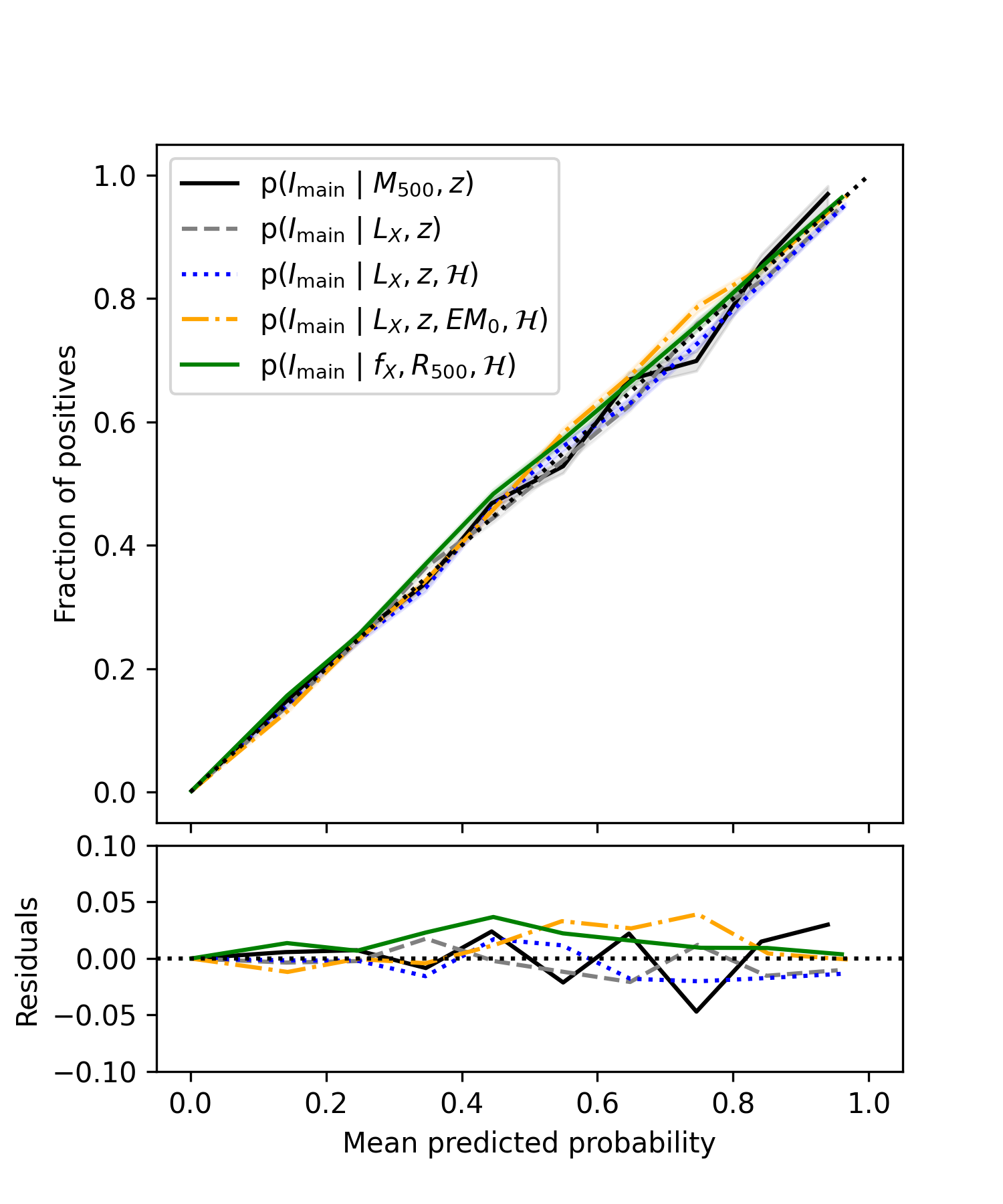}
   \caption{Evaluating the reliability of five models predicting the probability of a cluster to appear in the eRASS1 primary cluster catalog (x-axis). The y-axis represents the actual fraction of selected objects in the test sample. The bottom panel shows the residuals (difference) of each curve with respect to the one-to-one line. Shaded regions indicate 68\% confidence intervals in each bin of probability.}
    \label{fig:mainmetric_calib}
\end{figure}

\begin{figure}
   \centering
   \includegraphics[width=\linewidth]{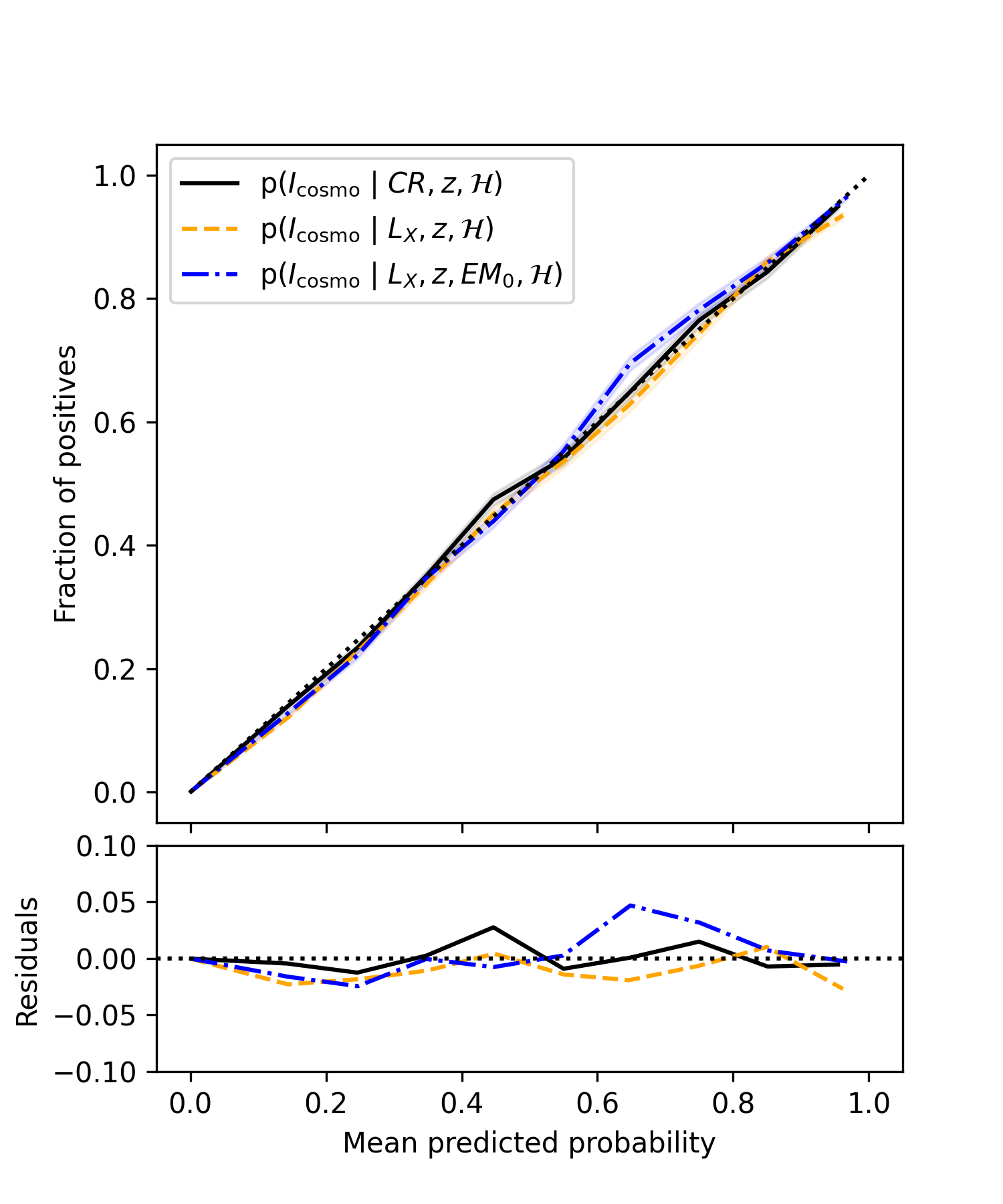}
   \caption{Similar as Fig.~\ref{fig:mainmetric_calib}, comparing three models predicting the presence of a cluster in the cosmology sample.}
    \label{fig:cosmometric_calib}
\end{figure}

We now assess the reliability of the probabilistic output of the models. In Fig.~\ref{fig:mainmetric_calib} we compare the same five models for the primary catalog selection function. All curves lie very close to the one-to-one line: the probability outputs reflect well the actual fraction of positive detections in the test catalog. The result for the cosmology sample classifiers appear in Fig.~\ref{fig:cosmometric_calib} and they all are very well aligned along the one-to-one line.

The tests depicted above illustrate good prediction performance of the models, in the sense that probability outputs correctly reflect the actual selection probabilities. This performance indicator is valid for the ensemble of clusters in the simulated eRASS1 sky.
In the course of our model production efforts, we have monitored several other indicators in order to assess the performance and reliability of the models. Among them were: the Receiver-Operator Curve (ROC) comparing recall and fall-out rates $FP/(FP+TN)$; the Detection-Error Tradeoff Curve (DET) comparing fall-out and miss rates $FN/(FN+TP)$ and the correlation coefficient suggested by \citet{1975Matthews}. All together, these performance indicators helped in selecting the best models to be used in the cosmological analysis.

%--------------------------------------------------------------------
\section{External validation of the selection function\label{sect:external_validation}}

We have described several models and we validated their performance on a simulated test set. This test set is produced in a very similar manner as the training sample, splitting the original twin simulation set in two parts. Additional tests are required using new samples independent from the twin simulation. We aim to test eRASS1 selection models with i)~a sample of clusters selected from a much deeper dataset (eFEDS) and ii)~two SZ-selected sample selected from South Pole Telescope surveys of different depths: SPTpol-100d and SPT-Early Cluster Sample (SPT-ECS). Selection biases affecting these samples must be taken into account; moreover we have to link their measured properties to eRASS1 selection variables. We present our formalism in the most generic manner, then we apply it to the specific cases of eFEDS and SPT-SZ.

\subsection{General presentation of the formalism}

Given two surveys, $R$ taken as a reference and the other $T$ to be tested, we wish to validate our selection function models for both $R$ and $T$. We do so by comparing the populations selected by each of the survey. In the optimal case where $R$ is much more complete (e.g.~deeper) than $T$, the problem does not require precise knowledge of the selection process leading to catalog $R$. In general both catalogs comprise a biased selection of the true underlying population of clusters, the latter we take as prior in the following analysis.

Extending the formalism laid out in previous work \citep[e.g.][]{2020Grandis} we calculate for each object $i$ in the reference catalog an associated probability $P_i = \pofc{\isdet{T}}{\isdet{R}, \tmes}$. There, $\isdet{T}$ and $\isdet{R}$ are random boolean variables indicating whether an object is listed in catalog $T$ and $R$ respectively. $\tmes$ is a collection of measured properties attached to the cluster: its measured flux, redshift, luminosity, extent, etc. are likely candidates. Using Bayes' rule we obtain:

\begin{equation}
\label{eq:P_i}
    P_i = \pofc{\isdet{T}^i}{\isdet{R}^i, \tmes} = \frac{\pof{\isdet{T}^i, \isdet{R}^i, \tmes^i}}{\pof{\isdet{R}^i, \tmes^i}} = \frac{N_i}{D_i}
\end{equation}
where:
\begin{equation*}
N_i = \int \dd \tmodel \pof{\tmodel} \pofc{\isdet{T}^i, \isdet{R}^i, \tmes^i}{\tmodel}
\end{equation*}
and
\begin{equation*}
D_i = \int \dd \tmodel \pof{\tmodel} \pofc{\isdet{R}^i, \tmes^i}{\tmodel}
\end{equation*}
From these equations, $D_i$ may be interpreted as the number of objects expected in the reference catalog with measured properties $\tmes^i$; $N_i$ may be interpreted as the number of objects expected in both the reference and test catalogs with these same measured properties. In this perspective, $P_i$ is a model for the fraction of reference objects listed with some measured properties $\tmes^i$ that are also present in the test sample.

These expressions may be expanded further by using selection variables $\tsel$ for each of the surveys. We have:
\begin{multline*}
\pofc{\isdet{T}^i, \isdet{R}^i, \tmes^i}{\tmodel} =  \int \dd \tselr \dd \tselt \dd \vec \alpha \\
\times \pofc{\tselt, \tselr, \vec\alpha}{\tmodel} \\
\times  \pofc{\isdet{R}^i}{\tselr} \pofc{\isdet{T}^i}{\tselt} \pofc{\tmes^i}{\vec \alpha}
\end{multline*}
and similarly:
\begin{multline*}
\pofc{\isdet{R}^i, \tmes^i}{\tmodel} =  \int \dd \tselr \dd \vec \alpha \pofc{\tselr, \vec\alpha}{\tmodel} \\
\times  \pofc{\isdet{R}^i}{\tselr} \pofc{\tmes^i}{\vec \alpha}
\end{multline*}
The last two expressions are valid as long as selection in the test survey only depends on variable $\tselt$ and selection in the reference survey only depends on variable $\tselr$. We have introduced a new, intermediate variable $\vec\alpha$ to reflect the dependence of the measurement vector. For instance, if the measurement vector is X-ray flux, $\vec\alpha$ may represent the cluster luminosity and redshift.

The interest of this decomposition lies in it involving the selection functions $\pofc{I}{\tsel}$, which are tabulated for each survey. It also accounts for covariances in the selection and measurement processes, through the multivariate distribution of $(\tselt, \tselr, \vec\alpha)$ given the values of $\tmodel$.

The probability $P_i$ (Eq.~\ref{eq:P_i}) is the main outcome of this model. It is calculated independently for each object listed in the reference catalog. By comparing this probability to the actual presence of a match in the test catalog, we obtain powerful diagnostics on the validity of the models, and the presence of outliers in the population. In particular, we will denote by `surprising detections' those objects listed in $R$ and detected in the test catalog despite their low forecast probability ($P_i < 0.025$ and $\isdet{T}^i=1$). We will denote `missed objects' those listed in $R$ and not detected in the test catalog despite their high forecast probability ($P_i > 0.925$ and $\isdet{T}^i=0$).

Our procedure to find matches between the reference and the test catalogs is relies on simple positional match. For efficiency reason we trim the reference catalog to the approximate sky footprint of the test sample and conversely. We then search for symmetric 2-arcmin matches between the samples, keeping only the best match, that is the closest distance. Such a procedure does not require redshifts to be identical in both catalogs. Our choice of the maximal separation corresponds to the most sensible cross-matching radius for surveys and instruments considered in this study \citep{2022Bulbul}.

Beyond the interpretation of individual $P_i$ values, we also base our evaluation on three aggregated diagnostics.
First, from repeated Bernoulli samples of $P_i$ values, we obtain a distribution for the expected number of matching entries between both catalogs, and its associated uncertainty. Comparing the actual number of matches to this distribution provides a bulk validation test of the model.
The second diagnostic consists in grouping clusters by their $P_i$ values in bins of finite width, and in computing the actual fraction of matches in each bin. This diagnostics informs on the reliability of the $P_i$ values,  hence on the underlying model.
Our third diagnostic empirically tests for leakage in our model, by introducing two numbers representing the fraction of objects that should be in the test catalog, but they are not ($\delta_1$) and the fraction of objects that should not be in the test catalog, but they are present ($\delta_2$). We compute posterior distributions for $\delta_1$ and $\delta_2$ by sampling the following likelihood (product of independent Bernoulli likelihoods):
\begin{multline}\label{eq:likelihood_leakage}
\mathcal{L}(\delta_1, \delta_2) = \prod_{i\, \in \, R} \left[ P_i (1-\delta_1) + (1-P_i) \delta_2 \right]^{\isdet{T}^i}\\
\times \left[ P_i \delta_1 + (1-P_i)(1-\delta_2) \right]^{1-\isdet{T}^i}
\end{multline}
The product runs over all clusters in the reference catalog, and $\isdet{T}^i$ equals 1 if the cluster is matched in the test catalog, it equals 0 otherwise. If the posterior distribution for $\delta_1$ significantly excludes zero, it indicates that part of the clusters predicted to be in the test catalog actually escape our model and are not detected for some reason. Similarly, a distribution of $\delta_2$ significantly away from zero hints towards our model not capturing properly the number of undetected objects. In this computation we assume flat prior distribution for $\delta_1$ and $\delta_2$ bounded in the [0, 1] interval. We sample the posterior with a Monte-Carlo Markov Chain algorithm provided in the \texttt{emcee} package \citep{emceepaper}.

Using the formalism depicted in this section, we have run several study cases:
\begin{enumerate}
\item eFEDS as reference, eRASS1-cosmo as test (Sect.~\ref{sect:baseefeds_checkerass1cosmo});
\item eRASS1-cosmo as reference, eFEDS as test (Sect.~\ref{sect:baseerass1cosmo_checkefeds});
\item eFEDS as reference, eRASS1-primary as test (Sect.~\ref{sect:baseefeds_checkerass1main});
\item SPTpol-100d as reference, eRASS1-cosmo as test  (Sect.~\ref{sect:basesptpol_checkerass1cosmo});
\item eRASS1-cosmo as reference, SPTpol-100d as test (Sect.~\ref{sect:baseerass1cosmo_checksptpol}).
\item SPTpol-ECS as reference, eRASS1-cosmo as test (Sect.~\ref{sect:basesptecs_checkerass1cosmo}).
\item eRASS1-cosmo as reference, SPTpol-ECS as test (Sect.~\ref{sect:baseerass1cosmo_checksptecs}).
\end{enumerate}

Table~\ref{table:summary_external_validation} provides a summary of the setup for each test, along with the number of matched entries between the catalogs.

\begin{table*}
\caption{\label{table:summary_external_validation}Setup summary of the external validation tests. The last column indicates the expected number of matching entries $\langle N \rangle$ predicted by our model, together with its uncertainty.}
\centering
\begin{tabular}{cccccccc}
\hline\hline
Reference\tablefootmark{a}	&	Test	&	$N_{\rm match}$\tablefootmark{b}	&	$\tselr$	&	$\tselt$	&	$\tmes$	&	$\alpha$	&	$\langle N \rangle$ \\
\hline
eFEDS (531)	&	Cosmo &	35	&	$L_X, z, EM_0, T_{\rm exp}$	&	$CR, z, \mathcal{H}$	&	\ftroiscents{}, $\zefeds$	&	$L_X, z$	&	$19.5 \pm 3.2$	\\	
Cosmo (85)	&	eFEDS &	36	&	$CR, z, \mathcal{H}$	&	$L_X, z, EM_0, T_{\rm exp}$	&		\fluxerassco{}, $\zerass$	&	$L_X, z$	&	$35.7 \pm 1.3$	\\
eFEDS (531)	&	Main	&	61 &	$L_X, z, EM_0, T_{\rm exp}$	&	$CR, z, \mathcal{H}$	&	\ftroiscents{}, $\zefeds$	&	$L_X, z$	&	$30.5 \pm 4.2$	\\	
\hline
SPTpol\tablefootmark{c} (65)	&	Cosmo	&	8	&	$\zetasptpol$	&	$CR, z, \mathcal{H}$	&	$\xi$, $\zspt$	&	$L_X, z, \zetasptpol$	&	$11.4 \pm 2.4$	\\
Cosmo (19)	&	SPTpol	&	8	&	$CR, z, \mathcal{H}$		&	$\zetasptpol$	&		\fluxerassco{}, $\zerass$	&	$L_X, z, \zetasptpol$	&	$7.0 \pm 1.3$	\\
SPTecs\tablefootmark{d} (163)	&	Cosmo	&	103	&	$\zetasptecs, \gamma_{\rm field}$	&	$CR, z, \mathcal{H}$	&	$\xi$, $\zspt$	&	$L_X, z, \zetasptecs$	&	$ 103 \pm 4$	\\
Cosmo (968)	&	SPTecs	&	101	&		$\zetasptecs, \gamma_{\rm field}$	&	$CR, z, \mathcal{H}$		&	\fluxerassco{}, $\zerass$	&	$L_X, z, \zetasptecs$	&	$65.4 \pm 5.8$	\\
\hline
\end{tabular}
\tablefoot{
\tablefoottext{a}{Number in parentheses indicate the number of entries in the reference catalog selected in the approximate common sky footprint.}\\
\tablefoottext{b}{The number of matching entries $N_{\rm match}$ results from $2\arcmin$ coordinate positional matching.}\\
\tablefoottext{c}{SPTpol is the SPTpol-100d catalog presented in \citet{2020Huang}, keeping only entries above $z=0.25$.}\\
\tablefoottext{d}{SPTecs is the SPTpol-ECS catalog presented in \citet{2020Bleem}, keeping only entries with $\xi>5$ and $z>0.25$.}
}
\end{table*}

%%%%%%%%%%%%%

	\subsection{Testing the eRASS1 cosmological cluster sample against eFEDS\label{sect:baseefeds_checkerass1cosmo}}

The deep eFEDS survey consists of a 140\,deg$^2$ area scanned by eROSITA during the CalPV phase with exposure about 10 times that of the average eRASS1 survey. Specifically, the average net exposure of eFEDS is about 1200\,s, while in eRASS1 it amounts to a median 80\,s in the same field (in both cases accounting for vignetting).

We opt for a population model described by $\tmodel = (M_{500}, z)$ representing the (true) mass within $R_{500c}$ and the (true) redshift of halos. The prior distribution $\pof{\tmodel}$ is the halo mass function $\dd n/\dd M_{500} \dd z$. Its exact shape in this specific experiment is not of critical relevance, given the wide sensitivity difference between eFEDS and eRASS1. We assume it follows the fit of \citet{2008Tinker}.

The eFEDS cluster sample \citep{2022LiuA} comprises 542 candidate galaxy groups and clusters, among them 531 are in the eRASS1 catalog footprint. Each entry is associated to a measured redshift $\zefeds$ (column \texttt{z} in the catalog) and to an estimated $0.5-2$\,keV flux within a 300\,kpc aperture (column \ftroiscents{}, expressed in units \fluxunit{}). These two elements constitute our vector of measured features $\tmes$.
We construct a generative model for flux and redshift that depends on $\vec\alpha = (L_X, z, M_{500})$, where $L_X$ is a cluster (true) 0.5-2 keV luminosity integrated in a cylinder of radius $R_{500}$.
Our generative model further assumes independence of flux and redshift measurements, i.e.:
\begin{equation}\label{eq:efeds_generative_model}
\pofc{\tmes}{\vec\alpha} = \pofc{\zefeds}{z} \times \pofc{\ln \ftroiscents{}}{L_X, z, M_{500}}
\end{equation}

We take redshift measurements to be unbiased, with uniform Gaussian scatter at fixed $z$, that is supposed equal to $0.01 (1+z)$:
\begin{equation}\label{eq:efeds_z_generative_model}
\pofc{\zefeds}{z} \sim \normaldist{z}{0.01 (1+z)}
\end{equation}
We suppose a log-normally distributed flux \ftroiscents{} given the values of $\vec\alpha$, that is:
\begin{equation}\label{eq:efeds_flux_generative_model}
\pofc{\ln \ftroiscents{}}{L_X, z, M_{500}} \sim \normaldist{\ln \mu_{\ftroiscents{}}}{\sigma_{\ln \ftroiscents{}}}
\end{equation}
In this expression, both $\mu_{\ftroiscents{}}$ and $\sigma_{\ln \ftroiscents{}}$ depend on $(L_X, z, M_{500})$ as follows.
We first convert any set of values $(L_X, z)$ into a $0.5-2$\,keV flux within $R_{500}$ by assuming an APEC plasma model with temperature set by a standard luminosity-temperature relation \citep{2019Bulbul}. Foreground absorption is taken into account using a position-dependent value of the hydrogen column density \citep{2016HI4PI}. Aperture correction (from $R_{500}$ to 300\,kpc) requires knowledge of an emissivity profile for the ICM. We take an isothermal, isometallicity gas density profile obtained from local X-COP clusters \citep{2019Ghirardini}. The aperture correction thus depends solely on the value of $R_{500}$ (itself deriving from $M_{500}$ and $z$). We assume unbiased flux measurements and set $\mu_{\ftroiscents{}}$ to the value of this aperture-corrected flux.
We build a simple model for measurement errors by setting a power-law model, with an extra term $\epsilon^i$:
\begin{equation}\label{eq:efeds_fluxerror_generative_model}
\sigma_{\ln \ftroiscents{}} = 0.14 \left(\frac{\mu_{\ftroiscents{}}}{10^{-13}\,\mathrm{erg\,s}^{-1}\,\mathrm{cm}^{-2}}\right)^{-0.47} + \epsilon^i
\end{equation}
Numerical coefficients in this expression were obtained by fitting a power-law model to the flux uncertainties reported in \citet{2022LiuA}. In order to account for cluster-to-cluster variability of the error, the extra term $\epsilon^i$ is specific to each cluster (labelled by $i$). As an assumption, $\epsilon^i$ is the deviation of the reported error to the value predicted by the empirical power-law model.
The eFEDS cluster catalog contains some clusters for which only an upper limit is reported on the measured 300\,kpc flux. For these objects, we still assume a log-normal model. Its central value $\mu_{\ftroiscents{}}$ is taken to be half of the upper limit value; the dispersion is such that $\epsilon^i$ is the deviation of the reported upper limit to the fixed value of $2.5 \times 10^{-14}$\,\fluxunit{} (median of all upper limits in the catalog).
Although imperfect, the generative model depicted here should capture the main trends underlying the flux measurement process. It is aimed at predicting what would be the reported flux \ftroiscents{} and redshift $\zefeds$ of an eFEDS-detected cluster, for any set of (true) values $\vec\alpha= L_X, z, M_{500}$.

The reference survey selection function $\pofc{\isdet{R}^i}{\tselr}$ is the eFEDS selection function $\pofc{\isdet{eFEDS}}{L_X, z, EM_0, T_{\rm exp}}$. It is constructed as described in \citet{2022LiuA}, with addition of the morphological parameter $EM_0$ characterizing the central emissivity of the clusters (App.~\ref{app:more_models}). The exposure time $T_{\rm exp}$ varies from one cluster to the other and its value is read from the eFEDS exposure map \citep{2022Brunner}.

The eRASS1 selection $\pofc{\isdet{cosmo}}{\tselt}$ is the main model ingredient we wish to validate in this work. We present results for the model used in the eRASS1 cosmological analysis \citep[][see also Fig.~\ref{fig:GP_RateFLwoNH0223_nH_simbkg_Texp_z_Cosmo_cut_6-0_SEED000}]{2024Ghirardini}, involving count-rate $CR$, redshift $z$ as well as position-dependent, cluster-independent, selection features $\mathcal{H}^i$. These features are galactic absorption column density, local background brightness and exposure time and their values are read from static sky maps at each cluster position. The eRASS1 cosmology sample is restricted to clusters with eRASS1-measured redshift $0.1 < \zerass < 0.8$. We assume Gaussian-distributed errors for the eRASS1 redshifts with standard deviation $\sigma_z/(1+z) = 0.015$, see also Eq.~\ref{eq:erass1cosmo_z_generative_model}. The criterion \texttt{IN\_ZVLIM=True}, recommended by \citet{2024Kluge} is present in our model through the implementation of a sky mask.

The last ingredient required in the model is the distribution $\pofc{\tselt, \tselr, \vec\alpha}{\tmodel}$. Given the assumptions enumerated in this section, it reduces to $\pofc{L_X}{M_{500}, z}$. We take the scaling relation of \citet{2019Bulbul} relating the core-included $0.5-2$\,keV luminosity within $R_{500}$ to the $M_{500}$ mass and redshift $z$. It follows a log-normal distribution with scatter 0.27 in the luminosity direction. In practice, this operation implies computing a probabilistic weight on a three-dimensional grid representing the values of mass, redshift and luminosity.

Putting together the model ingredients, we are able to predict for each cluster in eFEDS, associated to a given flux and redshift measurement, what would be its probability to appear in the eRASS1 cosmological catalog of clusters. By performing a 2\,arcmin match between the eFEDS and eRASS1 catalog, we find 35 actual clusters in common between both samples \citep[see also][]{2024Bulbul}, and 90\% of the matches are separated by less than $30\arcsec$. 

\begin{figure}
   \centering
   \includegraphics[width=\linewidth]{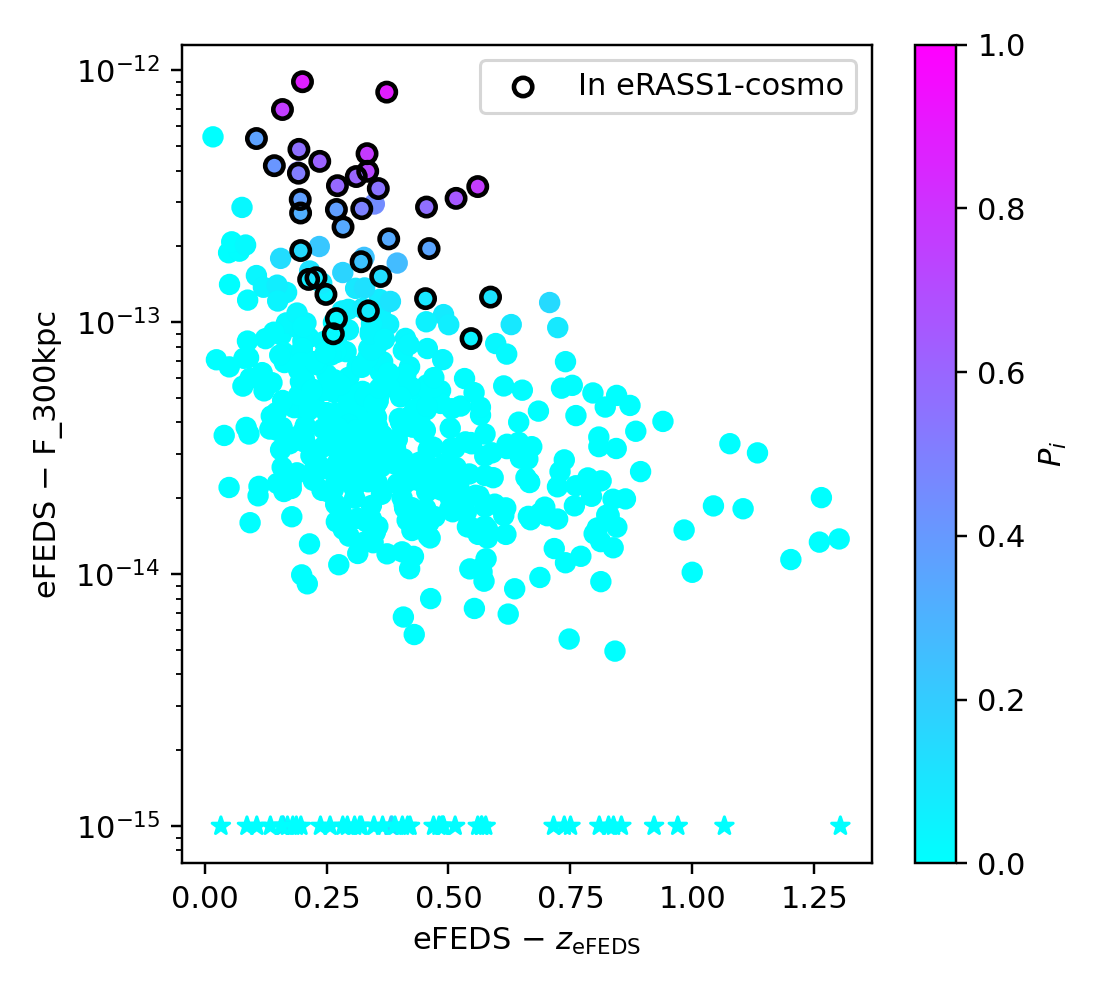}
   \caption{Testing the models with eRASS1-cosmo as a test catalog, eFEDS as a reference catalog. Each dot corresponds to one of the 531 eFEDS clusters in the plane of measured 300\,kpc flux (\ftroiscents{} in units \fluxunit{}) and measured redshift ($\zefeds$). eFEDS clusters with only flux upper limits are placed at the bottom with star symbols. Black circles indicate clusters also found in the eRASS1 cosmology sample. Colour reflects the model probability $P_i$ (Eq.~\ref{eq:P_i}) computed for each eFEDS cluster.}
    \label{fig:efeds_erass1_c_XCOPALL_feature_plot}
\end{figure}

\begin{figure}
   \centering
   	\includegraphics[width=0.9\linewidth]{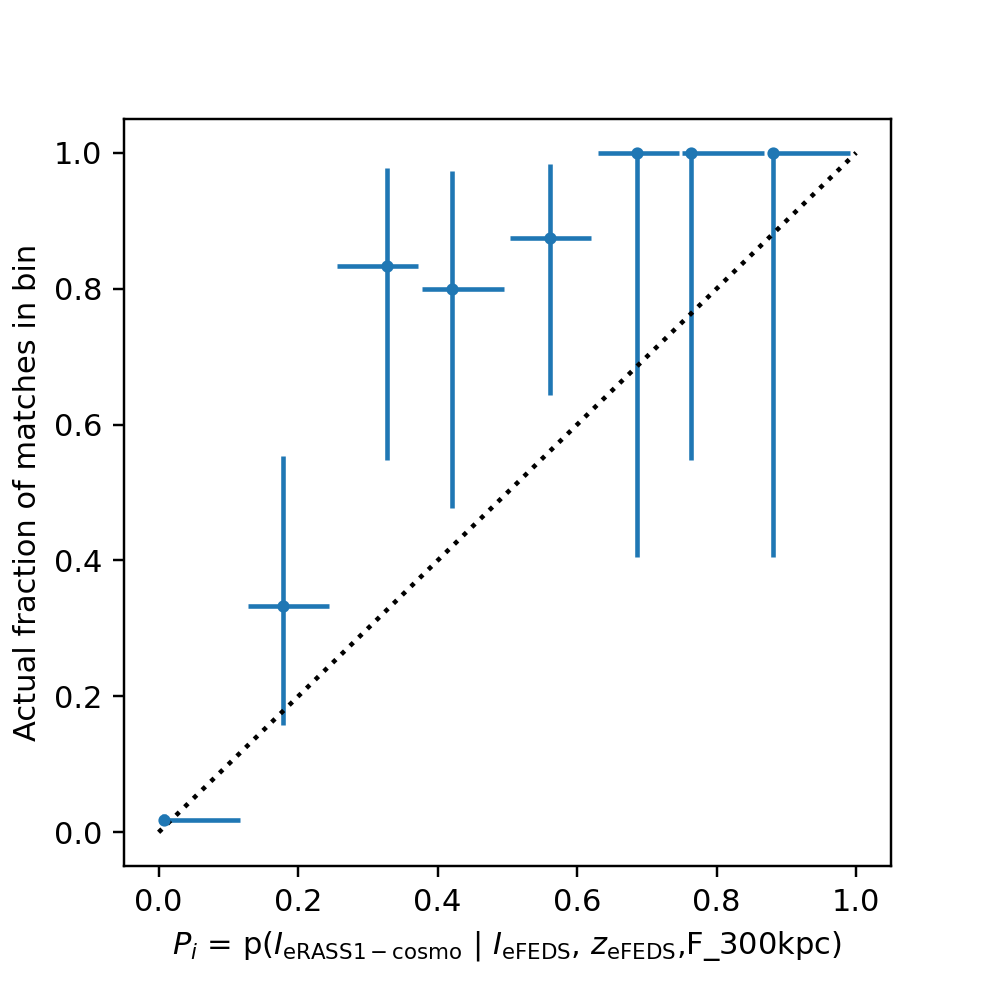}
   \caption{Testing the models with eRASS1-cosmo as test catalog, eFEDS as reference catalog. The 531 eFEDS entries are grouped in bins of the model output probabilities $P_i$ (horizontal axis). In each bin, the actual fraction of matches in the eRASS1 catalog is reported on the vertical axis. Error bars represent the estimated 68\% confidence range (App.~\ref{app:binomial_unc}). The dotted line indicates the one-to-one relation, that would follow a perfectly reliable model. In general, the model predicts slightly lower probabilities than actually observed.}
    \label{fig:efeds_erass1_c_XCOPALL_calibration}%
\end{figure}

Fig.~\ref{fig:efeds_erass1_c_XCOPALL_feature_plot} summarizes the outcome of this exercise. As expected, low-flux clusters are associated to a low $P_i$ value. Reassuringly, none of the `bright' (rigorously, the high-$P_i$) eFEDS clusters is missed in the eRASS1 cosmology sample.
The model predicts $19.5 \pm 3.2$ eFEDS clusters to be in the eRASS1 cosmology sample, while the actual number is 35 (that is, a 5-$\sigma$ underestimate).  We find the empirical parameters $\delta_1$ and $\delta_2$ have posterior distributions compatible with zero . However, the $P_i$ values appear systematically lower than the actual fraction of matches in each bin of $P_i$ values (Fig.~\ref{fig:efeds_erass1_c_XCOPALL_calibration}).
Overall these results point to a slight mismatch between the predicted detection rates and the actual number of matches. We will discuss this issue later in the paper.

	\subsection{Testing the eFEDS cluster sample against eRASS1 cosmology sample\label{sect:baseerass1cosmo_checkefeds}}

By exchanging the role of eFEDS and eRASS1 in the test presented above, we can use the eRASS1 cosmology catalog as a reference sample and test the eFEDS selection model. The interest of this operation is limited to sanity checks, because eRASS1 is 15 times shallower than eFEDS in their overlapping area. We consider only the 85 eRASS1 clusters extracted from the cosmology sample located in the fiducial eFEDS area bounded in Right Ascension by 8\,h and 10\,h and in Declination by $-5$ and $8 \deg$.

For describing each of the eRASS1 detections, we construct a model for the flux and redshift measurements of each entry. We assume independence of the estimated redshift $\zerass$ (column \texttt{Z\_LAMBDA} in the eRASS1 cosmology catalog) and the estimated $R_{500}$ flux in the band 0.5-2\,keV (\fluxerassco{}, units \fluxunit{}) and we form the model:
\begin{equation}\label{eq:erass1cosmo_z_generative_model}
\pofc{\zerass}{z} \sim \normaldist{z}{0.015 (1+z)}
\end{equation}
This model expresses normally distributed redshift uncertainties with error increasing with redshift. For flux measurements, we assume a log-normal distribution:
\begin{equation}\label{eq:erass1cosmo_flux_generative_model}
\pofc{\ln \fluxerassco{}}{L_X, z} \sim \normaldist{\ln \mu_{\fluxerassco{}}}{\sigma_{\ln \fluxerassco{}}}
\end{equation}
In this equation, $\mu_{\fluxerassco{}}$ and $\sigma_{\fluxerassco{}}$ depend on $L_X$ the true luminosity within $R_{500}$ and on the true redshift $z$.
Assuming unbiased flux measurements, $\mu_{\fluxerassco{}}$ is obtained by converting the cluster rest-frame $0.5-2$\,keV luminosity into the observer-frame $0.5-2$\,keV flux, assuming an isothermal ICM and a position-dependent hydrogen column density taken from the HI4PI survey \citep{2016HI4PI}. As for the uncertainty model, we proceed by fitting a simple model linking the catalog flux and errors reported in \citet{2024Bulbul}. We take:
\begin{equation}
\sigma_{\ln \fluxerassco{}} = 0.48 \left(\frac{\mu_{\fluxerassco{}}}{10^{-13}\,\mathrm{erg\,s}^{-1}\,\mathrm{cm}^{-2}}\right)^{-0.51} + \epsilon^i.
\end{equation}
The quantity $\epsilon_i$ is specific to each cluster in the eRASS1 sample and equals the deviation of the catalog uncertainty relative to the power-law model.

\begin{figure}
   \centering
   \includegraphics[width=\linewidth]{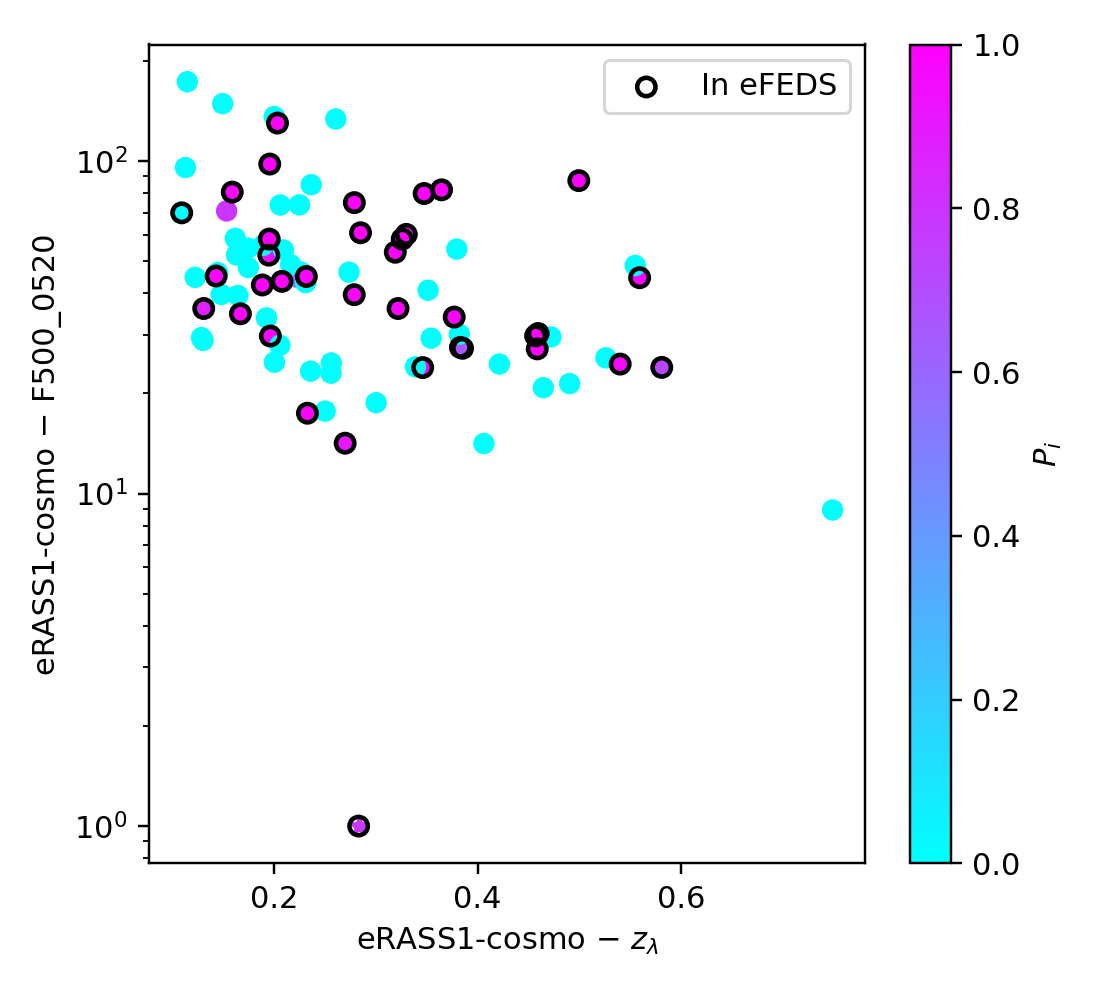}
   \caption{Figure similar to Fig.~\ref{fig:efeds_erass1_c_XCOPALL_feature_plot}, now taking eRASS1-cosmo as the reference sample (85 entries) and eFEDS as the test sample. The vertical axis represents \fluxerassco{}, the $0.5-2$\,keV flux within $R_{500}$ reported in the eRASS1 catalog (units $10^{-14}$\,\fluxunit{}). The horizontal axis is the catalog redshift $\zerass$. Clusters falling in zones of zero eFEDS exposure have vanishing probability $(P_i=0)$. Other eRASS1 clusters are very likely to be detected in eFEDS ($P_i \simeq 1$); in fact they are, as shown by the black circles. The isolated point at the bottom of the figure only has a flux upper limit $1.5 \times 10^{-13}$\,\fluxunit{}.}
    \label{fig:erass1_c_efeds_XCOPALL_feature_plot}
\end{figure}

Fig.~\ref{fig:erass1_c_efeds_XCOPALL_feature_plot} illustrates the outcome of this comparison exercise. The model predicts $35.7 \pm 1.3$ eRASS1 clusters to be found in eFEDS. The actual number of matches is 36\footnote{Border effects due to more or less restrictive sky masking lead to a different number of matches than mentioned in the previous section (35 matches). This effect is explicitly taken into account in our model.}. Note that a number of eRASS1 clusters fall in the non-exposed eFEDS area, hence their value of $P_i=0$. Most of the clusters covered by eFEDS have very high probabilities of being detected ($P_i \simeq 1$), as a consequence of its much deeper exposure. Reassuringly the eFEDS survey does not miss any eRASS1 cluster, highlighting in turn the low level of spurious contamination in the eRASS1 cosmology sample. In particular, it is not considered problematic that both clusters 1eRASS~J094023.3+022824 and 1eRASS~J084147.8-031154 have no corresponding match in eFEDS despite their respective predicted values $P_i=0.79$ and $0.55$. At their location on sky, the eFEDS survey is very shallow (with 140\,s and 10\,s effective exposure time, respectively) and the eFEDS selection model is not well calibrated in this regime.

The model explains correctly the outcome of matching eFEDS to eRASS1 cosmology catalog. There is no evidence for a bias in the value of the probabilities $P_i$; most of them are close to zero or one, in agreement with the cluster being absent or present in the eFEDS catalog (respectively). The values of $\delta_1$ and $\delta_2$ are themselves compatible with zero, indicating again consistency between the model and the present catalogs being tested.

	\subsection{Testing the eRASS1-main sample against eFEDS\label{sect:baseefeds_checkerass1main}}

We now turn to a test using the extended version of the eRASS1 catalog, comprising all X-ray clusters with $\extlike > 3$ \citep{2024Bulbul}. In contrast to the cosmology sample this catalog does not apply any redshift subselection. The model we construct is very similar to that presented in Sect.~\ref{sect:baseefeds_checkerass1cosmo}. In particular the distributions of eFEDS flux and redshift measurements are identical to that depicted in Eq.~\ref{eq:efeds_generative_model} to~\ref{eq:efeds_fluxerror_generative_model}.
The main modification consists in using a model selection function $\pofc{\isdet{T}^i = \isdet{main}^i}{\tselt}$. We take as selection variables the count-rate $CR$, true redshift $z$ and sky position-dependent parameters $\mathcal{H}$.

\begin{figure}
   \centering
   	\includegraphics[width=0.9\linewidth]{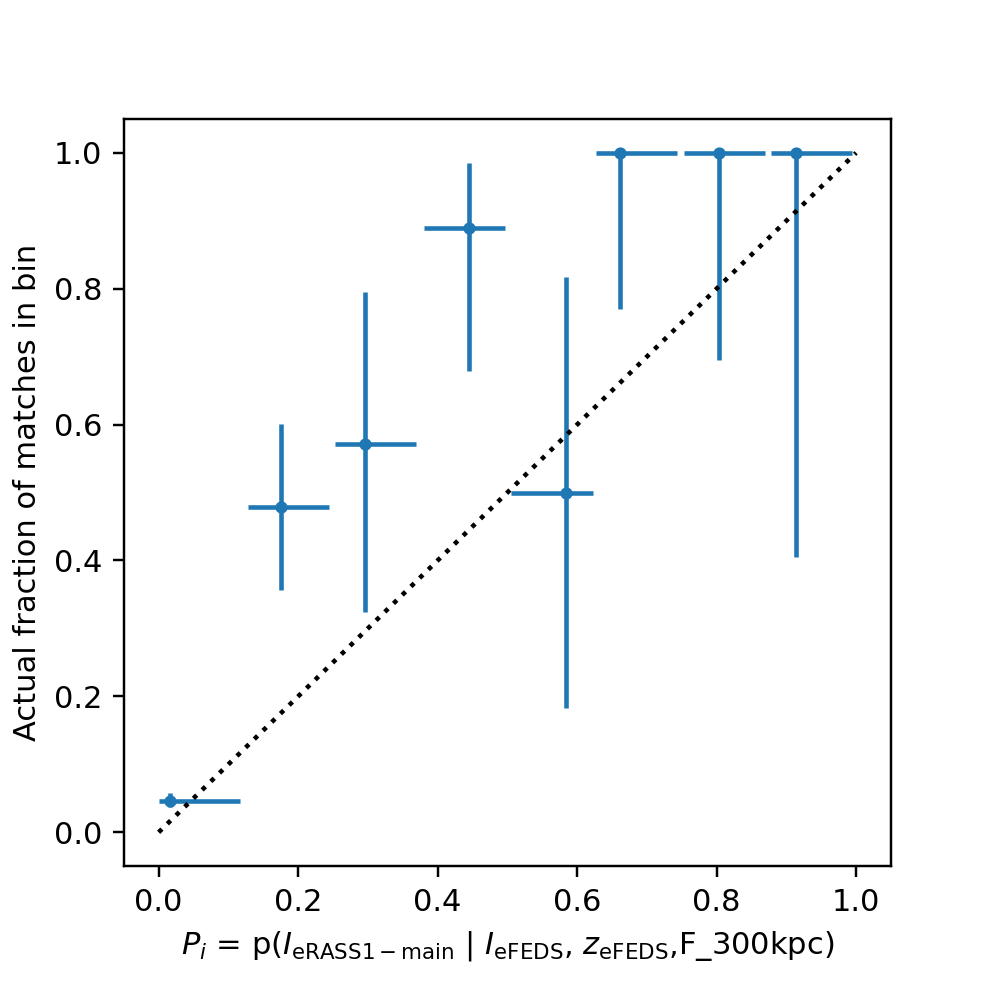}
   \caption{Figure similar as Fig.~\ref{fig:efeds_erass1_c_XCOPALL_calibration}, using the primary sample of clusters eRASS1-main as a test sample, instead of the eRASS1 cosmology sample. }
    \label{fig:efeds_erass1GT3_XCOPALL_calibration}%
\end{figure}

\begin{figure}
   \centering
   	\includegraphics[width=\linewidth]{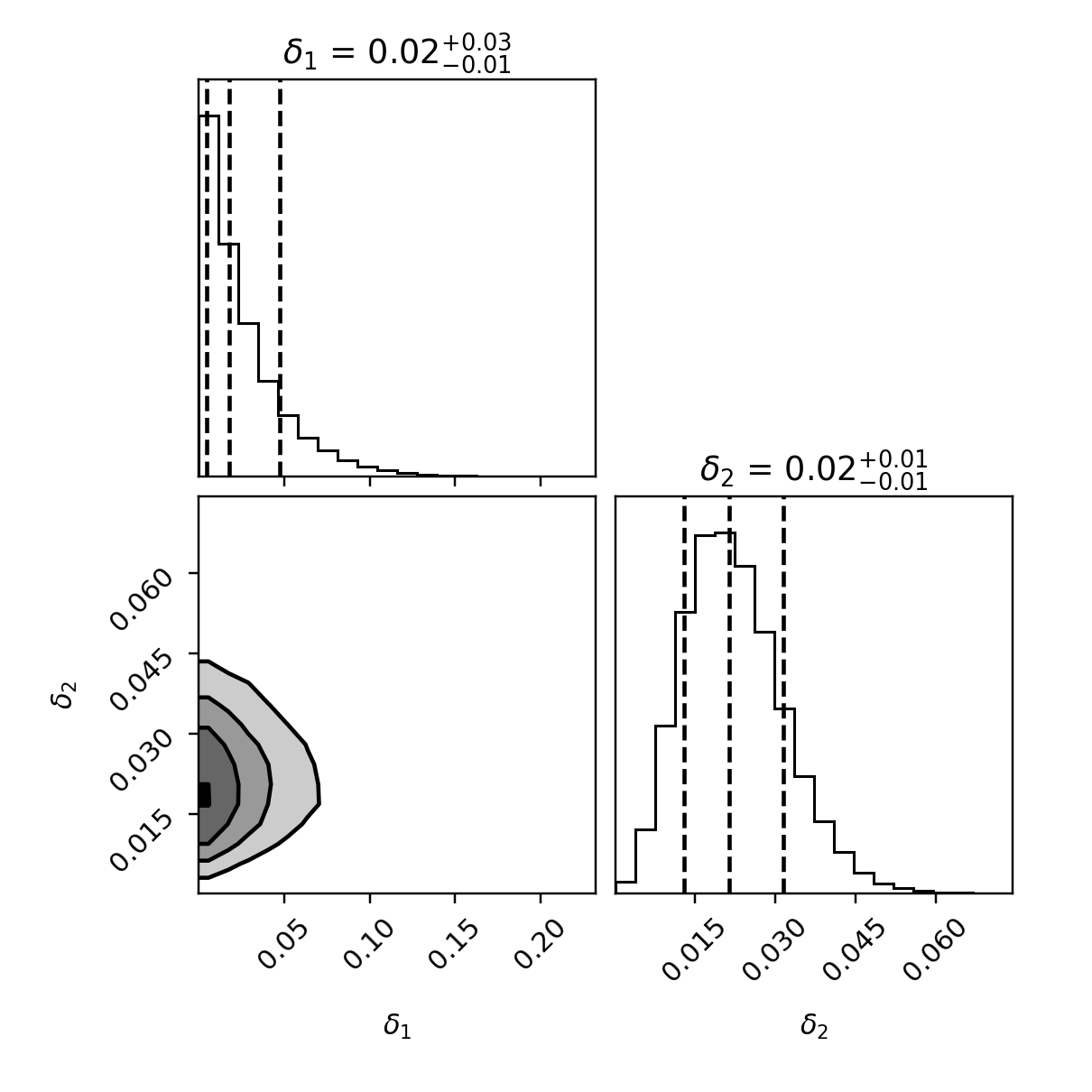}
   \caption{Posterior distribution for the pair of parameters $(\delta_1, \delta_2)$ introduced in Eq.~\ref{eq:likelihood_leakage}, for the experiment using eRASS1-main as a test and eFEDS as reference sample. The histograms represent the marginalized posterior distributions, with mean and 68\% range overprinted. The contours represent the equivalent 0.5-, 1-, 1.5- and 2-$\sigma$ distribution of the joint posterior. Clearly the parameter $\delta_2$ departs from zero, indicating that a fraction of the systems we predict not be detected leak into the actual eRASS1 sample.}
    \label{fig:efeds_erass1GT3_XCOPALL_leakage}%
\end{figure}

Our model predicts $30.5 \pm 4.2$ matches, while the actual number is 61 (hence a larger than 7-$\sigma$ difference). We show one diagnostic plot on Fig.~\ref{fig:efeds_erass1GT3_XCOPALL_calibration}, comparing the agreement between the $P_i$ values output of the model and the actual fraction of matches in the eRASS1 catalog. This validation experiment using the eRASS1-main sample highlights a higher discrepancy than that obtained in the experiment described in Sect.~\ref{sect:baseefeds_checkerass1cosmo}, using the eRASS1 cosmology catalog. The model predicts less clusters to be found in eRASS1 than actually observed, due to underestimated $P_i$ values. This is confirmed with the posterior distribution of $(\delta_1, \delta_2)$ shown in Fig.~\ref{fig:efeds_erass1GT3_XCOPALL_leakage}, significantly away from the zero-point.
Seven objects are qualified as `surprises' ($P_i<0.025$ and matched). According to our model, they are too faint to appear in the eRASS1 sample. For two cases (1eRASS~J093024.6-020635 and 1eRASS~J084558.5+031340), deeper eFEDS data reveals point-sources near the cluster emission, that were correctly excised to provide a eFEDS flux measurement, however in the eRASS1 data they are undistinguishable from the cluster.
We do not find any cluster missed by eRASS1 according to this model. Overall, this result indicates our selection function for the primary cluster sample is less well understood than for the cosmology sample. Alternatively, our generative model for eFEDS flux \ftroiscents{} may fail to exactly reproduce the measurement process described in the eFEDS cluster catalog paper \citep{2022LiuA}. We will return to this issue in Sect.~\ref{sect:discussion}.

	\subsection{Testing the eRASS1 cosmological cluster sample against SPTpol-100d\label{sect:basesptpol_checkerass1cosmo}}

\citet{2020Huang} present a catalog of 89 galaxy clusters discovered via the Sunyaev-Zeldovich (SZ) effect in the SPTpol-100d field, located at R.A.~=~23h30m and Dec~=~$-55 \deg$. In the area bounded by 23h < R.A. < 0h and $-60 \deg$ < Dec < $-55 \deg$, there are 65 SPTpol clusters at $z>0.25$. This lower redshift limit is chosen to account for growing incompleteness of the SPT catalog in the low-$z$ regime. In all that follows we will force the SPTpol-100d selection function to take value zero at redshifts below $z=0.25$.

The SPTpol-100d selection function depends on one single parameter $\zetasptpol$. We write:
\begin{equation}
\pofc{\isdet{SPTpol}}{\zetasptpol} = \int^{u(\zeta)}_{-\infty} \left( 2 \pi \right)^{-1/2} e^{-x^2/2} \dd x
\end{equation}
The upper integration bound is a function of $\zetasptpol$, namely $u(\zetasptpol) = \left[\left(3+\zetasptpol^2\right)^{1/2} - 4.5\right]$. This expression reflects the SPTpol-100d selection being a mere thresholding on the signal-to-noise parameter $\xi>4.5$. The parameter $\xi$ is biased and normally distributed with unit standard deviation \citep[e.g.][]{2020Bleem, 2020Huang}:
\begin{equation}\label{eq:p_xi_zeta}
\pofc{\xi}{\zetasptpol} \sim \normaldist{\sqrt{\zetasptpol^2+3}}{1}.
\end{equation}

The measurement vector $\tmes$ is taken from the SPTpol-100d catalog, namely the \texttt{xi} and \texttt{redshift} columns, standing respectively for the measured, uncorrected signal-to-noise $\xi$ and the measured redshift $\zspt$. We again assume approximate independence of both measurements and we take as model:
\begin{equation*}
\pofc{\zspt}{z} \sim \normaldist{z}{0.02 (1+z)}
\end{equation*}
This expression broadly reflects the redshift measurement uncertainty. We use again Eq.~\ref{eq:p_xi_zeta} to generate the distribution for $\xi$, now seen as a measurement feature.

Involving the selection model for the eRASS1 cosmology sample (Sect.~\ref{sect:baseefeds_checkerass1cosmo}) requires incorporation of the $0.5-2$\,keV luminosity $L_X$ as a latent variable in the model. Consequently we need to construct a joint model for the distribution of luminosities and $\zetasptpol$. We combine the $L_X-M_{500}-z$ scaling relation mentioned previously \citep{2019Bulbul} and the $\zeta-M_{500}-z$ relation\footnote{We multiply their value of $A_{SZ}$ by $\gamma_{\rm field}=2.66$, as prescribed by the authors. We also account for a misprinted $h$ in their Eq.~10 (S.~Bocquet, priv. comm.)} of \citet{2020Huang} into a bivariate log-normal distribution with covariance matrix:
\begin{equation*}
\mathrm{Cov}\left(\ln L_X, \ln \zetasptpol \right) = \begin{pmatrix}
\sigma_{\ln L_X}^2 & \rho \sigma_{\ln L_X} \sigma_{\ln \zeta} \\
\rho \sigma_{\ln L_X} \sigma_{\ln \zeta} & \sigma_{\ln \zeta}^2
\end{pmatrix}
\end{equation*}
Here, $\sigma_{\ln L_X}$ and $\sigma_{\ln \zeta}$ are the log-normal scatters of both distributions at fixed mass $M_{500}$ and redshift $z$ (with values respectively 0.27 and 0.18). The value of $\rho$ ranges between $-1$ and $1$, it reflects correlated scatter between the SZ signal-to-noise and X-ray luminosity at fixed mass and redshift. We take a fiducial value $\rho=0.2$, we will investigate the impact of varying this parameter in Sect.~\ref{sect:discussion}.

\begin{figure}
   \centering
   	\includegraphics[width=\linewidth]{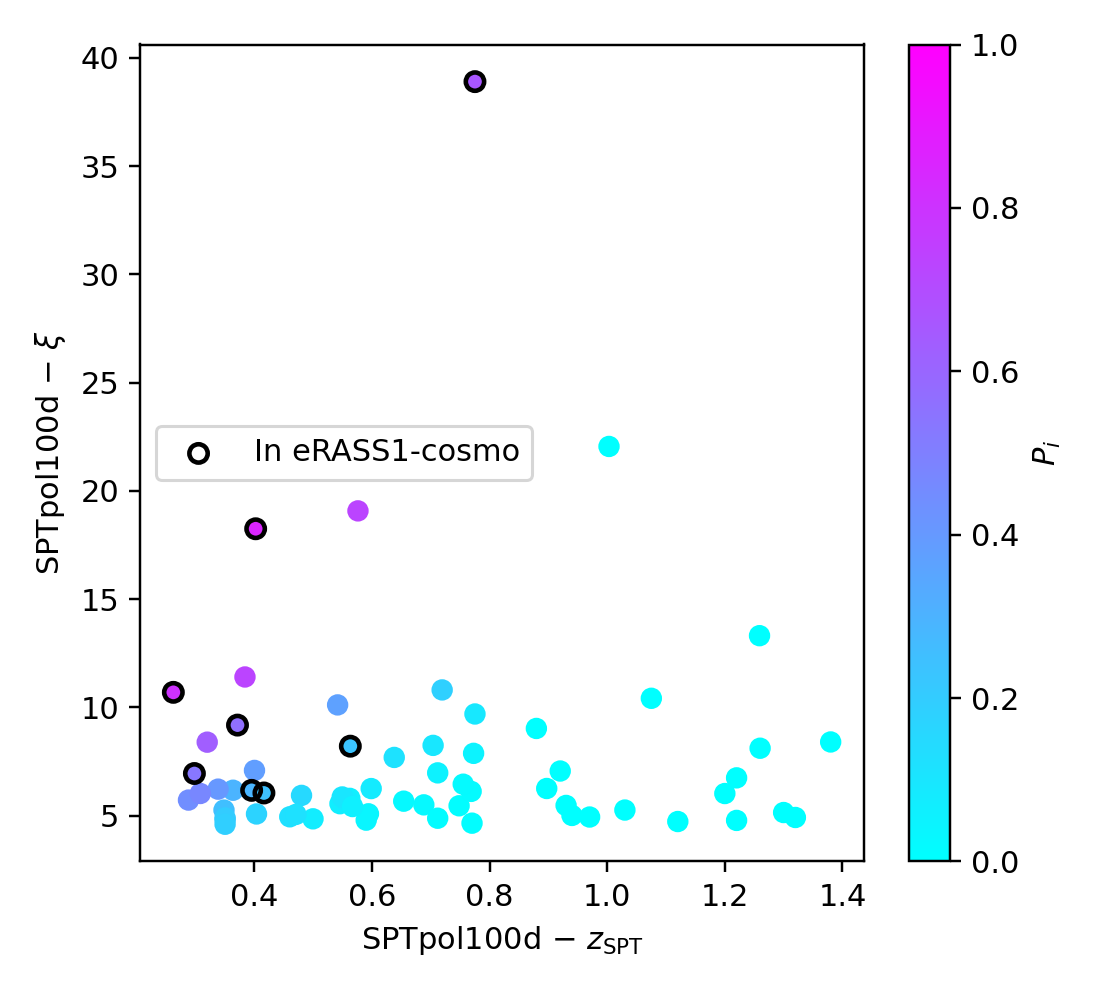}
   \caption{Figure similar as Fig.~\ref{fig:efeds_erass1_c_XCOPALL_feature_plot}, using the cosmology sample of eRASS1 clusters as a test sample, and the SPTpol-100d catalog as a reference sample (65 objects at $z>0.25$, coloured dots). The vertical axis represents the column \texttt{xi} in the SPTpol catalog, standing for the measured cluster signal-to-noise (unitless). The horizontal axis represents the measured redshift in the SPT catalog. The model predicts only a handful of eRASS1 detections among those objects, consistently with the observed number of 8 matches.}
    \label{fig:sptpolhuang_erass1_c_XCOPALL_feature_plot}%
\end{figure}

The model predicts $11.4 \pm 2.4$ matches, while the actual number of matches is 8. Fig.~\ref{fig:sptpolhuang_erass1_c_XCOPALL_feature_plot} shows the distribution of values $P_i$ output of the model, in the plane of the measured SPT mass and redshift. A small number of SPTpol-100d clusters have their $P_i$ high enough to be likely detectable in eRASS1. In fact, the matched instances are those with high $P_i$ values in general.

\begin{figure}
   \centering
   	\includegraphics[width=0.9\linewidth]{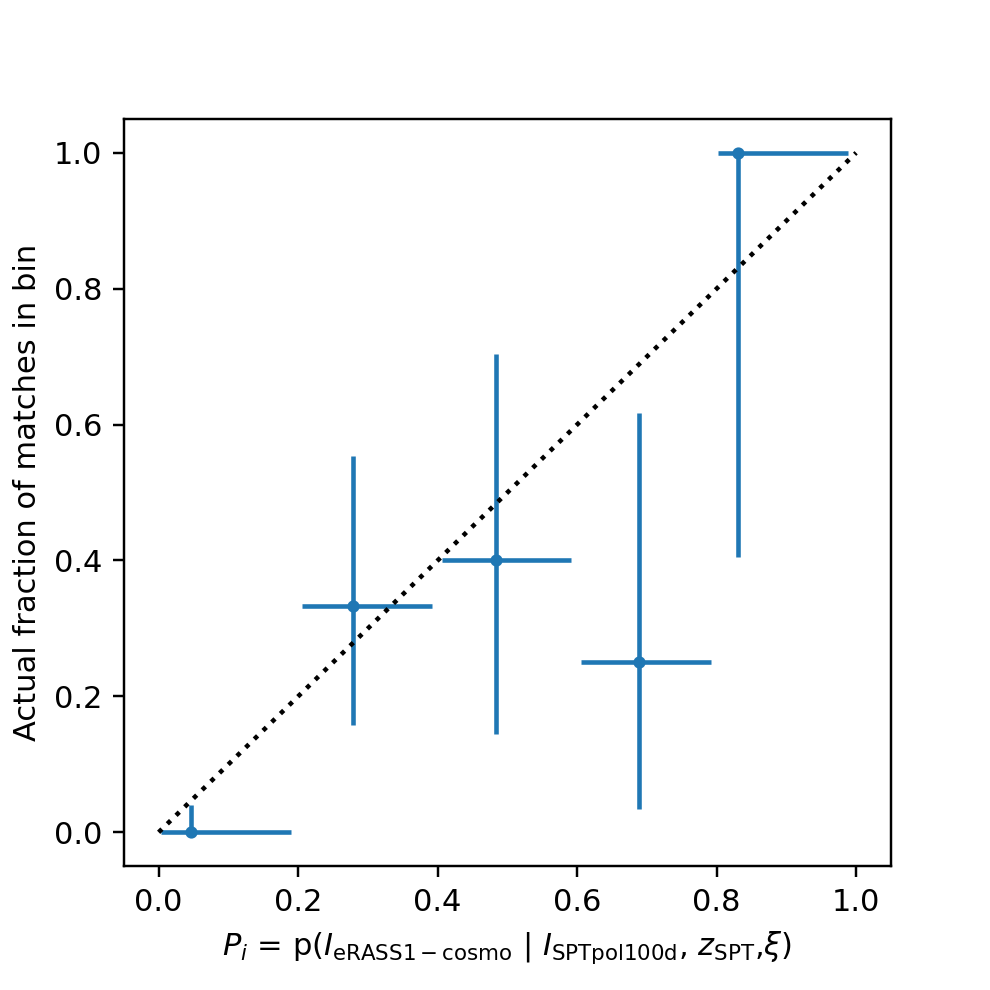}
   \caption{Figure similar as Fig.~\ref{fig:efeds_erass1_c_XCOPALL_calibration}, using the cosmology sample of eRASS1 clusters as a test sample, and the SPTpol-100d cluster sample as reference. Within the large uncertainties, there is good agreement between the predicted eRASS1 detection probability and the actual fraction of matched systems.}
    \label{fig:sptpolhuang_erass1_c_XCOPALL_calibration}%
\end{figure}

Fig.~\ref{fig:sptpolhuang_erass1_c_XCOPALL_calibration} demonstrates the fair agreement between the 65 $P_i$ values outcome of the model and the actual fraction of matches (in bins of $P_i$ values). From these diagnostics we conclude there is no evidence for inconsistencies either in the catalogs nor in the model. In particular, the selection function models account correctly for the observed matches between SPTpol-100d and eRASS1-cosmo.

	\subsection{Testing the SPTpol-100d cluster sample against eRASS1-cosmo\label{sect:baseerass1cosmo_checksptpol}}

\begin{figure}
   \centering
   	\includegraphics[width=\linewidth]{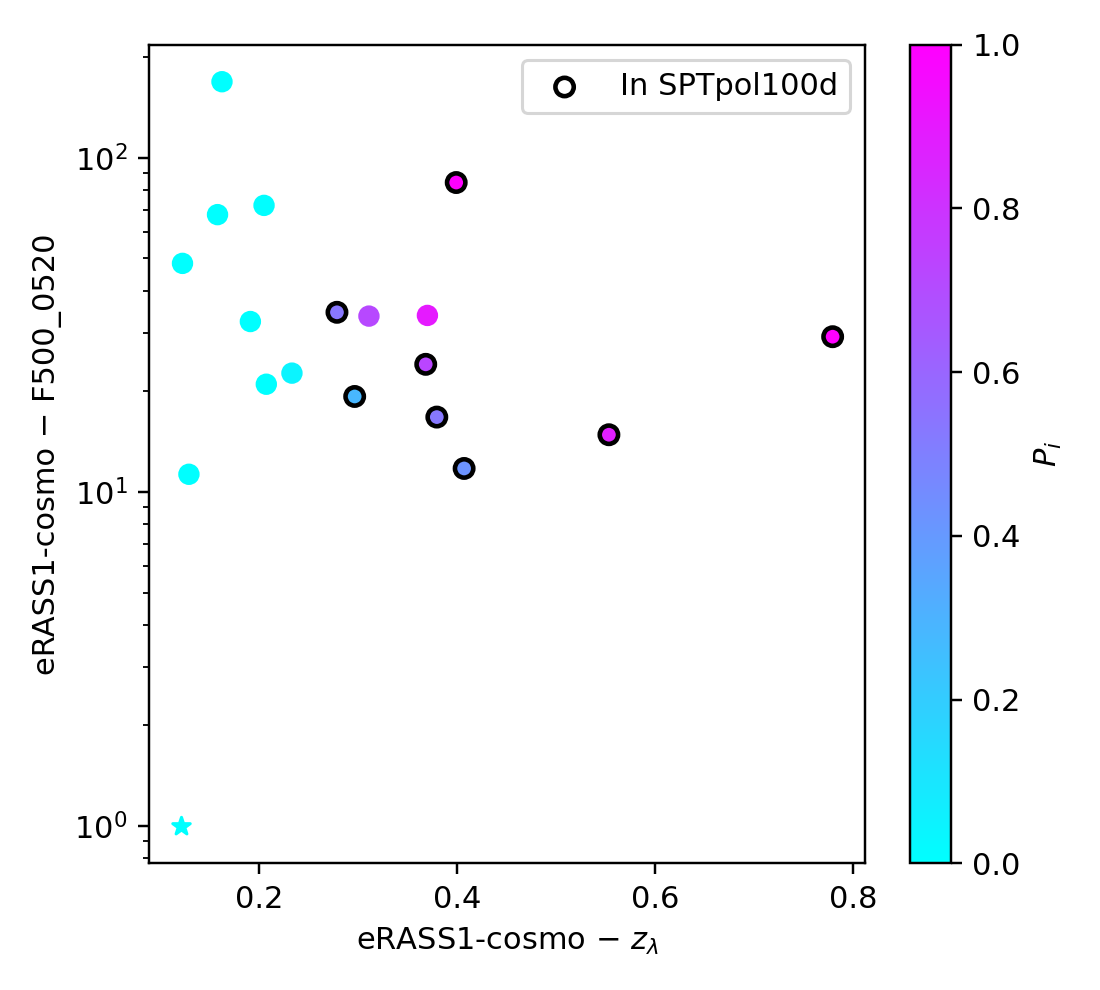}
   \caption{Figure similar as Fig.~\ref{fig:erass1_c_efeds_XCOPALL_feature_plot}, using the cosmology sample of eRASS1 clusters as a reference sample and the SPTpol-100d cluster sample as a test sample. Two eRASS1 clusters with high-$P_i$ values and absent from the SPTpol-100d catalog (hollow magenta dots) are discussed in the text. The star symbol at the bottom of the figure corresponds to a eRASS1 entry with upper flux limit $7 \times 10^{-14}$\,\fluxunit{}.}
    \label{fig:erass1_c_sptpolhuang_XCOPALL_feature_plot}%
\end{figure}

The previous test is reverted by exchanging the roles of the SPTpol-100d catalog with that of the eRASS1 cosmology sample. We thus test the SPTpol catalog, taking eRASS1 as reference. There are 19 eRASS1 clusters in the common footprint between both surveys. Construction of the model involves both selection functions and the same joint probability distribution $\pofc{L_X, \zetasptpol}{M_{500}, z}$ as described in Sect.~\ref{sect:basesptpol_checkerass1cosmo}. We also use a similar generative model as in Sect.~\ref{sect:baseerass1cosmo_checkefeds} for the flux \fluxerassco{} and redshift $\zerass$ listed in the eRASS1 catalog. The 19 output probabilities $P_i$ returned by the model are shown in Fig.~\ref{fig:erass1_c_sptpolhuang_XCOPALL_feature_plot}. The model predicts $7.0 \pm 1.3$ matches, while the actual number is 8, indicating consistency between the results and the predictions.

\begin{figure}
   \centering
   	\includegraphics[width=0.9\linewidth]{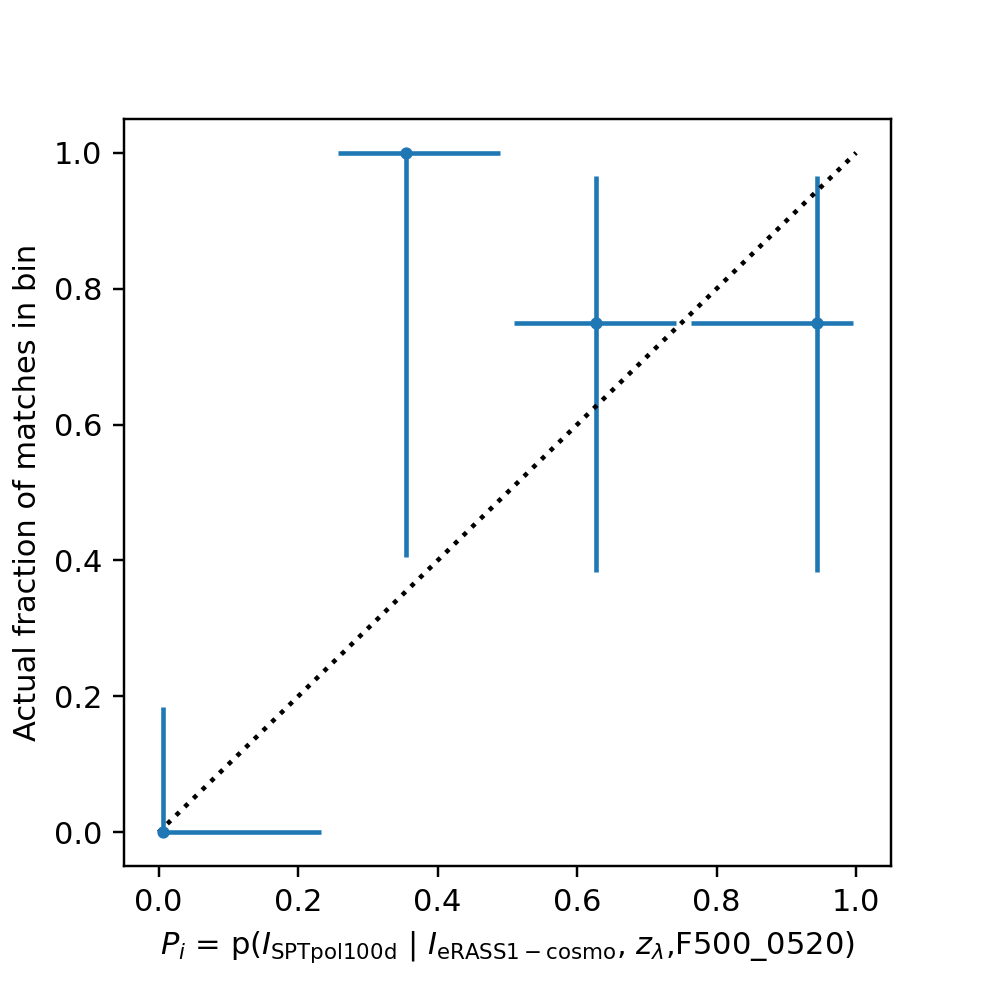}
   \caption{Figure similar as Fig.~\ref{fig:efeds_erass1_c_XCOPALL_calibration}, using the cosmology sample of eRASS1 clusters as a reference sample and the SPTpol-100d cluster sample as a test sample. The low number statistics (only 19 systems) prevents from finer binning in $P_i$ values (horizontal axis). Nevertheless, there is satisfactory agreement with the dotted (one-to-one) line.}
    \label{fig:erass1_c_sptpolhuang_XCOPALL_calibration}%
\end{figure}

Fig.~\ref{fig:erass1_c_sptpolhuang_XCOPALL_calibration} demonstrates adequacy between the 19 values of $P_i$ predicted by the model and the actual fraction of matches in each bin. Low number statistics prevents from gaining more insight into the model, calling for further experiments with larger samples of clusters.

As apparent from Fig.~\ref{fig:erass1_c_sptpolhuang_XCOPALL_feature_plot}, one cluster is not found in the SPTpol-100d survey, although our model predicts $P_i=0.90$. The eRASS1 cluster 1eRASS~J230029.2-510022 is measured at $\zerass = 0.37$ with $\fluxerassco = 3.4 \times 10^{-13}$\,\fluxunit{} and should be massive enough to appear in the SPTpol survey. It is missed by SPTpol, very likely due to its location close to the very edge of the surveyed field (at 23h Right Ascension).

Similarly, the eRASS1 cluster 1eRASS~J231306.7-550417 measured with $\zerass = 0.31$ with $\fluxerassco = 3.4 \times 10^{-13}$\,\fluxunit{} is not detected by SPTpol despite the model predicted $P_i = 0.72$.

	\subsection{Testing the eRASS1 cosmological cluster sample against SPTpol-ECS\label{sect:basesptecs_checkerass1cosmo}}

The model described for the SPTpol-100d catalog is straightforwardly adapted to the SPTpol-ECS catalog presented in \citet{2020Bleem}. The ECS sample is collected over $2770 \deg^2$ area in two regions of the Southern Hemisphere. A total of 163 clusters with significance $\xi > 5$ and located at redshift $z>0.25$ fall within the footprint of the eRASS1 cosmology sample. It is shallower than the SPTpol-100d survey, hence we expect a majority of the clusters to be found in the eRASS1 sample, up to $z=0.8$.

Our model follows exactly the same steps as in Sect.~\ref{sect:basesptpol_checkerass1cosmo}, with following modifications:
\begin{itemize}
\item the measurement features $\xi$ and $\zspt$ are taken from columns \texttt{XI} and \texttt{REDSHIFT} of the SPT-ECS catalog;
\item the unbiased significance parameter ruling SZ detections is denoted $\zetasptecs$. Its logarithm scales as a power-law with mass, as described in Eq.~4 of \citet{2020Bleem}. In particular we multiply the normalisation $A_{SZ}$ of this scaling relation by a factor $\gamma_{\rm field}$ specific to each sub-field of the SPT-ECS survey. An additional  multiplicative factor $\gamma_{\rm ECS} = 1.124$ rescales all values of $A_{SZ}$, thus reflecting calibration updates  \citep[Table~1 and Sect.~5 in][]{2020Bleem};
\item we only select entries with $\xi$ greater than 5.
\end{itemize}

\begin{figure}
   \centering
   	\includegraphics[width=\linewidth]{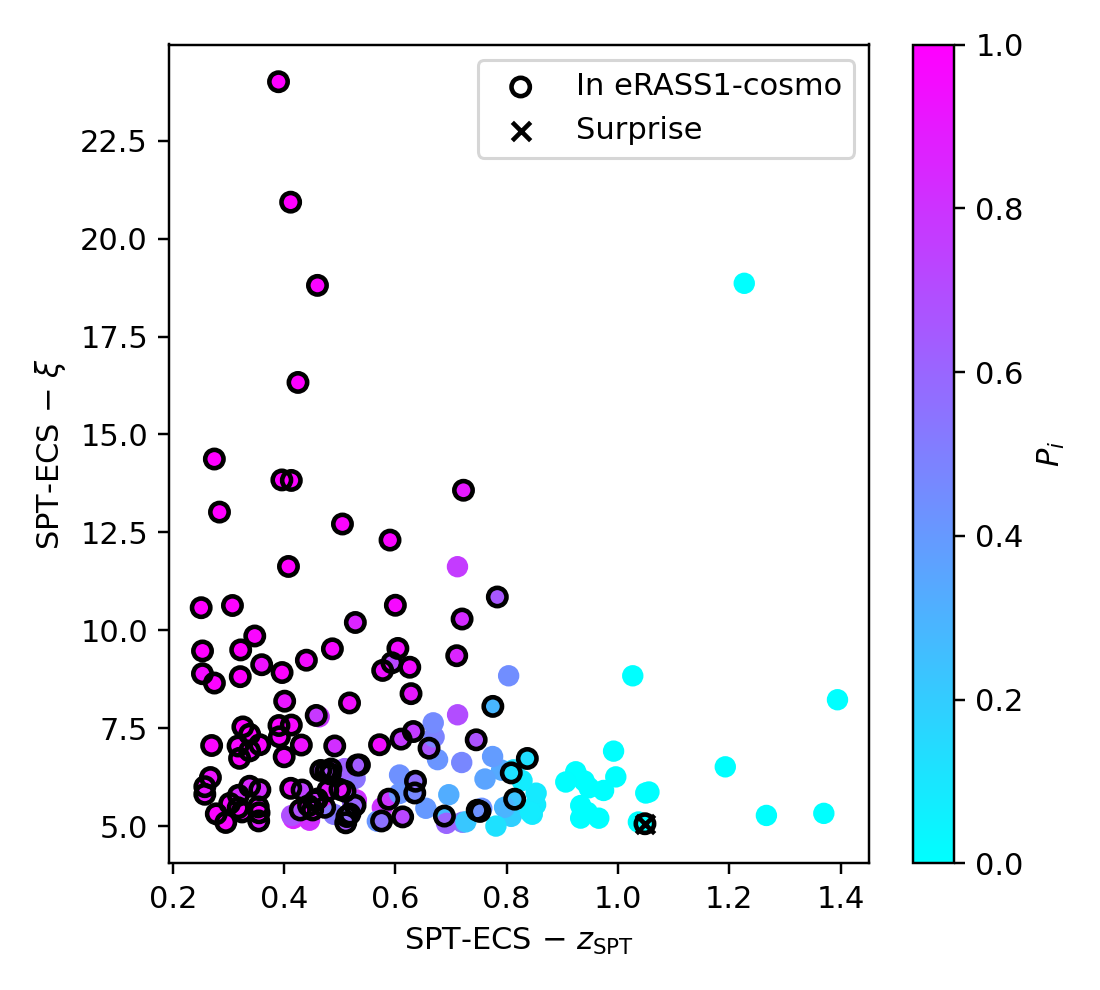}
   \caption{Figure similar as Fig.~\ref{fig:sptpolhuang_erass1_c_XCOPALL_feature_plot}, using 163 clusters in SPTpol-ECS catalog as a reference sample and the eRASS1 cosmology sample as test.}
    \label{fig:sptecs_erass1_c_XCOPALL_feature_plot}%
\end{figure}

In the following we keep a fixed value $\rho=0.2$ for the correlated scatter between $\ln L_X$ and $\ln \zetasptecs$.
There are 103 matching instances between both catalogs, based on a $2\arcmin$ positional match. Our model predicts $103 \pm 4$ matches, in excellent agreement with the observed number. Fig.~\ref{fig:sptecs_erass1_c_XCOPALL_feature_plot} shows the distribution of the 163 entries in the (measured) significance and redshift plane. Most of the SPT-ECS clusters at $z < 0.8$ are detected by eRASS1 (black circles). There is no obvious missing cluster among the 163 SPT detections. However we identified one `surprising' detection, SPT-CLJ0333-3707 (black `X' in the figure, $P_i \simeq 0$). It is listed in the ECS catalog with significance 5.1 and photometric redshift $\zspt=1.05 \pm 0.04$. The matching instance in eRASS1 is 1eRASS~J033323.9-370744, listed at $\zerass = 0.52 \pm 0.01$. This cluster is one of the 3\% ambiguous cases discussed in \citet{2024Kluge}, where two unrelated clusters overlap along the same line of sight.

\begin{figure}
   \centering
   	\includegraphics[width=0.9\linewidth]{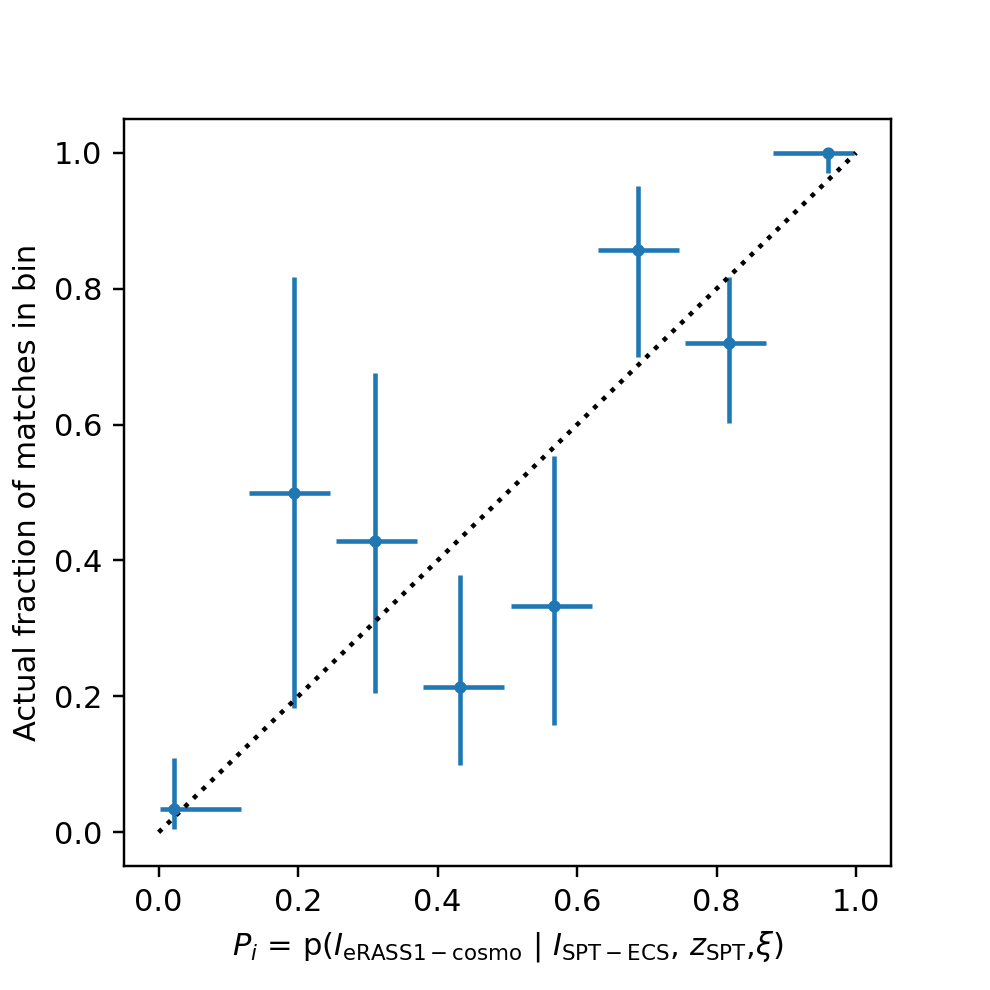}
   \caption{Figure similar as Fig.~\ref{fig:sptpolhuang_erass1_c_XCOPALL_calibration}, obtained by using the SPT-ECS sample as reference (163 entries) and using the cosmology sample of eRASS1 clusters as test.}
    \label{fig:sptecs_erass1_c_XCOPALL_calibration}%
\end{figure}

Fig.~\ref{fig:sptecs_erass1_c_XCOPALL_calibration} demonstrates the reliability of the $P_i$ values as predicted by the model. We conclude to satisfactory agreement between the model and the observed matches and non-matches in the two catalogs.

	\subsection{Testing the SPTpol-ECS sample against eRASS1 cosmology sample\label{sect:baseerass1cosmo_checksptecs}}

We conclude this series of tests by reversing the experiment of Sect.~\ref{sect:basesptecs_checkerass1cosmo}. There are 958 eRASS1 clusters in the cosmology catalog falling within the sky footprint of SPTpol-ECS survey. The model takes as measurement features \fluxerassco{} and $\zerass$ and it assumes identical generative distributions as in Sect.~\ref{sect:baseerass1cosmo_checkefeds}.

\begin{figure}
   \centering
   	\includegraphics[width=\linewidth]{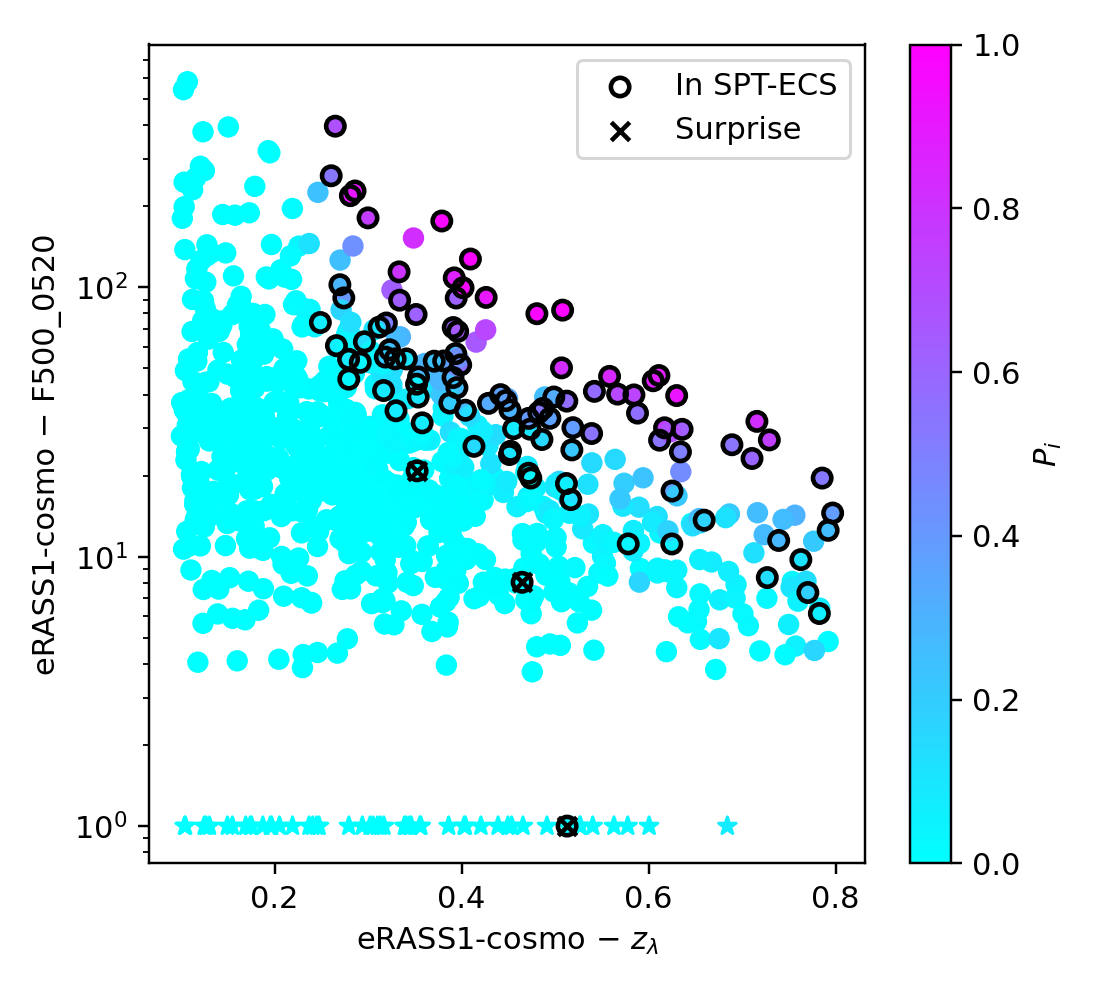}
   \caption{Figure similar as Fig.~\ref{fig:erass1_c_sptpolhuang_XCOPALL_feature_plot}, using 958 eRASS1 clusters from the cosmology sample as reference, and SPTpol-ECS catalog as a test sample. Star symbols at the bottom of the figure indicate eRASS1 objects with no flux measurement, possibly with an upper flux limit.}
    \label{fig:erass1_c_sptecs_XCOPALL_feature_plot}%
\end{figure}

\begin{figure}
\begin{tabular}{c}
	\includegraphics[width=0.8\linewidth]{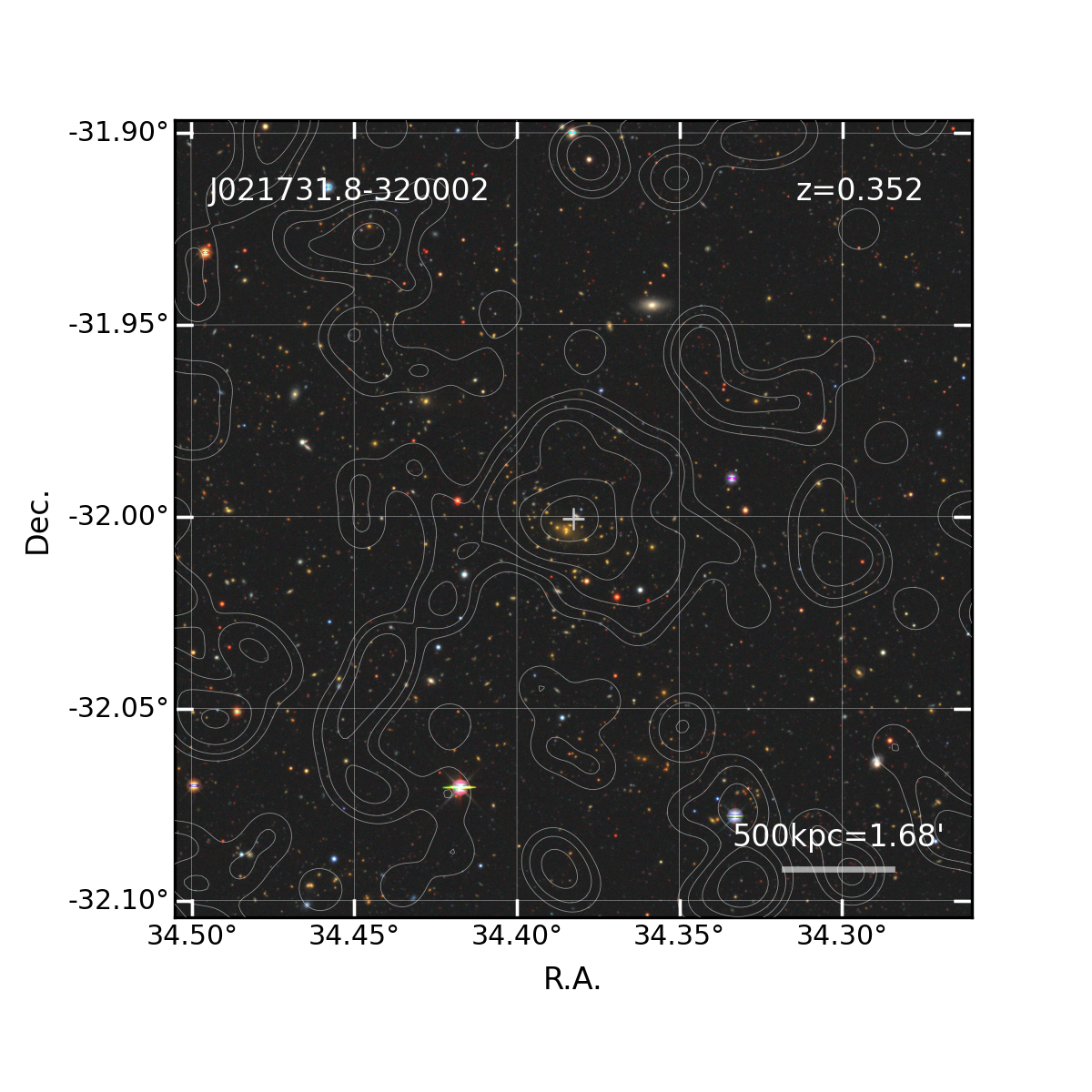} \\
	\includegraphics[width=0.8\linewidth]{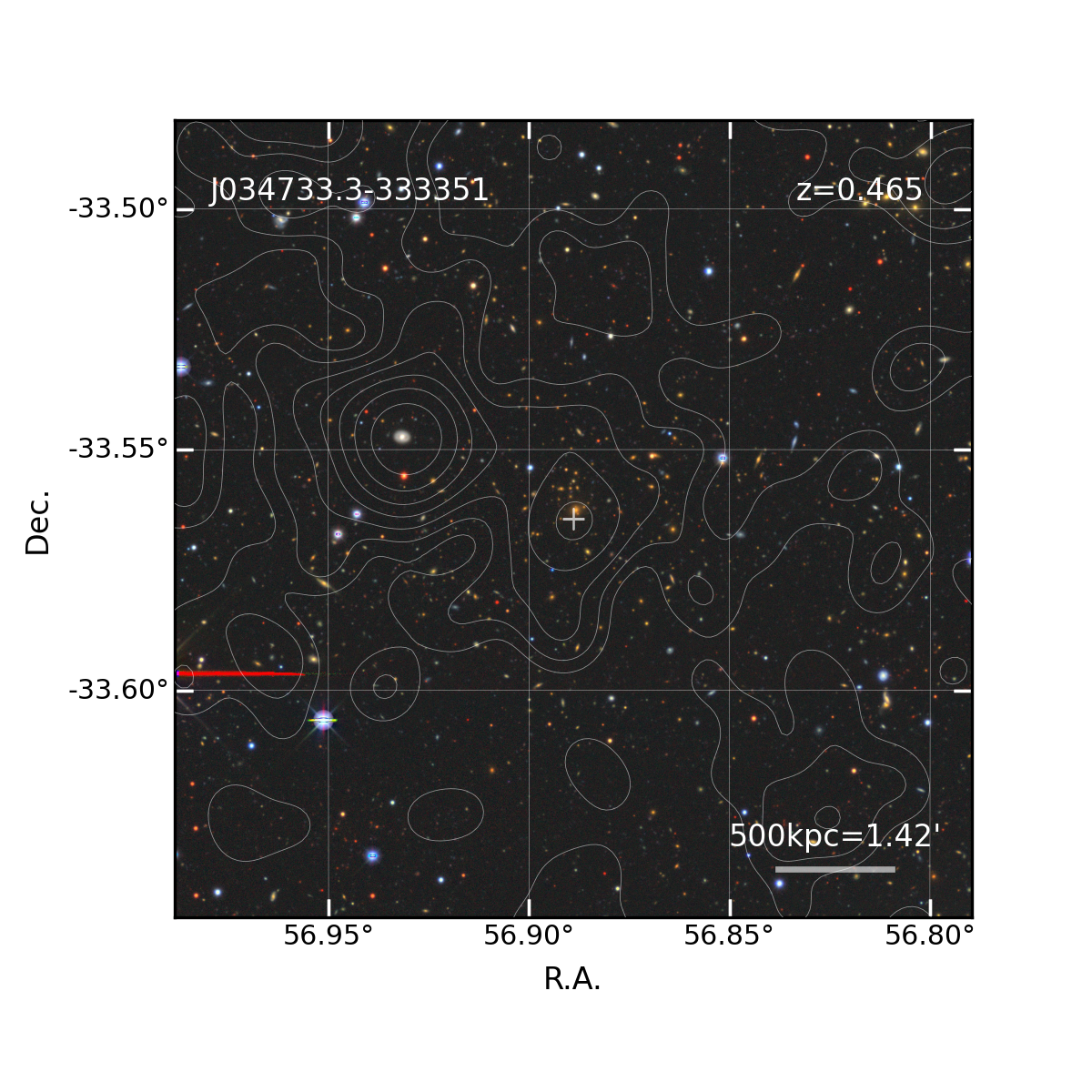} \\
	\includegraphics[width=0.8\linewidth]{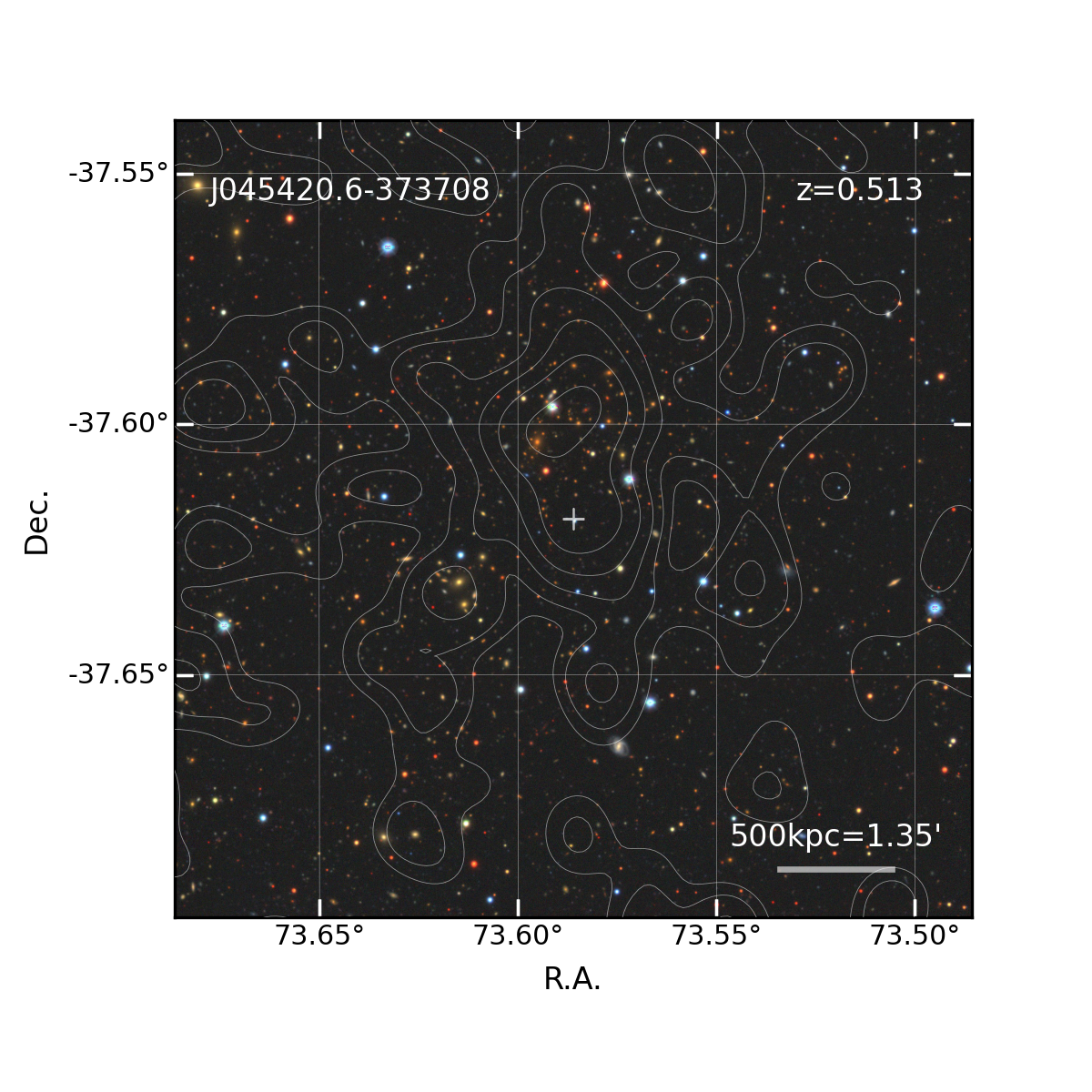} \\
\end{tabular}
	\caption{Optical cutouts (DR10 Legacy survey) and eRASS1 X-ray contours overlaid. These eRASS1 clusters are present in the cosmology sample and are unexpectedly matched to SZ detected clusters in the SPT-ECS sample. Our model indeed predicts a probability value $P_i = \pofc{\isdet{\rm SPT-ECS}}{\isdet{\rm cosmo}, \fluxerassco, \zerass } < 0.025$ for each of these entries.}
    \label{fig:sptecs_surprises}%
\end{figure}

Fig.~\ref{fig:erass1_c_sptecs_XCOPALL_feature_plot} shows the outcome of this calculation. Our model predicts $65.4 \pm 5.8$ matches, significantly less than the observed 101 matches observed in the common footprint.
In particular we find three unexpected SPT-ECS detections ($P_i < 0.025$). Secondary components along the line of sight of these three clusters may be responsible for the discrepant mass estimates, either by boosting the SZ significance or by decreasing the measured X-ray flux.
They are illustrated in Fig.~\ref{fig:sptecs_surprises}:
\begin{itemize}
\item 1eRASS~J021731.8-320002 is listed at $\zerass = 0.352 \pm 0.009$ with flux $2.1 \pm 0.4 \times 10^{-14}$\,\fluxunit{} and extent likelihood 14.4. The estimated X-ray mass is $M_{500, X} \simeq 3 \times 10^{14}\,M_{\odot}$. The matched SPT-ECS detection is at $\zspt = 0.354 \pm 0.009$ (from Magellan/FourStar) and $\xi=5.1$. Its estimated SZ mass is $M_{500, {\rm SPT}} \simeq 4 \times 10^{14}\,M_{\odot}$.
\item 1eRASS~J034733.3-333351 is listed at $\zerass = 0.465 \pm 0.009$ with flux $8_{-3}^{+10}\times 10^{-14}$\,\fluxunit{} and extent likelihood 7.5. The estimated X-ray mass is $M_{500, X}\simeq 2\times 10^{14}$\,M$_{\odot}$. The matched SPT-ECS detection is at $\zspt=0.45 \pm 0.01$ (photometric, from Magellan/FourStar) and $\xi=5.4$. Its estimated mass is $M_{500, {\rm SPT}} \simeq 4\times 10^{14}$\,M$_{\odot}$.
\item 1RASS~J045420.6-373708 is listed at $\zerass = 0.513 \pm 0.008$ without any flux measurement, nor an upper limit and extent likelihood 6.3. The corresponding SPT detection is at $\zspt = 0.509 \pm 0.006$ (photometric) and significance $\xi = 5.9$, with estimated mass $M_{500, {\rm SPT}} \simeq 4.8 \times 10^{14}$\,M$_{\odot}$.
\end{itemize}

\begin{figure}
   \centering
   	\includegraphics[width=0.9\linewidth]{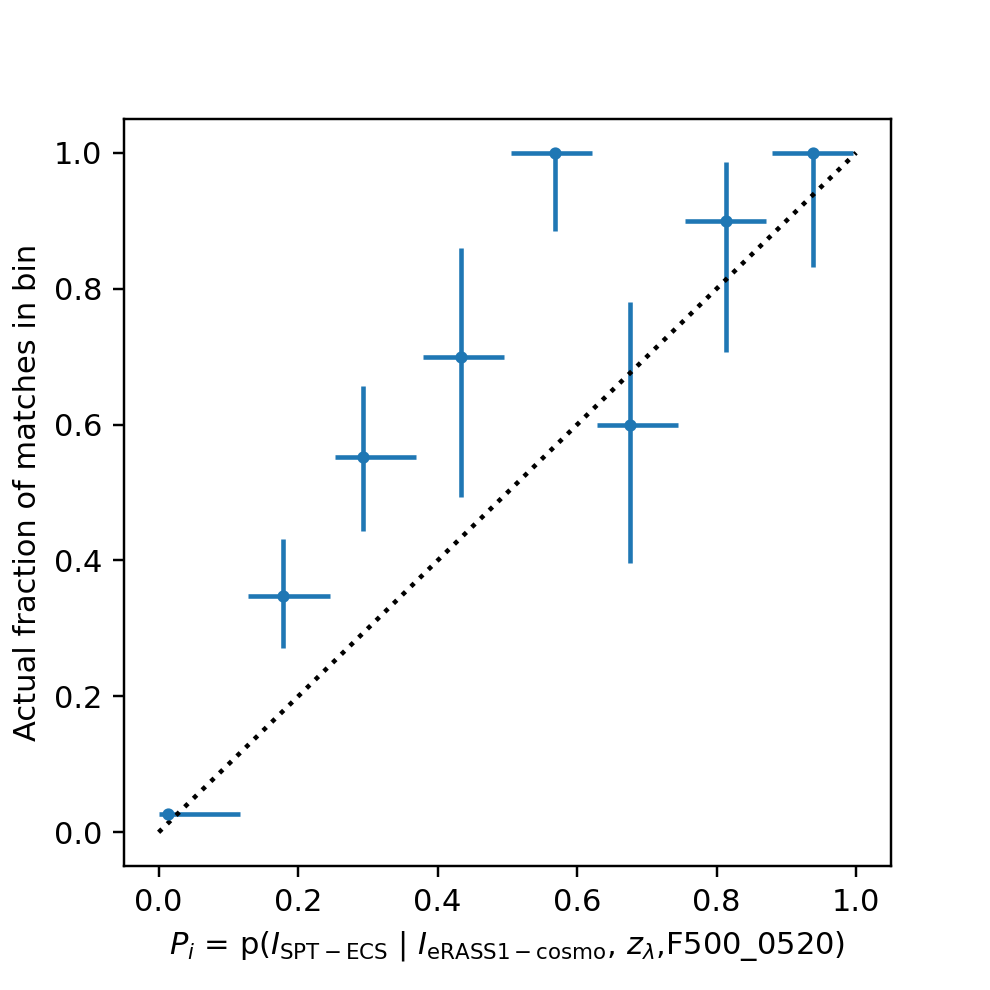}
   \caption{Figure similar as Fig.~\ref{fig:erass1_c_sptpolhuang_XCOPALL_calibration}, obtained by using the eRASS1 cosmology sample (958 entries) as reference and using the SPTpol-ECS as test sample.}
    \label{fig:erass1_c_sptecs_XCOPALL_calibration}%
\end{figure}

Fig.~\ref{fig:erass1_c_sptecs_XCOPALL_calibration} assesses the reliability of the $P_i$ values as returned by the model. The predicted probabilities are generally lower than the actual fraction of matches. Several causes for this shift may include: our fiducial $L_X-M_{500}-z$ relation \citep[here taken from][]{2019Bulbul} predicting too high values of X-ray luminosity at fixed halo mass; or an underestimate of eRASS1 flux measurements; or an overestimate of some SPT significance values $\xi$ due, e.g.~to Malmquist bias.

%%%%%%%%%%%%%%%%%%%%%%%%%%%%%%%%%%%%%%%%%%%%%%%%%%%%%%%%%%%%%%%%%%%%%%%%%%%%%%%%%%
\section{Discussion\label{sect:discussion}}

	\subsection{Linking our models with $\extlike$ and $\detlike$}

In processing real and simulated data, eSASS assigns several measured quantities to each detected source. Among the most relevant for understanding selection effects are detection ($\detlike$) and extent ($\extlike$) likelihoods. Applying thresholds on these quantities leads to defining a new sample, characterized by a balance between contamination and completeness \citep[e.g.][]{2022Seppi}.
In the specific case of the eRASS1 cluster catalog, a source is detected if it shows $\detlike$ greater than 5. The threshold on $\extlike$ for selecting this source as a cluster is either 3 or 6, the latter being chosen for the cosmology sample.

Such parameters are absent from all models discussed in this paper, neither as an input nor as an output of their predictions. It may seem at odds with studies \citep[e.g.][]{2021Rix} suggesting to model selection function based on catalog parameters cut only (here collectively designed under a generic name, $\xi_X$). In fact, let us acknowledge that:
\[
\pofc{I}{\tmodel} = \pofc{\xi_X > \xi_X^{\rm min}}{\tmodel} = \int_{\xi_X^{\rm min}}^{+ \infty} \dd \xi_X \pofc{\xi_X}{\tmodel}
\]
Building a selection model then translates into constructing a distribution of $\xi_X$ given a set of model parameters. This may involve auxiliary variables \citep[such as observed flux, see e.g.][]{2020Grandis} and a combination of empirical scaling laws between $\xi_X$ and those variables, involving some parameterized form of scatter.
An appealing method consists in letting all or a subset of these parameters free in the scientific analysis, together with the cosmological parameters \citep[e.g.][]{2010Vanderlinde}. These scaling relation parameters would then be `learnt' directly from data themselves; they would self-adjust to catalog parameters and contribute in increasing the global cosmology likelihood, via calibration through weak gravitational lensing or other mass proxies. In this approach, the exact shape of distributions (e.g.~log-normal or Gaussian scatter) is a strong prior assumption. In absence of other information (e.g.~external or simulated catalogs), we extrapolate our knowledge on brighter systems to fainter ones, potentially leading to interpretation errors close to the detection limit \citep{2023Gallagher}. We may even imagine extreme situations in which these assumptions are not verifiable, e.g.~if the unimodal scatter in scaling relations does not account for a separate class of faint, flat surface-brightness systems that escapes detection in our catalog.

% Smooth alternative
A safer alternative may consist in constructing a model $\xi_X$ directly from $\extlike$ and $\detlike$ as recorded in simulated datasets. There are also difficulties associated with this approach. Most notably, only detected (resp. selected) objects have an associated value of $\detlike$ (resp. $\extlike$). A replacement value needs to be assigned to those simulated systems missing a detection. One may also attempt to run the source detection algorithm with lower thresholds (\texttt{detlikemin} and \texttt{extlikemin} in eSASS) in order to assign a value to fainter objects; however this has a number of non-trivial side-effects from the algorithmic point of view, such as masking, source splitting, background estimates, etc. As a consequence, there is no guarantee that the samples selected with $\detlike>5$ in both runs (the one with \texttt{detlikemin}=5 and the other with \texttt{detlikemin}<5) are truly identical.

\begin{figure*}
   \centering
   \begin{tabular}{cc}
	   \includegraphics[width=0.48\linewidth]{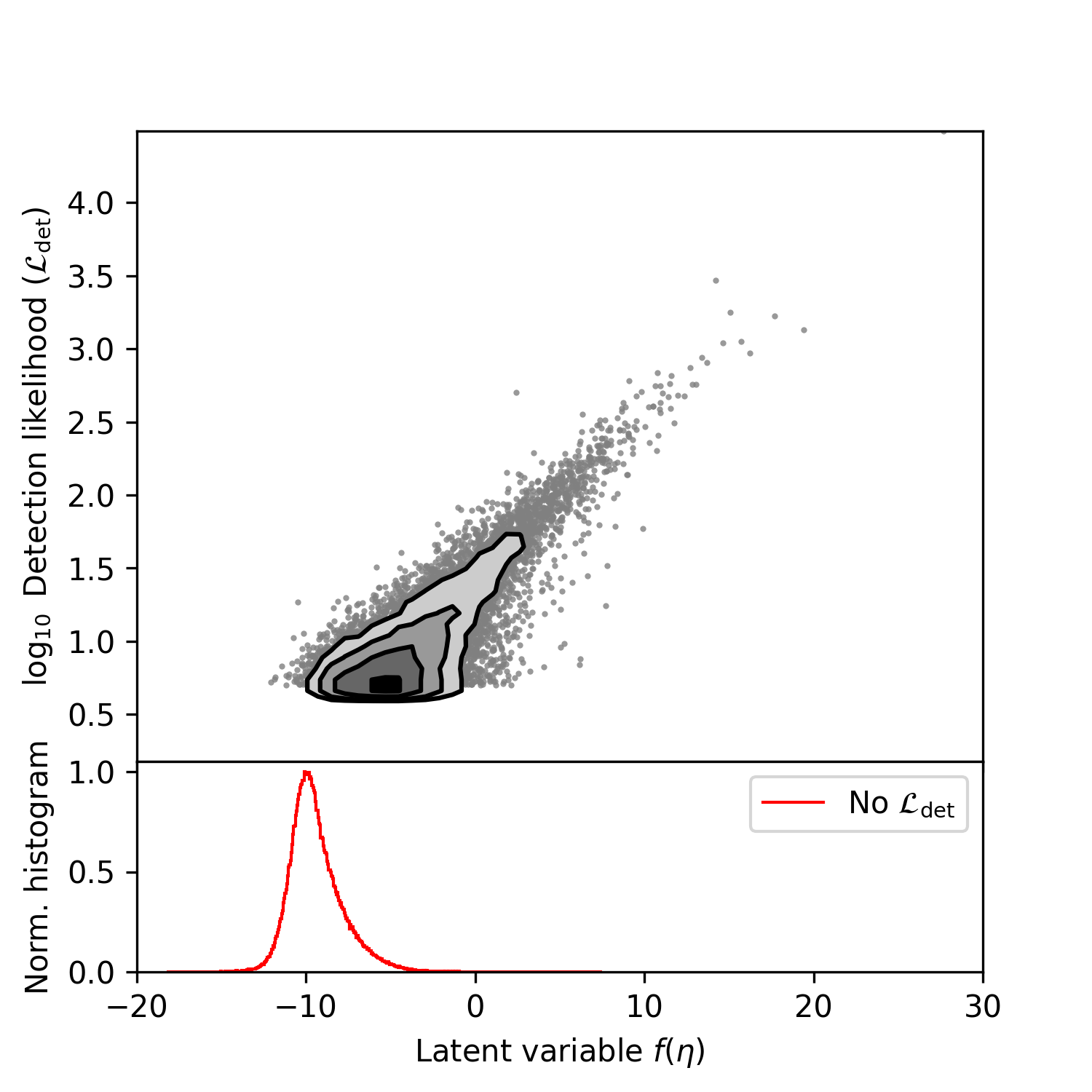} &
   		\includegraphics[width=0.48\linewidth]{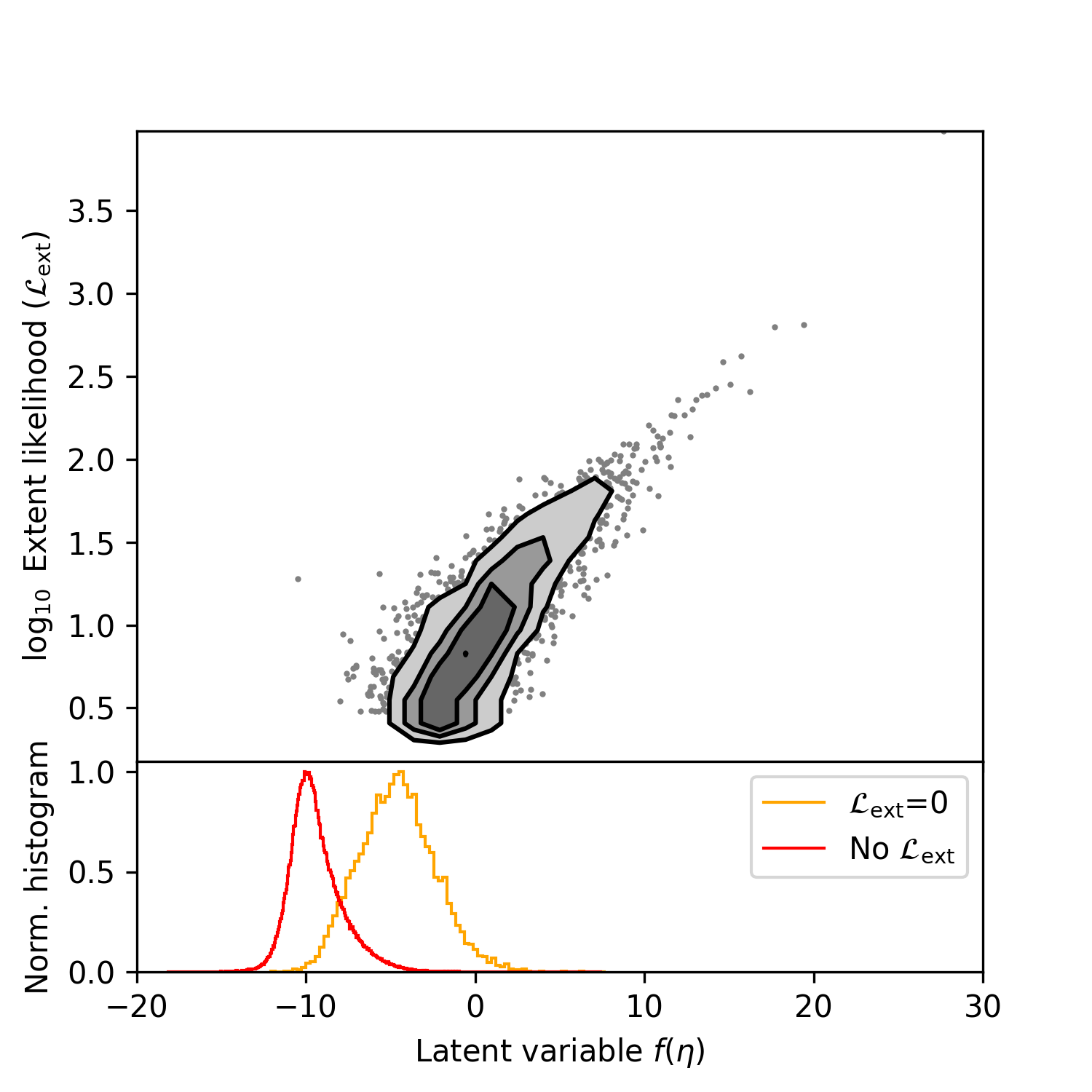} \\
   \end{tabular}
   \caption{Relation between the latent variable $f(\sbrightmod)=w_0 + \sum_j w_j \sbrightmod_j$ calculated for each simulated cluster in the test sample after performing logistic regression with 35 surface brightness features, and the values of $\detlike$ and $\extlike$ as provided by eSASS for those matched to a detection. Each dot stands for a simulated cluster, density contours ease visualisation in crowded regions. Undetected clusters have no associated $\detlike$ value (red histogram in the lower panel). Clusters detected but not categorised as extended have zero value for $\extlike$ (orange histogram). The apparent correlation between $f$ and both measurements indicate that the model has `discovered' the importance of $\detlike$ and $\extlike$ in the selection process.}
    \label{fig:latent_logistic_like}
\end{figure*}

% Role of latent variable
Interestingly, the logistic regression formalism presented in Sect.~\ref{sect:logistic_profile} uses a real-valued function $f$, whose expression is given in Eq.~\ref{eq:f_zeta}. In our analysis $f$ is a latent variable, without predefined physical meaning. The selection probability monotonically increases with $f$, through the sigmoid function of Eq.~\ref{eq:expit_formula}. As a useful cross-check, we show the correlation between $f$ and the value of $\detlike$ and $\extlike$ for the logistic regression involving 35 features (Fig.~\ref{fig:latent_logistic_like}). Reminding that only binary flags where exposed to the model in the training phase, it may be reassuring that the probability output is driven by a quantity behaving similarly as detection and extent likelihood.

\begin{figure*}
   \centering
   \begin{tabular}{cc}
	   \includegraphics[width=0.48\linewidth]{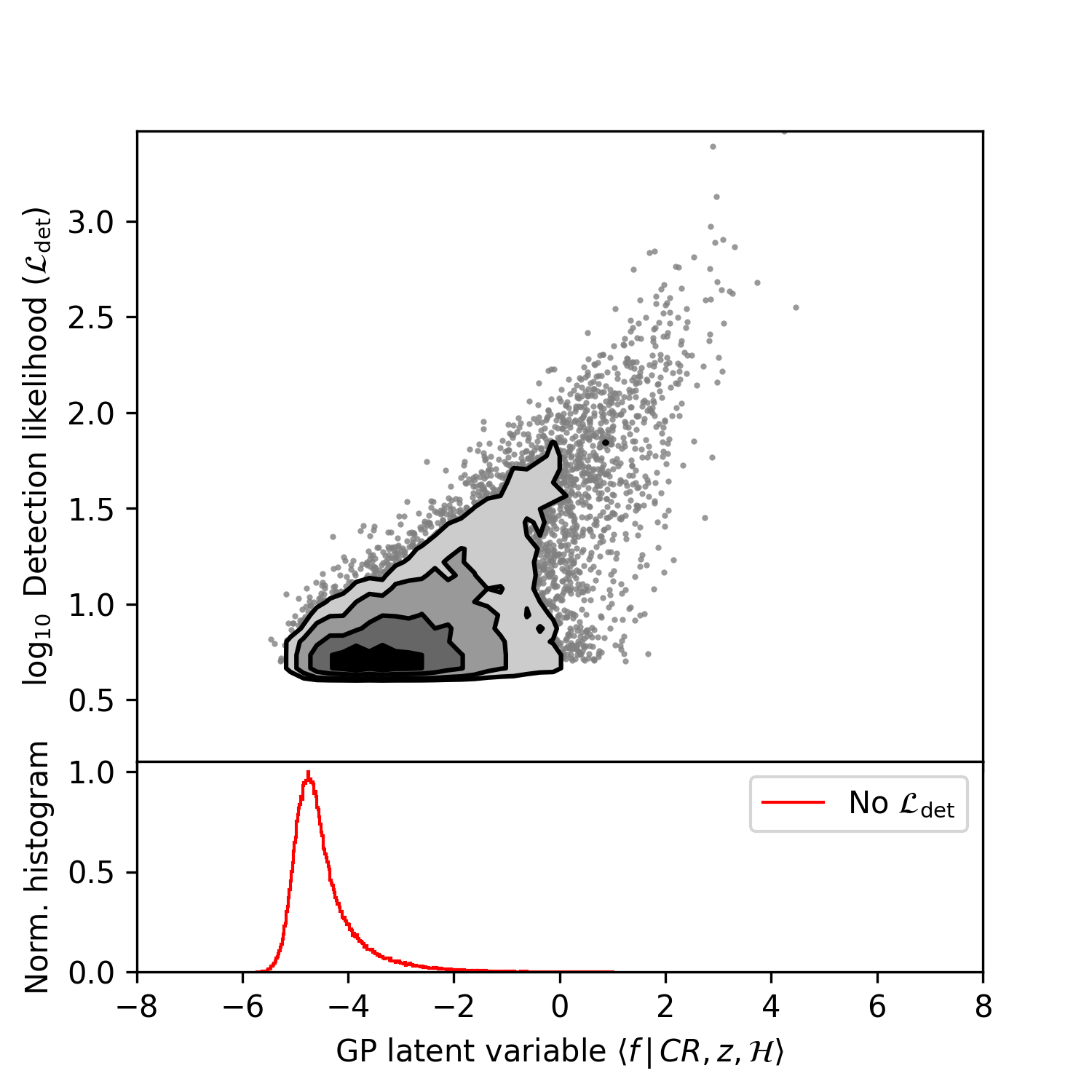} &
   		\includegraphics[width=0.48\linewidth]{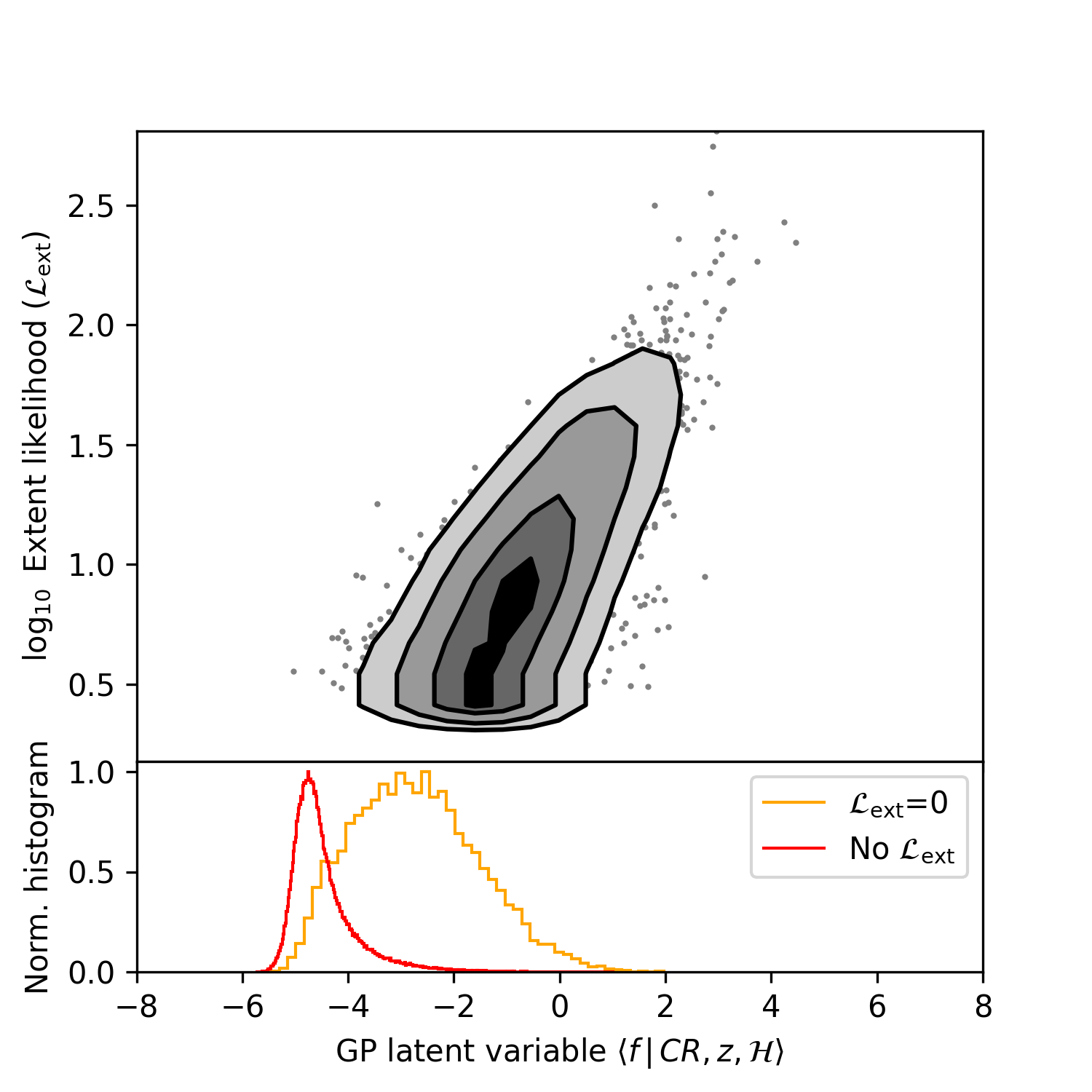} \\
   \end{tabular}
   \caption{Similar as Fig.~\ref{fig:latent_logistic_like}, replacing the x-axis by the mean of the latent variable implicit in the Gaussian Process classifier $\pofc{\isdet{\rm cosmo}}{CR, z, \mathcal{H}}$.}
    \label{fig:latent_gp_like}
\end{figure*}

We find a similar result with the mean of the latent variable $\vec{f}$ governing the predictions output of the Gaussian Process classifier (see~App.~\ref{app:gaussian_process_classifier}). Fig.~\ref{fig:latent_gp_like} in particular shows the a posteriori correlation between $\langle f \rangle$ and the detection and extent likelihood obtained in a test sample of simulated clusters.

	\subsection{Underestimate of the eRASS1 selection model?}
	
Our external validation results in Sect.~\ref{sect:external_validation} suggest an inconsistency between the predicted number of eRASS1-detected objects among the eFEDS sample and the actual number of matching entries. The difference is about 5-$\sigma$ when considering the eRASS1 cosmology sample and above 7-$\sigma$ with the primary eRASS1 cluster sample. Our validation against the much brighter SPT-selected samples does not reveal such inconsistency. This suggests that our model somehow fails at describing the fainter population of eFEDS clusters.

		\subsubsection{Uncertainties in the generative model}
% Discussing the generative model
We may first incriminate our simple generative model for eFEDS-measured fluxes. Our model for \ftroiscents{} (Eq.~\ref{eq:efeds_flux_generative_model} and~\ref{eq:efeds_fluxerror_generative_model}) assumes unbiased flux measurements regardless of the brightness of the source. If the reported eFEDS fluxes \citep{2022LiuA} of faint systems were known to be biased low for some reason, our models would predict a higher number of matched entries in the eRASS1 sample and we would conclude to no discrepancy. On the one hand, eFEDS luminosity measurements appear to agree with \emph{XMM-Newton} measurements \citep[e.g.][]{2022Turner}, as also shown by the luminosity functions in \citet[][]{2022LiuA}. On the other hand, we have found a 15\% offset between eRASS-measured fluxes and \emph{Chandra} measurements, the latter showing higher value, likely attributed to calibration issues \citep{2024Bulbul}.

We obtained very similar predictions, and thus similarly discrepant results, when modelling the measured $R_{500}$ X-ray luminosity of eFEDS clusters \citep[column \luminbahar{} in the catalog of][]{2022Bahar} in lieu of the measured flux. This operation somewhat simplifies our generative model, and it suppresses any potential pitfall in deriving aperture corrections and spectrally-dependent K-corrections. There we assumed unbiased, normally distributed luminosity measurements, with an empirical uncertainty model following:
\begin{equation}
\sigma_{\luminbahar} = 0.2 \times 10^{43}\,{\rm erg\,s}^{-1} \left(\frac{L_{X}}{10^{43}\,{\rm erg\,s}^{-1}} \right)^{0.9} + \epsilon^i
\end{equation}
In this expression $\epsilon^i$ is a constant for each eFEDS cluster indexed by $i$. We obtained $\epsilon^i$ by subtracting the value predicted by the power-law model (without $\epsilon^i$) in the above equation to the uncertainty reported in the catalog of \citet{2022Bahar}.

		\subsubsection{Uncertainties in trained models}
% Uncertainties in the selection models
Our selection models inherently embed some uncertainty, due to the limited size of the training sample and to the complex algorithm leading to optimize many hundreds of Gaussian Process hyperparameters at once (App.~\ref{app:gaussian_process_classifier}). We have taken benefit of our multiple Poisson realizations of the eRASS1 twin simulations to produce seven selection models, each relying on a distinct training set as large as eight times the actual eRASS1 sky. We folded these selection functions through our eFEDS/eRASS1 comparison algorithm. In all cases we retrieved the under-prediction of eRASS1 clusters, the discrepancy varying between 5 and 7-$\sigma$ among the seven models. The variability and uncertainty attached to the training phase are thus too small to explain the observed discrepancy.

		\subsubsection{Uncertainties in the training sample}
% Trainig set problems?
We now discuss whether our eRASS1 training set may be biased in some way, which would offset the models from the truth. Such offsets would not be visible in our internal validation tests (Sect.~\ref{sect:gp_internal_validation}) since the test sample is constructed in exactly the same way as the training sample.

% a) Matching
Our baseline procedure associating simulated clusters to detected sources leads to fewer selected halos than with the positional matching procedure (see Sect.~\ref{sect:matching_catalogs}). It is especially visible at cluster fluxes around $10^{-14}$\,\fluxunit{}, where the number of matched sources can differ up to a factor two (Fig.~\ref{fig:compa_phbased_nway}). We have trained a selection function model with a new set of detection flags constructed with the alternative matching technique. This model indeed predicts slightly higher detection probabilities at low flux, in accordance with the update in the training set. However, the $5-7$-$\sigma$ discrepancy remains in the comparison with the eRASS1 clusters detected in eFEDS. This is not surprising, since the eRASS1 flux limit in the eFEDS field is around $10^{-13}$\,\fluxunit{}, well above the regime where the two matching methods depart from each other (Fig.~\ref{fig:compa_phbased_nway}).

% b) Processing
After completion of this work, we found minor differences in the eSASS software configuration parameters used in processing the simulations \citep[version eSASSusers\_201009,][]{2022Seppi} and in processing the actual eRASS1 data \citep[version 010,][]{2024Merloni}. These changes mainly impact the background map creation and source splitting into multiple entries. In particular, in real data the \texttt{ermldet} algorithm would attempt to split detections into two sources only if the number of counts is greater than 25; while in our simulations no specific threshold has been set on this value.
We have tried and processed a subset of the twin simulation with the updated algorithm. We find a tentative higher detection probability than in the original mock catalog, which may explain part of the deficit in the predicted number of eRASS1 sources. However, this finding relies only on a subset of the entire twin simulations, insufficient to firmly close the issue.

% c) Models
Our final test consists in using a selection model that explicitly involves the cluster sizes $R_{500}$ in their features (Fig.~\ref{fig:GP_nH_Texp_simbkg_fx_R500am_AllXGOOD_cut_0-0_SEED0189}). By using a selection model in the form $\pofc{\isdet{\rm main}}{f_X, R_{500}, \mathcal{H}}$, we do not rely any longer on the distribution of sizes imprinted in the simulation. Rather, we assume our $L_X-M_{500}$ is correct and we distribute our modelled flux within the radius associated to $M_{500}$. Folding this new model into our eFEDS/eRASS1 analysis, we find better agreement between predictions and actual number of matches. Hence, using this alternative selection model instead of our baseline model, the discrepancy reduces from 7- to 5-$\sigma$ for the eRASS1-main catalog and from 5- to 3-$\sigma$ for the cosmology catalog. This finding corroborates Fig.~6 in \citet{2020Comparat}, showing that the average $R_{500}$ radius of our simulated clusters at fixed luminosity is larger than in several published samples\footnote{Different cosmological models between samples and simulations entering this figure may induce deviations of a few percent in the luminosity and angular distance values, which is not enough to account for the shift.}.

		\subsubsection{Sensitivity of the external validation test}
% Conclusion on this
In this section we have attempted to address the apparent mismatch between the number of eRASS1 systems predicted in the eFEDS field and the actual number of matches. We could not identify one single cause for the discrepancy, rather we have found several potential difficulties. Overall, our experiments reveal the high sensitivity of our external validation tests to small variations in the model assumptions. They also highlight the amount of effort needed to properly understand selection effects at low fluxes and close to the detection threshold.

	\subsection{X-ray and SZ correlated detection}

\begin{figure}
   \centering
   \includegraphics[width=\linewidth]{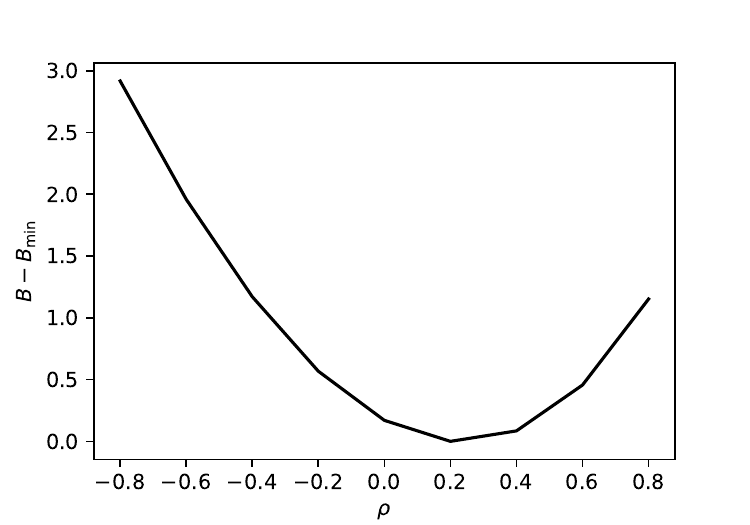}
   \caption{Value of the combined Bernoulli negative log-likelihood (Eq.~\ref{eq:bernoulli_loglike}) in the comparison of the SPT-ECS sample (reference) and the eRASS1 cosmology sample (test) for various values of the correlation coefficient $\rho$. This coefficient expresses the level of covariance between $\ln L_X$ and $\ln \zeta$ at fixed halo mass and redshift. Assuming all other model parameters are correct, the constraint obtained from the catalog comparison exercise depicted in this paper is relatively loose: $\rho \simeq 0.2 \pm 0.6$ (68\% confidence level) as deduced from $(B-B_{\rm min}) \sim \chi^2_1$.}
    \label{fig:rhoanaecs3}
\end{figure}

In Sect.~\ref{sect:external_validation} we have introduced a joint distribution for the X-ray luminosity and SZ significance $\pofc{\ln L_X, \ln \zeta}{M_{500}, z}$ as a bivariate normal distribution with correlation coefficient $\rho = 0.2$. We now investigate its impact on the results. We varied $\rho$ between $-0.8$ and $0.8$, while repeating the experiment with the SPT-ECS sample as reference $R$ and the eRASS1-cosmo sample as test $T$ (Sect.~\ref{sect:basesptecs_checkerass1cosmo}). We form the quantity:
\begin{equation}\label{eq:bernoulli_loglike}
B = -2 \ln \mathcal{L} = -2 \sum_{i \in R} \isdet{T}^i \ln P_i + (1-\isdet{T}^i) \ln (1-P_i)
\end{equation}
Its mimimum $B_{\rm min}$ is obtained by varying $\rho$, all other model parameters held fixed at their (supposedly) true value. We removed the cluster SPT-CLJ0333-3707 whose redshift measurements disagree with the matched entry 1eRASS~J033323.9-370744 (Sect.~\ref{sect:basesptecs_checkerass1cosmo}). There are 162 entries entering the sum. Following \citet{1979Cash}, $B-B_{\rm min}$ approximately follows a $\chi^2$ distribution with one degree of freedom.
Fig.~\ref{fig:rhoanaecs3} shows the curve $B-B_{\rm min}$ for various values of $\rho$, from which we derive an approximate 68\% confidence interval $-0.4 < \rho < 0.8$. Consistently with previous studies \citep[e.g.][]{2016deHaan, 2019Dietrich} our data does not favour strongly correlated or anti-correlated scatter between the X-ray and SZ signals.

The fiducial value $\rho = 0.2$ chosen in Sect.~\ref{sect:external_validation} is in fact the best-fit, with $B_{\rm min} = 99.9$. We have checked the goodness of fit by multiple random Bernoulli draws from the $P_i$ values returned by the best-fit model; only 24\% of the simulations provide a value of $B$ smaller than $B_{\rm min}$.
Given the many other assumptions made (e.g.~fixed scaling relation parameters, etc.), we cannot firmly exclude specific values of $\rho$ from our dataset. A comparison involving larger samples may put stronger constraints on $\rho$ through this method.

It is interesting to note that $\rho$ is intimately related to the ICM distribution for clusters in the reference survey. A physical link should exist between $L_X$ and $\zeta$ that allows computing $\rho$ or at least a prior distribution, without using the halo mass $M_{500c}$. However, our implementation of the SPT-SZ selection function through $\zeta$ imposes to rely on an empirical scaling relation with mass, which prevents from investigating further in this direction.

%%%%%%%%%%%%%%%%%%%%%%%%%%%%%%%%%%%%%%%%%%%%%%%%%%%%%%%%%%%%%%%%%%%%%%%%%%%%%%%%%%
\section{Conclusions\label{sect:conclusion}}

This work presents an investigation of X-ray selection function models aimed at describing the completeness of the primary and cosmological eRASS1 cluster samples. Driven by our astrophysical and cosmological objectives, we have developed a framework inspired by classification problems in statistics and especially machine learning research. Such empirical models are well-suited to our purpose, since selection of sources in astronomical catalogs is (at least in part) resulting from human decisions, whose detailed physical modelling may be cumbersome. The models we selected are intentionally simple, enabling their interpretability and explainability.

Our main findings are summarized as follows:
   \begin{enumerate}
      \item We have performed extensive twin simulations of the eRASS1\_DE sky under certain cosmological and physical assumptions. These simulations reproduce to a large extent the chain of selection filters along the path from sources to final catalogs.
      \item We have identified which simulated halos are detected and selected by means of custom matching algorithms. One algorithm relies on tracking the origin of simulated photons, the other algorithm on Bayesian positional matching. We have highlighted relevant differences in the outcome of the two algorithms, mostly below cluster fluxes a few times $10^{-14}$~\fluxunit.
      \item By means of a simple logistic (linear) model, we have asked which surface brightness features govern the detectability and selection of clusters. In particular, we find the central $20\arcsec$ counts help in detecting a cluster, but not so in characterizing it as extended. Cluster counts beyond approximately $1.5\arcmin$ from the cluster centre play a minor role in the selection of clusters. This model may be seen as a simplified emulator of the eSASS source detection algorithm.
      \item Acknowledging that selection function based on surface brightness features require important modelling effort on the cosmological likelihood side, we have developed models relying on intermediate variables. We identified a hierarchy of models, some strongly depending on assumptions made in the twin simulations, e.g.~$\pofc{I}{M_{500}, z}$, others enabling less dependence on such assumptions e.g.~$\pofc{I}{CR, z, T_{\rm exp}, N_H, {\rm bkg}}$. Using the latter class of models comes at a cost, namely a heavier modelling effort in scientific analyses.
      \item We have tested all our models with left-over simulation samples. In particular we established a ranking of models based on their precision, their sensitivity (also known as recall), and the reliability of their predicted probabilities. Models accounting for at least one morphological feature (e.g.~the cluster $R_{500}$ radius or its central emission measure $EM_{0}$) perform better.
      \item Using real data, we have tested our models by means of a custom population model. Comparing to eFEDS or SZ-based surveys, we find no obvious missing cluster from our eRASS1 samples. Conversely, there is no eRASS1 cluster in the cosmology sample that is strongly missing in other tested surveys, in agreement with the known low contamination of the sample \citep{2022Seppi, 2024Ghirardini}.
      \item While comparison with the (brighter) SPT-SZ population reveals no specific issue, we find more eRASS1 clusters in the eFEDS sample than our models would predict. We have highlighted a series of putative causes for the discrepancy. We will pursue work to deepen our understanding of the selection of faint clusters.
   \end{enumerate}

In future work we will refine the simulation set with updated cluster models and halo simulations. An alternative assesment of selection systematics may be gained through the use of `inject-and-retrieve' methods \citep[e.g.][]{2006Pacaud, 2016Suchyta, 2020Kong}, where mock cluster images are distributed in real data. 
Moreover, ultimate validation of selection models should rely on a performance metric that relates directly to cosmological parameters, especially since our purpose is to constrain those parameters from data. Developing multiple eRASS simulations spanning a range of cosmological models is a promising route in this direction, although computationally intensive.
Interestingly, forward models of the observed large scale structure using Effective field theory (EFT) have progressed tremendously \citep[e.g.][]{JascheLavaux_2019A&A...625A..64J, Schmidt_2021JCAP...04..033S}. These models predict the cosmological matter density field at percent (and even sub percent) accuracy. In the future, these models will enable tight constraints on cosmological parameters by using not only linear but also mildly non-linear information contained in the density field \citep{Schmidt_2021JCAP...04..032S, KosticNguyenSchmidt_2023JCAP...07..063K} and by marginalizing fully over the selection function.
Finally, pursuing the comparison exercise depicted in this paper with multiple other surveys across a range of wavelengths will provide interesting constraints on model parameters and reveal any significant population outliers worth a deep follow-up study.

\begin{acknowledgements}
This work is based on data from eROSITA, the soft X-ray instrument aboard SRG, a joint Russian-German science mission supported by the Russian Space Agency (Roskosmos), in the interests of the Russian Academy of Sciences represented by its Space Research Institute (IKI), and the Deutsches Zentrum für Luft- und Raumfahrt (DLR). The SRG spacecraft was built by Lavochkin Association (NPOL) and its subcontractors, and is operated by NPOL with support from the Max Planck Institute for Extraterrestrial Physics (MPE).

The development and construction of the eROSITA X-ray instrument was led by MPE, with contributions from the Dr. Karl Remeis Observatory Bamberg \& ECAP (FAU Erlangen-Nuernberg), the University of Hamburg Observatory, the Leibniz Institute for Astrophysics Potsdam (AIP), and the Institute for Astronomy and Astrophysics of the University of Tübingen, with the support of DLR and the Max Planck Society. The Argelander Institute for Astronomy of the University of Bonn and the Ludwig Maximilians Universität Munich also participated in the science preparation for eROSITA.

The eROSITA data shown here were processed using the eSASS software system developed by the German eROSITA consortium.

\\

N.~Clerc acknowledges financial support from CNES.
E.~Bulbul, A.~Liu, V.~Ghirardini, C.~Garrel, S.~Zelmer, and X.~Zhang acknowledge financial support from the European Research Council (ERC) Consolidator Grant under the European Union’s Horizon 2020 research and innovation program (grant agreement CoG DarkQuest No 101002585). 

\\

Some of the results in this paper have been derived using the following packages: \texttt{scikit-learn} \citep{scikitlearn}, \texttt{MOCpy\footnote{\url{https://github.com/cds-astro/mocpy/}}} \citep{2014MOCpy}, \texttt{GPy}\footnote{\url{https://github.com/SheffieldML/GPy}} \citep{gpy2014}, \texttt{astropy} \citep{2022astropy}, \texttt{matplotlib} \citep{2007Hunter}, \texttt{healpy} and HEALPix\footnote{\url{http://healpix.sourceforge.net}} \citep{2005Gorski, 2019Zonca},  \texttt{HEASoftpy}\footnote{\url{https://heasarc.gsfc.nasa.gov/lheasoft/heasoftpy/}}, \texttt{climin}\footnote{\url{https://github.com/BRML/climin}}\citep{2015climin}.

\end{acknowledgements}

\bibliographystyle{aa}
\bibliography{biblio}

\begin{appendix} %First appendix

%%%%%%%%%%%%%%%%%% Appendix : logistic regression uncertainties
\section{Uncertainties on logistic regression coefficients}
\label{app:uncertainties_logistic}
Considering a galaxy cluster indexed by $i$ among $n$ simulated systems in the training set, we denote by $\vec{\sbrightmod}_i$ the vector of its $m$ associated features. Simulations indicate whether this cluster is detected as extended ($I_i=1$) or not ($I_i=0$).
Our implementation of logistic regression writes:
\begin{equation}
    \pofc{I_i}{\vec{\sbrightmod}_i} = \varphi\left( f(\vec{\sbrightmod}_i) \right) = \varphi\left(\vec{w}^{\prime \intercal} \vec{\sbrightmod}_i\right)
\end{equation}
We used $\varphi(t) = \left( 1 + e^{-t}\right)^{-1}$ and $\vec{w} = (w_0, w_1, w_2, \dotsc, w_m)$ the vector of coefficients ($w_0$ is the intercept of the underlying linear model). 
We have defined $\vec{\sbrightmod}_i^{\prime} = (1, \sbrightmod_{i, 1}, \sbrightmod_{i, 2}, \dotsc, \sbrightmod_{i, m})$.

The fit is performed by minimizing the following cost function (known as log-loss) over all possible values for $\vec{w}$, where we have written $q_i = \pofc{I_i}{\vec{\sbrightmod}_i}$:
\begin{equation}\label{eq:log_loss}
    \mathcal{C}(\vec{w}) = \sum_{i=1}^n \left(-I_i \ln(q_i) - (1 - I_i) \ln(1 - q_i)\right)
\end{equation}

A formula for the Hessian matrix $H$ derives by noticing that:
\begin{equation}
    \partial q_i/\partial w_j = q_i (1-q_i) \sbrightmod^{\prime}_{i, j}
\end{equation}
We obtain:
\begin{equation}
    H_{jk} = \frac{\partial \mathcal{C}}{\partial w_j \partial w_k} = \sum_{i=1}^{n}  q_i (1-q_i) \sbrightmod^{\prime}_{i, j} \sbrightmod^{\prime}_{i, k}
\end{equation}

We obtain an estimate of the covariance matrix of best-fit coefficients $\vec{w}$ by inverting the $(m+1) \times (m+1)$ matrix $H$. This approximation is roughly valid as long as the best-fit model corresponds to the minimum of the cost function and in the limit of large $n$.

%%%%%%%%%%%%%%%%%% Appendix : logistic regression of EXT>=0 clusters
\section{Logistic regression predicting detected clusters}

\begin{figure}
   \centering
   \includegraphics[width=\linewidth]{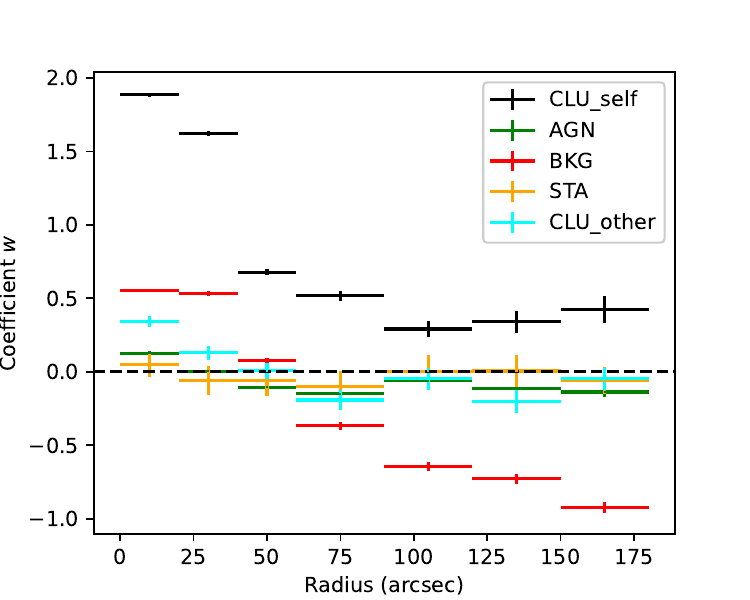}
   \caption{Similar figure as Fig.~\ref{fig:coeff_profile}, but for a logistic model $\pofc{\isdet{eRASS1}}{\rm counts}$, trained to predict whether a cluster is detected among all simulated clusters.}
    \label{fig:coeff_profile_onlydet}%
\end{figure}

We have shown results in Sect.~\ref{sect:logistic_profile} for a model trained to predict whether a cluster is selected (that is, detected and characterized as extended), given $7\times5$ surface brightness features (7 radial ranges and 5 components). We reiterate this exercise with a model trained to predict whether a cluster is detected. Figure~\ref{fig:coeff_profile_onlydet} shows the 35 coefficients associated to each of these features. As discussed in Sect.~\ref{sect:logistic_profile}, these coefficients can be interpreted as the sensitivity of the detection to a small variation in a given feature. Fig.~\ref{fig:coeff_profile} and Fig.~\ref{fig:coeff_profile_onlydet} differ notably through the coefficients associated to the cluster emission (black crosses). The central $0-20\arcsec$ surface brightness plays a major role in the detection model. Counts deposited at radii larger than 2\,arcmin are relatively more important for detecting than for selecting a cluster.

%%%%%%%%%%%%%%%%%% Appendix : binomial uncertainties
\section{Confidence range for proportions $n/N$\label{app:binomial_unc}}

Some figures in the paper incorporate uncertainties on a certain ratio $n/N$ with $n$ and $N$ both integer numbers, $0\leq n \leq N$. We estimated the 68\% confidence range $[L, U]$ by means of the score interval \citep{1927Wilson} with continuity correction \citep{1993Vollset}:
\begin{eqnarray}
L & = & \frac{n_{-} + \lambda^2/2}{N + \lambda^2} - \lambda \frac{\sqrt{n_{-} - n_{-}^2/N + \lambda^2/4}}{N+\lambda^2} \\
U & = & \frac{n_{+} + \lambda^2/2}{N + \lambda^2} + \lambda \frac{\sqrt{n_{+} - n_{+}^2/N + \lambda^2/4}}{N+\lambda^2}
\end{eqnarray}
where $n_{-} = n-1/2$ and $n_{+}= n+1/2$. The bounds are set to $L=0$ if $n=0$ and to $U=1$ if $n=N$. The value for $\lambda$ is the $1-\alpha/2$ quantile of the normal distribution, in our case we set $\lambda=1$ for $\alpha=0.32$ (hence, we show the 68\% confidence intervals). Without continuity correction, we would have used $n$ in place of $n_{-}$ and $n_{+}$ in those formulas.

These intervals should be adequate to represent uncertainties in the proportion $n/N$ \citep{1998Newcombe}, although the confidence interval may be too conservative in general \citep{1998AgrestiCoull}.
We emphasize that we do not use such uncertainties while computing our models, nor in performing calculations. They only serve for visualisation purposes.

%%%%%%%%%%%%%%%%%% Appendix : gaussian process classification
\section{Gaussian Process classification}
\label{app:gaussian_process_classifier}

The selection function models presented in Sect.~\ref{sect:gp_models} rely on Gaussian Process classification (GPC). To overcome the well-known computational bottleneck encountered in standard GPC we have applied the Stochastic Variational Gaussian Process (SVGP) algorithm \citep[][]{2015Hensman} implemented in the \texttt{GPy} library \citep{gpy2014}. This section presents briefly the main characteristics of the method and some relevant references. For an application in the context of cluster astrophysics, see e.g.~\citet{2022Debackere}, in particular their Appendix~B provides concise descriptions of the method for binomial likelihood (we use instead Bernoulli likelihood).

	\subsection{Defining a Gaussian Process prior}

We refer the reader to \citet{2006RasmussenWilliams} for a pedagogical introduction of GPC and for the notations followed in this Appendix. The principle of Gaussian Process binary classification builds upon a latent variable $f(\vec x)$, function of the $m$ features $\vec x = \{x_i\}_{1 \dotsc m}$ and on the assumption of a Gaussian Process (GP) prior on $f$. Values of the latent function are then mapped to probability values ranging between 0 and 1 by means of a bijective link function. In our specific case we choose the inverse probit for the link function, that is the cumulative Gaussian distribution $\Phi(f)$. Other sigmoid-shaped link functions would convene (e.g.~the inverse logit).

By definition, any finite subset of random variables constituting a Gaussian Process follows a joint multivariate normal distribution. The mean function $\mu(\vec x)$ and the covariance function $k$ suffice to fully specify the Gaussian Process. We write:
\begin{equation}
f \sim \mathcal{GP}\left( \mu(\vec x) \mathrel{} ; \mathrel{} k(\vec{x}, \vec{x^{\prime}}) \right)
\end{equation}
Using $\mathbb{E}$ to denote the expectation value of a random variable, we have defined:
\begin{equation}
\mu(\vec{x}) = \mathbb{E}\left[ f(\vec{x} ) \right]
\end{equation} 
\begin{equation}
k(\vec{x}, \vec{x^{\prime}}) = \mathbb{E}\left[ \left(f(\vec{x}) - \mu(\vec{x}) \right) \left(f(\vec{x^{\prime}}) - \mu(\vec{x^{\prime}}) \right) \right]
\end{equation}

In our setup the mean function $\mu$ is constantly zero and the covariance function is the Radial Basis Function kernel (RBF), also known as stationary squared exponential kernel. This latter choice ensures extremely smooth properties of the underlying Gaussian Process. One introduces $m$ hyperparameters $l_i$ entering the definition of the RBF kernel, defining the length-scales of variation along each feature dimension:
\begin{equation}\label{eq:rbf_kernel}
k(\vec{x}, \vec{x^{\prime}}) = \sigma^2 \exp \left[ -\frac{1}{2} \sum_{i=1}^m \left(\frac{x_i - x^{\prime}_i}{l_i} \right)^2 \right]
\end{equation}
In this expression $\sigma$ is an additional hyperparameter governing the amplitude of the correlation. The RBF correlation function is intuitively easy to understand: close pairs of points in the $m$-dimensional feature space are associated to strongly covariant random variables, the actual value of covariance being close to $\sigma^2$. Points far from each other are uncorrelated, with zero covariance. The length-scales govern the typical distance correlation cut-off along each dimension.

Given a GP prior and a set of $n^*$ test points $X^* = (\vec{x}^*_1, \vec{x}^*_2,\dotsc,\vec{x}^*_{n^*})$ one may generate a random vector $\vec{f^*}$ by sampling the multivariate normal distribution $\vec{f^*} \sim \normaldist{\vec{0}}{K(X^*, X^*)}$. Here $K(X^*, X^*)$ is a shorthand notation for the $n^* \times n^*$ covariance matrix whose coefficients are the $k(\vec{x}^*_p, \vec{x}^*_q)$.

	\subsection{Predicting and training in standard GPC}

In order to finely model our predictions, we wish to condition the GP on a set of known measurements (observations), also named training set in this paper.
A set of $n$ known input data points $X = \{\vec{x}_i\}_{i =1 \dotsc n}$ is associated to $n$ binary observations $\vec{I} = \{I_i\}_{i=1 \dotsc n}$, with $I_i$ taking either value 0 or 1 (compliant with our notations in Table~\ref{table:summary_notations}). The ensemble $(X, \vec{I})$ constitutes the training set. We assume a Bernoulli likelihood to describe the data given the model values, i.e.:
\begin{equation}
\pofc{I_i}{f_i} = \Phi(f_i)^{I_i} \left(1-\Phi(f_i)\right)^{1-I_i}
\end{equation}

Given a new test point $\vec{x}^*$, the model should provide a probabilistic prediction, that is interpreted as the cluster detection probability given our knowledge of the training dataset. Our desired model output writes:
\begin{equation}\label{eq:prediction_finale_gp}
\pofc{I^*}{X, \vec{I}, \vec{x}^*} = \int \Phi(f^*) \pofc{f^*}{X, \vec{I}, \vec{x}^*} \dd f^*
\end{equation}
It is therefore required to obtain a distribution of the latent function $f^*$ at the point $\vec{x}^*$. We have that:
\begin{equation}\label{eq:prediction_gp}
\pofc{f^*}{X, \vec{I}, \vec{x}^*} = \int \pofc{f^*}{X, \vec{x}^*, \vec{f}} \pofc{\vec{f}}{X, \vec{I}} \dd \vec{f}
\end{equation}

The first term in the integral derives from the GP prior directly. It can be expressed in a standard way by conditioning the joint Gaussian distribution $\pofc{\vec{f}, f^*}{X, \vec{x}^*}$ on the latent function variables $\vec{f}$:
\begin{equation}\label{eq:gp_conditioning}
f^* \mathrel{} | \mathrel{} X, \vec{x}^*, \vec{f} \sim \normaldist{\vec{k}^{* \intercal} K^{-1} \vec{f}}{k(\vec{x}^*, \vec{x}^*) - \vec{k}^{* \intercal} K^{-1} \vec{k}^*}
\end{equation}
We have defined $K \equiv K(X, X)$ and $\vec{k}^* \equiv K(\vec{x}^*, X)$, both are obtained through the RBF kernel, see Eq.~\ref{eq:rbf_kernel}. If the number of training points $n$ is large, the inversion of the $n \times n$ matrix $K$ may be cumbersome and this represents a common numerical bottleneck in standard GP analysis.
The term $\pofc{\vec{f}}{X, \vec{I}} = \pofc{\vec{I}}{\vec{f}} \pofc{\vec{f}}{X}/\pofc{\vec{I}}{X}$ is the posterior over latent variables. It is generally not analytically tractable and often it must be approximated.

In GP classification, the marginal likelihood (or evidence) $\pofc{\vec{I}}{X} = \int \pofc{\vec{I}}{\vec{f}, X} \pofc{\vec{f}}{X} \dd \vec{f}$ plays an important role. The hyperparameters of the model (including those from the kernel) are optimally found by maximizing the value of the marginal likelihood. Throughout this paper, we identify this operation with the `training phase' of a GP classifier. In our case, an approximation is again needed as the marginal likelihood does not have a simple analytic expression.

%% Now describe sparse GP and then SVGP.... with words

	\subsection{Principles of the selected SVGP algorithm}

Several works propose schemes to approximate the distributions and make the associated computations numerically tractable. We only briefly depict here the principles associated to the SVGP algorithm used in our modelling. Details and related references are found in \citet{2015Hensman}.

It is useful to introduce a new set of hyperparameters made of $k$ pairs of input data points and their associated output values $(\vec{Z}, \vec{u})$. The $k \times m$ coordinates of these points, called inducing inputs, will be optimized together with other hyperparameters during the training phase. Loosely speaking, these points will approximately `summarize' the GP at locations that are most relevant.
The second stage of approximations uses searching variational bounds \citep[e.g.][for a review]{2018Blei}, leading to the following inequality:
\begin{equation}\label{eq:equation_variational}
\ln \pofc{\vec{I}}{X} \geq \sum_{i=1}^{n} \mathbb{E}_{q(f_i)} \left[ \ln \pofc{I_i}{f_i} \right] - {\rm KL} \left[ q(\vec{u}) \mathrel{} \middle| \middle| \mathrel{} p(\vec{u}) \right]
\end{equation}
In this expression ${\rm KL}$ represents the Kullback-Leibler divergence, it measures the proximity between the two densities $q(\vec{u})$ and $p(\vec{u})$. The distribution $q(\vec{u})$ belongs to a predefined family of distributions. Selecting a suitable family of distributions is at the core of variational inference, since it amounts to finding the most suitable $q$ among this family. In practice, this procedure is equivalent to resolving an optimization problem. In our case the selected family is constituted by the normal multivariate distributions, hence $q(\vec{u}) \sim \normaldist{\vec{m}}{S}$. This formulation requires the introduction of an additional set of hyperparameters indexing the members in the family: the vector $\vec m$, with $k$ coefficients, and the symmetric matrix $S$, with $k (k+1)/2$ independent coefficients. The algorithm specifically uses lower Cholesky factorization $S = LL^{\top}$ for numerical stability reasons. The expectation under the sum sign is taken with respect to the marginals $q(f_i)$ of the distribution $q(\vec{f})$, defined such that $q(\vec{f}) \equiv \int \pofc{\vec{f}}{\vec{u}} q(\vec{u}) \dd \vec{u}$. The computation of $\pofc{\vec{f}}{\vec{u}}$ follows directly from the GP prior assumption, in a similar way as in Eq.~\ref{eq:gp_conditioning} (replacing $\vec{f}$ with $\vec{u}$ and $f^*$ with $\vec{f}$).

In the training phase, the SVGP algorithm consists in finding the set of hyperparameters maximizing the bound on the marginal likelihood, that is the right-hand side of Eq.~\ref{eq:equation_variational}. Gradient descent algorithms are well-suited to such optimization tasks. Our algorithm uses stochastic optimization techniques, by drawing random mini-batches of the input data in a large number of successive steps.

The interest of the method depicted here consists in a reduced numerical complexity, as compared to standard GPC. Ensuring that the number of inducing points $k$ is much smaller than the training set (i.e.~$k \ll n$) puts much of the cost on the calculation of the expectation $\mathbb{E}_{q(f_i)}\left[\dotsc\right]$ and of the gradients with respect to the hyperparameters. In any case, there is no requirement for inverting of a $n \times n$ matrix any longer. Model predictions are also numerically efficient, thanks to the approximation of the posterior $\pofc{\vec{f}, \vec{u}}{X, \vec{I}} \rightarrow \pofc{\vec{f}}{\vec{u}} q(\vec{u})$. In contrast to the standard GPC method (Eq.~\ref{eq:prediction_gp}), computing a prediction does not need knowledge of the values of the training set $(X, \vec{I})$; only the inducing points (along with all other hyperparameters) are required. Saving a trained model on disk is thus very memory efficient.

It is important to remind here that several approximations lead to the method outlined above, which results from a compromise between efficiency and accuracy. In particular,  \citet{2018Blei} discuss open problems related to variational inference and the use of KL divergence to set a lower bound to the evidence. For our own application we rely on a posteriori validation of the models (e.g.~Sect.~\ref{sect:gp_internal_validation}). In general, the expectation values of the GPC models reflect well the cluster detection rates, but we find the associated variance to be slightly underestimated.

	\subsection{Setup of our GPC model construction}

Let us now relate the previous notations to the setup of our numerical experiments, as described in Sect.~\ref{sect:gp_models}.
The number of training points is the number of simulated clusters in the training set\footnote{This number reduces to $n = 4.8 \times 10^5$ when training models for the cosmology sample, as we restrict simulated clusters to the range $0.05<z<0.85$.}, $n=5.7 \times 10^5$. The number of features may take values ranging from $m=2$ (e.g.~in the case of models based on $M_{500}$ and $z$) to $m=6$ (e.g.~for models with $L_X, z, EM_0, N_H, T_{\rm exp}$ and bkg). We choose a small number of inducing points $k=30$, thus reaching an empirical trade-off between representativeness of the input data space, computational efficiency and smoothness of the model.
The baseline model for cosmological studies has $m=5$, this leads to optimizing 651 hyperparameters. This number decomposes into: 6 for the kernel ($\sigma$ and the five length-scales), $30 \times 5$ coordinates of the inducing points $\vec{Z}$, 30 for the values of the $\vec{m}$ parameter and $30 \times 31/2$ for the coefficients of the $L$ matrix.
Before training, each feature in the training set $X$ is rescaled to the $[0, 1]$ interval using simple functions (e.g.~linear or logarithmic), selected according to the dynamic range and distribution of each feature.

Given the large number of free hyperparameters, we used the ADADELTA algorithm \citep{adadelta} as implemented in the \texttt{climin} library. Our training set is extremely unbalanced, with only a few percent of the simulated clusters listed as detected or selected. In this context, we have found that stochastic optimization does provide satisfactory results only if mini-batches are of large enough size. In our case we choose random mini-batches containing $2^{16} = 65536$ simulated clusters, among them a few hundreds are detected systems. Therefore only about 7 different mini-batches are available. We secured a sufficiently high number of iterations (typically of order $10^4$) before stopping the training phase. In the course of our model production, we have noticed some models converge to undesired solutions, as apparent from e.g.~unusual best-fit length-scales. This behaviour was in many cases due to unlucky draws of the first mini-batches, containing an uncommon fraction of detected systems. In such failure cases, launching a new training with a fresh random seed would generally provide a satisfactory solution.

The training phase of a classifier took typically a few hours on a standard multi-core machine. Once finalized, each model went through the list of diagnostic tests listed in Sect.~\ref{sect:gp_internal_validation}. Each satisfactory model was shared in the eRASS1 collaboration after dumping onto disk the values of the hyperparameters and of the factors rescaling the features to the $[0, 1]$ interval. Prediction of the probabilistic values $\pofc{I^*}{\vec{x}^*}$ is a very quick operation, even for large numbers of test points $\vec{x}^*$ (Eq.~\ref{eq:prediction_finale_gp}). This  is in fact a requirement of the cosmological modelling \citep[][and Eq.~\ref{eq:number_counts_equation}]{2024Ghirardini} to obtain computationally efficient response of the selection function models. A crude estimate of the uncertainties on the model outputs may be obtained by folding the $\pm 1\sigma$ enveloppe of the latent function posterior (given by the variance in Eq.~\ref{eq:gp_conditioning}) into the link function $\Phi$.

%%%%%%%%%%%%%%%%%% Appendix : mode models
\section{Additional selection models with intermediate variables}
\label{app:more_models}

This section presents additional models discussed throughout this paper, constructed from intermediate variables as discussed in Sect.~\ref{sect:gp_models}.

		\subsection{Absorbed flux and redshift}

\begin{figure}
   \centering
   \includegraphics[width=\linewidth]{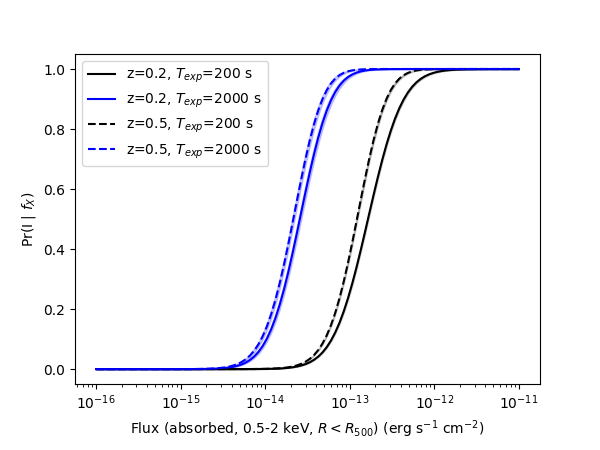}
   \caption{Representation of a model $\pofc{\isdet{main}}{f_X, z, N_H, T_{\rm exp}, {\rm bkg}}$ for fixed values of cluster redshift $z=0.2$ and $z=0.5$, local galactic absorption $N_H$, a nominal background level, two values of exposure times $T_{\rm exp}$ (units s) and a range of flux values (x-axis). The shaded regions indicate the approximate 68\% confidence range output of the model.}
    \label{fig:GP_nH_Texp_simbkg_fx_z_AllXGOOD_cut_0-0_SEED0189}
\end{figure}

The absorbed $R_{500}$ flux in the 0.5--2\,keV band is an alternative to the unabsorbed count-rate. It has units \fluxunit{} and it is available for all simulated clusters from the twin simulation. Fig.~\ref{fig:GP_nH_Texp_simbkg_fx_z_AllXGOOD_cut_0-0_SEED0189} shows slices through this model $\pofc{\isdet{main}}{f_X, z, N_H, T_{\rm exp}, {\rm bkg}}$. It has very similar behaviour as a selection expressed with $CR$ (Fig.~\ref{fig:GP_RateFLwoNH0223_nH_simbkg_Texp_z_Cosmo_cut_6-0_SEED000}).

		\subsection{Integrating morphological features}

Closer inspection of twin simulations reveals that cluster flux (or count rate, or luminosity) is not the only determinant of the detection probability. Also its extent, characterized by $R_{500}$ and its central emission measure, characterized by $EM_0$, play a secondary role. We have \citep[see also][]{2020Comparat}:
\begin{equation}
	EM(R) = - \log_{10} \int_{- \infty}^{+\infty} n_e n_p \dd l
\end{equation}
The integration is along the line of sight (coordinate $l$, expressed in units Mpc) at distance $R$ from the cluster centre, $n_e$ and $n_p$ are the electron and proton densities in the galaxy cluster (expressed in units cm$^{-3}$). For numerical stability reasons, the central emission measure $EM_0$ is defined as the average of $EM(R)$ in the radial range $0<R<0.025\,R_{500}$ and the integration bounds are set to $\pm 20\, R_{500}$ instead of infinity.

\begin{figure}
   \centering
   \includegraphics[width=\linewidth]{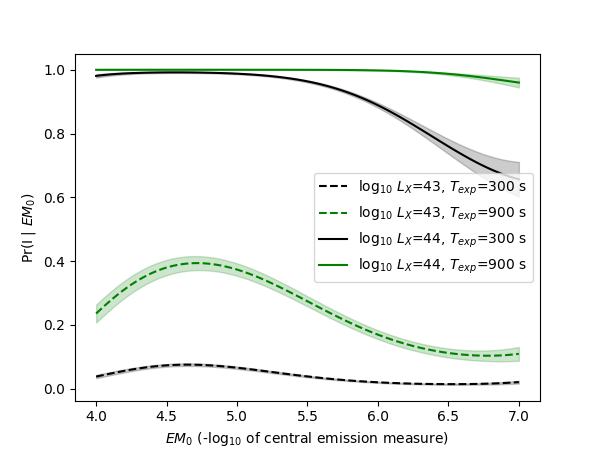}
   \caption{Similar figure as Fig.~\ref{fig:GP_nH_Texp_simbkg_fx_z_AllXGOOD_cut_0-0_SEED0189}, representing the outcome of a model $\pofc{\isdet{main}}{L_{X}, z, EM_0, N_H, T_{\rm exp}, {\rm bkg}}$ for fixed values of cluster redshift $z$, two different values for the luminosity $L_X$, and local galactic absorption $N_H$, a nominal background level and two values of exposure time $T_{\rm exp}$. The shaded regions indicate the approximate 68\% confidence range output of the model.}
    \label{fig:GP_nH_Texp_simbkg_lx_z_EM0_AllXGOOD_cut_0-0_SEED0189}
\end{figure}

Morphological features intervene non-trivially in the selection function. At fixed $R_{500}$ luminosity, a larger cluster size $R_{500}$ dilutes the photons over a larger area, thus making detection of faint objects increasingly difficult for eSASS; very bright extended objects are affected differently, since detection is split into a number of smaller sources, not necessarily classified as extended by eSASS and requiring thorough cleaning \citep{2024Bulbul}.
%But very small extents are also problematic since eSASS would classify the source as point-like and it will not be selected as a cluster.
Similarly, low values of $EM_0$ indicate peaked surface brightness profiles, decreasing the probability of selecting a cluster as extended. Large values of $EM_0$ indicate very flat cores, and thus difficulties in detecting such faint objects against the background. It is clear however that these trends depend also on the background level, cluster flux, redshift, exposure time. The GPC formalism comes particularly handy in constructing a model without imposing strong priors on the exact role of these parameters.
Fig.~\ref{fig:GP_nH_Texp_simbkg_lx_z_EM0_AllXGOOD_cut_0-0_SEED0189} shows four slices through this multi-dimensional model, as a function of the parameter $EM_0$ and for two different cluster luminosities, two different exposure times.

\begin{figure}
   \centering
   \includegraphics[width=\linewidth]{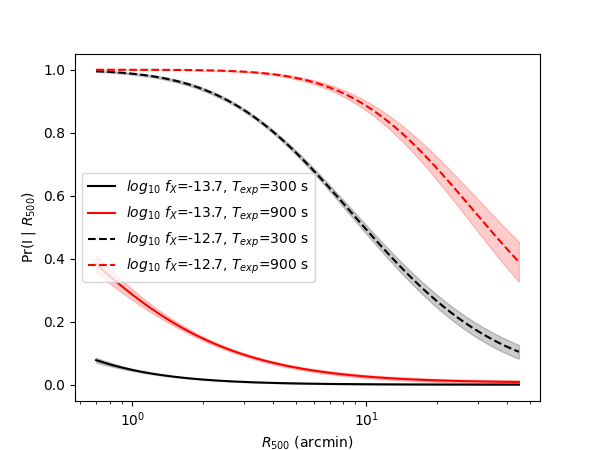}
   \caption{Similar as Fig.~\ref{fig:GP_nH_Texp_simbkg_fx_z_AllXGOOD_cut_0-0_SEED0189}, representing the outcome of a model $\pofc{\isdet{main}}{f_{X}, R_{500}, N_H, T_{\rm exp}, {\rm bkg}}$ for two different values for the flux $f_X$, fixed local galactic absorption $N_H$, a nominal background level and two values of exposure time $T_{\rm exp}$. The shaded regions indicate the approximate 68\% confidence range output of the model.}
    \label{fig:GP_nH_Texp_simbkg_fx_R500am_AllXGOOD_cut_0-0_SEED0189}
\end{figure}

Also on Fig.~\ref{fig:GP_nH_Texp_simbkg_fx_R500am_AllXGOOD_cut_0-0_SEED0189} we show the dependence on $R_{500}$ at fixed flux. The larger the apparent cluster extent (here the projected $R_{500}$ in units arcmin), the lower the probability of selecting the cluster in the sample. This is a consequence of diluting the photons emitted by the cluster over a larger area, hence decreasing the average surface brightness and therefore the ability to distinguish extended emission from background fluctuations.

\end{appendix}

\end{document}